\documentclass[aps,prd,11pt,tightenlines,final,superscriptaddress,
   preprintnumbers,showpacs,showkeys]{revtex4}    

\usepackage[latin1]{inputenc}                    
\usepackage{graphicx}                            
\usepackage{latexsym}                            
\usepackage{amsfonts}                            
\usepackage{amssymb}                             
\usepackage{amsmath}                             
\usepackage[mathscr]{eucal}                      
\usepackage{dcolumn}                             
\usepackage{theorem}                             
\usepackage{epsfig}
\usepackage{slashed}
\newcommand{\cO}{{\cal O}}
\newcommand{\half}{\mbox{\small $\frac{1}{2}$}}
\newcommand{\third}{\mbox{\small $\frac{1}{3}$}}
\newcommand{\quarter}{\mbox{\small $\frac{1}{4}$}}
\newcommand{\Dd}[1]{\overset{\leftrightarrow}{D}_{#1}}
\newcommand{\Dl}[1]{\overset{\leftarrow}{D}_{#1}}
\newcommand{\Dr}[1]{\overset{\rightarrow}{D}_{#1}} 
\newcommand{\MS}{{\overline{\mathrm{MS}}}}
\newcommand{\RI}{{\mathrm{RI}^\prime - \mathrm {MOM}}}
\newcommand{\MOM}{{\mathrm {MOM}}}
\newcommand{\momt}{\widetilde{\mathrm {MOM}}\mathrm{gg}}
\newcommand{\csw}{c_{\mbox{\tiny SW}}}

\begin{document}

\preprint{
\vbox{
\hbox{DESY 10-040} 
\hbox{Edinburgh 2010/08}
\hbox{Liverpool LTH 870}
}}

\title[Perturbative and Nonperturbative Renormalization]%
{Perturbative and Nonperturbative Renormalization in Lattice QCD}

\author{M.~G\"ockeler}\affiliation{Institut f\"ur Theoretische Physik,
  Universit\"at Regensburg, 93040 Regensburg, Germany}
\author{R.~Horsley}\affiliation{School of Physics and Astronomy, 
  University of Edinburgh, Edinburgh EH9 3JZ, UK}
\author{Y.~Nakamura}\altaffiliation{Present address: 
  Center for Computational Sciences, University of Tsukuba, 
  Tsukuba, Ibaraki 305-8577,Japan}
  \affiliation{Institut f\"ur Theoretische Physik,
  Universit\"at Regensburg, 93040 Regensburg, Germany}
\author{H.~Perlt}\affiliation{Institut f{\"u}r Theoretische Physik,
  Universit{\"a}t Leipzig, 04109 Leipzig, Germany}
\author{D.~Pleiter}\affiliation{Deutsches Elektronen-Synchrotron DESY \\
  and John von Neumann-Institut f\"ur Computing NIC, 15738  Zeuthen, Germany}
\author{P.E.L.~Rakow}\affiliation{Theoretical Physics Division, 
  Department of Mathematical Sciences, University of Liverpool, 
  Liverpool L69 3BX, UK}
\author{A.~Sch\"afer}\affiliation{Institut f\"ur Theoretische Physik,
  Universit\"at Regensburg, 93040 Regensburg, Germany}
\author{G.~Schierholz}\affiliation{Institut f\"ur Theoretische Physik,
  Universit\"at Regensburg, 93040 Regensburg, Germany}
\affiliation{Deutsches Elektronen-Synchrotron DESY, 22603 Hamburg, Germany}
\author{A.~Schiller}\affiliation{Institut f{\"u}r Theoretische Physik,
  Universit{\"a}t Leipzig, 04109 Leipzig, Germany}
\author{H.~St\"uben}
\affiliation{Konrad-Zuse-Zentrum f\"ur Informationstechnik Berlin, 
             14195 Berlin, Germany}
\author{J.M.~Zanotti}\affiliation{School of Physics and Astronomy, 
University of Edinburgh, Edinburgh EH9 3JZ, UK}
\collaboration{QCDSF/UKQCD Collaborations}\noaffiliation

\begin{abstract}
We investigate the perturbative and nonperturbative renormalization 
of composite operators in lattice QCD restricting ourselves to operators
that are bilinear in the quark fields (quark-antiquark operators).
These include operators which are relevant to the calculation of 
moments of hadronic structure functions. The nonperturbative 
computations are based on Monte Carlo simulations with two flavors
of clover fermions and utilize the
Rome-Southampton method also known as the RI-MOM scheme. We compare
the results of this approach with various estimates from lattice 
perturbation theory, in particular with recent two-loop calculations.
\end{abstract}




\maketitle

\section{Introduction}

The investigation of hadron structure has become a central topic 
of lattice QCD. In many cases this involves the evaluation of
matrix elements of local operators between hadron states.
For example, (moments of) generalized parton distributions can be
extracted from matrix elements of quark-antiquark operators, i.e., 
operators composed of a quark field, its adjoint and a number of 
gluon fields entering through covariant derivatives which act on 
the quark fields.
In general such operators have to be renormalized. In this process
the operator of interest may receive contributions also from other 
operators, i.e., it may mix with these additional operators.
On the lattice, mixing occurs more frequently than in the continuum 
due to the restricted space-time symmetries. 
Since in the end one wants to make contact with phenomenological 
studies, which almost exclusively refer to operators renormalized 
in the $\MS$ scheme of dimensional regularization, one needs the 
renormalization factors leading from the bare operators on the 
lattice to the $\MS$ operators in the continuum.

The most straightforward approach towards the calculation of
renormalization factors is based on lattice 
perturbation theory (for a review see Ref.~\cite{stefano}). 
Unfortunately, this method meets with some difficulties.
First, perturbation theory on the lattice is computationally much
more complex than in the continuum and therefore the calculations
rarely extend beyond one-loop order (see, however, 
Refs.~\cite{mason,panagopoulos1,panagopoulos2}). 
Second, lattice perturbation theory usually converges rather slowly 
so that the accuracy of perturbative renormalization factors
is limited. Identifying one
source of these poor convergence properties, Lepage and Mackenzie
proposed as a remedy the so-called tadpole improved perturbation
theory~\cite{lepmac}. Still, considerable uncertainty remains. 
Third, mixing with operators of lower dimension cannot be treated by
perturbation theory.

In special cases, when the renormalization factors contain no
ultraviolet divergences, a nonperturbative determination is possible
with the help of Ward identities~\cite{bochi}. However,
there are many interesting operators that cannot be renormalized by
this method.

A general nonperturbative approach to renormalization
has been developed within the Schr\"odinger functional 
scheme (see, e.g., Refs.~\cite{sf1,sf2}, reviews are given in
Refs.~\cite{sommer1,sommer2}). In this method the finite 
size of the lattices employed in the simulations (along with 
appropriate boundary conditions in Euclidean time) is used to set 
the renormalization scale. 
In the end continuum perturbation theory is employed to convert the 
results from the Schr\"odinger functional scheme to the $\MS$ scheme. 
Though theoretically appealing the practical implementation of the 
procedure requires a lot of effort and has to be repeated for every 
new operator again from the very beginning.

Another nonperturbative method for computing renormalization
coefficients of arbitrary quark-antiquark operators is the 
Rome-Southampton method (also known as the RI-MOM scheme)
introduced in Ref.~\cite{marti}.
It mimics the procedure used in continuum perturbation theory.
The basic objects are quark two-point functions with an insertion 
of the operator under consideration at momentum zero. These
are computed in a suitable gauge, e.g., the Landau gauge.
In continuum perturbation theory the two-point functions are 
calculated order by order in an expansion in
powers of the strong coupling constant while in the Rome-Southampton 
method they are evaluated within a Monte Carlo simulation on the lattice. 
In order to extract from these data renormalization factors 
which yield renormalized operators in the $\MS$ 
scheme in the continuum limit one needs a
renormalization condition which is applicable to lattice
data as well as to perturbative continuum results. A suitable condition
has been given in Ref.~\cite{marti}. 

Compared with the Schr\"odinger functional approach the 
Rome-Southampton method is distinguished by its relatively simple 
implementation. Furthermore,
one can deal with all desired operators in a single
simulation. On the other hand, the 
Schr\"odinger functional method is explicitly gauge invariant, while
the Rome-Southampton method requires gauge fixing. 

In a previous publication~\cite{reno} we have performed an 
extensive study of nonperturbative renormalization for a 
variety of quark-antiquark operators using the Rome-Southampton 
method, motivated by our investigations of hadron structure
functions. This was done with Wilson fermions in the quenched 
approximation for two values of the lattice spacing. Later on,
these studies have been extended to improved Wilson fermions,
based on quenched simulations at three values of the lattice 
spacing~\cite{timid}. Meanwhile we are using gauge field 
configurations generated with two flavors of dynamical quarks,
which has made a reconsideration of renormalization necessary.

In this paper we present results for renormalization factors
obtained within the Rome-Southampton approach with $n_f=2$ 
dynamical quarks. We work with nonperturbatively $O(a)$-improved 
Wilson fermions (clover fermions). The operators are, 
however, not (yet) improved. We continue to apply the momentum sources
introduced in Ref.~\cite{reno}. In addition, we have refined 
our approach, subtracting lattice artefacts through one-loop 
boosted perturbation theory. As in Ref.~\cite{reno} we consider
only flavor-nonsinglet quark-antiquark operators. For some 
thoughts concerning flavor-singlet operators see
Refs.~\cite{qmass,fermilab}.

The paper is organized as follows: After introducing in 
Sec.~\ref{sec.operators} the operators to be studied we explain the
method of nonperturbative renormalization in Sec.~\ref{sec.method}.
Our implementation of this method employing momentum sources
is described in Sec.~\ref{sec.numerics}. After a brief overview
over our gauge field configurations in Sec.~\ref{sec.configs} we
discuss the chiral extrapolation of our data in Sec.~\ref{sec.chiex}.
Section~\ref{sec.pert} reviews formulae from continuum perturbation 
theory that will be needed in the analysis. Results from lattice 
perturbation theory are compiled in Sec.~\ref{sec.latpert}. 
Section~\ref{sec.subtraction} explains how we apply lattice 
perturbation theory in order to subtract lattice artefacts.
In Sec.~\ref{sec.fit} we describe our method of 
extracting the renormalization factors from the Monte Carlo data.
The results (perturbative as well as nonperturbative) are then 
presented and discussed in Sec.~\ref{sec.results}. Finally,
we present our conclusions in Sec.~\ref{sec.concl}. Some technical
details are explained in the Appendices.

\section{The operators} \label{sec.operators}

In the Euclidean continuum we want to study the operators 
\begin{eqnarray}
  \cO _{\mu \mu_1 \cdots \mu_n} & = &  
        \bar{u}\gamma_{\mu} \Dd{\mu_1} \cdots \Dd{\mu_n} d \,, \\
  \cO ^5_{\mu \mu_1 \cdots \mu_n} & = &   
        \bar{u}\gamma_\mu \gamma_5 \Dd{\mu_1} \cdots \Dd{\mu_n} d \,, \\
  \cO ^T_{\mu \nu \mu_1 \cdots \mu_n} & = &   
        \bar{u}\sigma_{\mu \nu} \Dd{\mu_1} \cdots \Dd{\mu_n} d 
\end{eqnarray}
(with $\sigma_{\mu \nu} = (\mathrm i/2) [\gamma_\mu , \gamma_\nu]$
and $\Dd{\mu}= \Dr{\mu} - \Dl{\mu}$) 
or rather O(4) irreducible multiplets with definite charge conjugation 
parity. In particular, we obtain twist-2 operators by 
symmetrizing the indices and subtracting the traces. 
We have given the quark fields definite flavors (assumed to be 
degenerate) in order to make apparent that
we are considering the flavor-nonsinglet case. Hence the twist-2
operators do not mix and are multiplicatively renormalizable. 

Working with Wilson fermions it is straightforward to write
down lattice versions of the above operators. One simply replaces the
continuum covariant derivative by its lattice analogue. However, O(4)
being restricted to its finite subgroup H(4) (the hypercubic group) on the
lattice, the constraints imposed by space-time symmetry are less
stringent than in the continuum and the possibilities for mixing increase
\cite{capi,roma,grouptheory,pertz}. 

While the H(4) classification for operators $\cO _{\mu \mu_1 \cdots \mu_n}$
and $\cO ^5_{\mu \mu_1 \cdots \mu_n}$ with $n \leq 3$ has been treated in 
detail in Ref.~\cite{grouptheory}, we have to refer to Ref.~\cite{tensor}
for the operators $\cO ^T_{\mu \nu \mu_1 \cdots \mu_n}$. Note however that
the classification of the latter operators for $n \leq 2$ can be derived from
the results presented in Ref.~\cite{grouptheory}.

In our investigations of hadronic matrix elements we have considered
the following operators whose renormalization factors have already been
studied in Ref.~\cite{reno}:
\begin{eqnarray}
\label{opdefv2a}
\cO_{v_{2,a}}  & = &   \cO _{\{14\}} \,,  \\
\cO_{v_{2,b}}  & = &   \cO _{\{44\}} - \third ( \cO _{\{11\}}
                    + \cO _{\{22\}} + \cO _{\{33\}} ) \,,   \\    
\cO_{v_3}      & = &   \cO _{\{114\}} - \half ( \cO _{\{224\}}
                    + \cO _{\{334\}} )  \,,    \\    
\cO_{v_4}      & = &   \cO _{\{1144\}} + \cO _{\{2233\}}
                    - \cO _{\{1133\}} - \cO _{\{2244\}} \,, \\    
\cO_{a_2}      & = &   \cO ^5_{\{124\}} \,,   \\
\cO_{r_{2,a}}  & = &   \cO ^5_{\{14\}} \,,  \\
\cO_{r_{2,b}}  & = &   \cO ^5_{\{44\}} - \third ( \cO ^5_{\{11\}}
                    + \cO ^5_{\{22\}} + \cO ^5_{\{33\}} ) \,,   \\ 
\cO_{r_3}      & = &   \cO ^5_{\{114\}} - \half ( \cO ^5_{\{224\}}
                    + \cO ^5_{\{334\}} )  \,.
\end{eqnarray}
Their labels refer to the structure function moments that they determine. 
These operators have been selected such that they have a definite 
transformation behavior under H(4) (i.e., belong to an irreducible 
multiplet) with as little mixing as possible. Moreover we have tried
to minimize the number of nonzero momentum components required 
in the evaluation of the hadronic matrix elements. Note, however,
that in the numerical simulations reported in Ref.~\cite{timid} 
different operators (though from the same H(4) multiplets) 
have been used for the matrix elements $v_3$ and $v_4$.

The operators $\cO_{v_{2,a}}$ and $\cO_{v_{2,b}}$     
transform according to inequivalent representations of H(4), although
they belong to the same irreducible O(4) multiplet in the continuum.
Therefore their renormalization factors calculated on the lattice need not
coincide. The same remark applies to $\cO_{r_{2,a}}$ and $\cO_{r_{2,b}}$.

Since our matrix element calculations now involve additional operators,
not considered in Ref.~\cite{reno},
we have extended the above list by the following operators, again guided 
by the H(4) classification given in Refs.~\cite{grouptheory,tensor}:
\begin{eqnarray}
\cO_{v_{3,a}}  & = &  \cO _{\{124\}} \,,   \\
\cO_{h_{1,a}}  & = & \cO ^T_{1 \{23\}} \,,  \\ 
\cO_{h_{1,b}}  & = & \cO ^T_{122} - \cO ^T_{133} \,,  \\ 
\cO_{h_{2,a}}  & = & \cO ^T_{4 \{123\}} \,,  \\ 
\cO_{h_{2,b}}  & = & \cO ^T_{1 \{122\}} - \cO ^T_{1 \{133\}} 
                      + \cO ^T_{2 \{233\}}  \,,  \\ 
\cO_{h_{2,c}}  & = & \cO ^T_{13 \{23\}} + \cO ^T_{23 \{13\}} 
                    + \cO ^T_{41 \{24\}} + \cO ^T_{42 \{14\}} \,,  \\ 
\cO_{h_{2,d}}  & = & \cO ^T_{1211} - \cO ^T_{1222}
                    +  \cO ^T_{13 \{23\}} + \cO ^T_{23 \{13\}} 
                    - \cO ^T_{41 \{24\}} - \cO ^T_{42 \{14\}} \,.  
\end{eqnarray}
The operator $\cO_{v_{3,a}}$ yields the same structure function moment
as $\cO_{v_3}$. However, in contrast to $\cO_{v_3}$ it cannot mix 
with any operator of the same or lower dimension. On the other hand,
it has the disadvantage that one needs spatial momenta with two 
nonvanishing components in order to extract the moment $v_3$ from its
matrix elements. The latter fact is the reason why we did not employ
it in our previous investigations of nucleon structure. The operators
constructed from $\cO ^T_{\cdots}$ are relevant for
transversity.

Furthermore, we have studied the following operators without derivatives
(``currents''):
\begin{eqnarray}
\cO^S  & = &   \bar{u} d \,, \\
\cO^P  & = &   \bar{u} \gamma_5 d \,, \\
\cO^V_\mu  & = &   \bar{u} \gamma_\mu d \,, \\
\cO^A_\mu  & = &   \bar{u} \gamma_\mu \gamma_5 d \,, \\
\cO^T_{\mu \nu}  & = &   \bar{u} \sigma_{\mu \nu} d \,,
\end{eqnarray}
where all quark fields are taken at the same lattice point. 
Finally we have also considered the quark wave function renormalization
constant $Z_q$.

In Table~\ref{tab.ops} we list all 
operators studied, along with the H(4)
representation they belong to and their charge conjugation parity $C$.

While in the evaluation of hadronic matrix elements the members of a 
given operator multiplet require different momentum components to be
nonzero and hence are of different usefulness, such distinctions do not
matter in our computation of renormalization factors. Therefore we
consider not only individual operators but also complete operator 
bases for the representations under study. The representations studied
and the chosen bases are given in Appendix~\ref{sec.bases}.

Concerning the mixing properties a few remarks are in order. 
Mixing with operators of equal or lower dimension is excluded for the
operators $\cO_{v_{2,a}}$, $\cO_{v_{2,b}}$, $\cO_{v_{3,a}}$,
$\cO_{a_2}$, $\cO_{r_{2,a}}$, $\cO_{r_{2,b}}$, 
$\cO_{h_{1,a}}$, $\cO_{h_{1,b}}$, 
$\cO_{h_{2,a}}$, $\cO_{h_{2,b}}$, $\cO_{h_{2,c}}$
as well as for the currents. 

The case of the operator $\cO_{v_3}$, 
for which there are two further operators with the same
dimension and the same transformation behavior, is discussed in 
Refs.~\cite{grouptheory,pertz}. Similarly, $\cO_{h_{2,d}}$ could mix
with another operator of the same dimension.
The operators $\cO_{v_4}$, $\cO_{r_3}$,
on the other hand, could in principle mix not only with operators of 
the same dimension but also with an operator of one dimension less 
constructed from $\cO ^T_{\cdots}$. A few more details on the mixing
issue can be found in Ref.~\cite{timid}, in particular in Appendix B.

Our analysis ignores mixing completely. This seems to be 
justified for $\cO_{v_3}$. Here a perturbative calculation gives a rather
small mixing coefficient for one of the mixing operators \cite{roma,pertz}, 
whereas the other candidate for mixing does not appear at all in a one-loop
calculation of quark matrix elements at momentum transfer zero, 
because its Born term vanishes in forward direction. 
The same is true for all operators of dimension less or equal to 6
which transform identically to $\cO_{v_4}$: Their Born terms vanish
in forward matrix elements, hence they do not show up in a one-loop
calculation at vanishing momentum transfer. 
In the case of $\cO_{r_3}$, however, the mixing with an
operator of lower dimension is already visible at the one-loop level
even in forward direction. Nevertheless, the nucleon matrix elements 
of the operators mixing with $\cO_{v_3}$ and $\cO_{v_4}$ seem to be 
small, at least in the quenched approximation~\cite{timid}.

\begin{table}
\caption{Operators and their transformation 
behavior~\cite{grouptheory,tensor}. The charge conjugation parity
is denoted by $C$.}
\label{tab.ops}
\begin{ruledtabular}
\begin{tabular}{cccccc}
Op.  & Repr.  & $C$ & Op.  & Repr.  & $C$ \\
\hline
$\cO^S$          &  $\tau^{(1)}_1$  &  $+1$ & $\cO_{h_{1,b}}$  &  $\tau^{(8)}_1$  &  $+1$  \\
$\cO^P$          &  $\tau^{(1)}_4$  &  $+1$ & $\cO_{v_3}$      &  $\tau^{(8)}_1$  &  $-1$  \\
$\cO^V_\mu$      &  $\tau^{(4)}_1$  &  $-1$ & $\cO_{v_{3,a}}$  &  $\tau^{(4)}_2$  &  $-1$  \\
$\cO^A_\mu$      &  $\tau^{(4)}_4$  &  $+1$ & $\cO_{r_3}$      &  $\tau^{(8)}_2$  &  $+1$  \\
$\cO^T_{\mu \nu}$&  $\tau^{(6)}_1$  &  $-1$ & $\cO_{a_2}$      &  $\tau^{(4)}_3$  &  $+1$  \\
$\cO_{v_{2,a}}$  &  $\tau^{(6)}_3$  &  $+1$ & $\cO_{h_{2,a}}$  &  $\tau^{(3)}_2$  &  $-1$  \\
$\cO_{v_{2,b}}$  &  $\tau^{(3)}_1$  &  $+1$ & $\cO_{h_{2,b}}$  &  $\tau^{(3)}_3$  &  $-1$  \\
$\cO_{r_{2,a}}$  &  $\tau^{(6)}_4$  &  $-1$ & $\cO_{h_{2,c}}$  &  $\tau^{(6)}_2$  &  $-1$  \\
$\cO_{r_{2,b}}$  &  $\tau^{(3)}_4$  &  $-1$ & $\cO_{h_{2,d}}$  &  $\tau^{(6)}_3$  &  $-1$  \\
$\cO_{h_{1,a}}$  &  $\tau^{(8)}_2$  &  $+1$ & $\cO_{v_4}$      &  $\tau^{(2)}_1$  &  $+1$  
\end{tabular}
\end{ruledtabular}
\end{table}

\section{The method} \label{sec.method}

We calculate our renormalization constants with the help of the 
procedure proposed by Martinelli et al.~\cite{marti} 
(the Rome-Southampton approach). It follows closely the
definitions used in (continuum) perturbation theory. We work
on a lattice of spacing $a$ and volume $V$ in Euclidean space.
For a fixed gauge let
\begin{equation} \label{Gdef}
 G_{\alpha\beta} (p) = \frac{a^{12}}{V} \sum_{x,y,z} {\rm e}^{- {\mathrm i} 
      p \cdot (x-y) } \langle u_\alpha (x) \cO (z) \bar{d}_\beta (y) \rangle
\end{equation}
denote the nonamputated quark-quark Green function with one insertion
of the operator $\cO$ at momentum zero. It is to be considered as a
$12 \times 12$ matrix in the combined color and Dirac space. 
The corresponding vertex function (or amputated Green function) is
given by 
\begin{equation}
\Gamma (p) = S^{-1} (p) G(p) S^{-1} (p) \,, 
\end{equation}
where for $q=u$ or $q=d$
\begin{equation} \label{Sdef}
 S_{\alpha\beta} (p) = \frac{a^8}{V} \sum_{x,y} {\rm e}^{- {\mathrm i} 
      p \cdot (x-y) } \langle q_\alpha (x) \bar{q}_\beta (y) \rangle 
\end{equation}
denotes the quark propagator.
We define the renormalized vertex function by 
\begin{equation}
\Gamma_{\mathrm R} (p) = Z_q^{-1} Z \Gamma (p)
\end{equation}
and fix the renormalization constant $Z$ by imposing the
renormalization condition
\begin{equation} \label{defz}
 \mbox{\small $\frac{1}{12}$} {\rm tr} \left( \Gamma_{\mathrm R} (p)
   \Gamma_{\mathrm {Born}} (p) ^{-1} \right) = 1 
\end{equation}
at $p^2 = \mu_p^2$, where $\mu_p$ is the renormalization scale. 
So $Z$ can be calculated from the relation
\begin{equation} \label{calcz} 
 Z_q^{-1} Z \mbox{\small $\frac{1}{12}$} {\rm tr} \left( \Gamma (p)
   \Gamma_{\mathrm {Born}} (p) ^{-1} \right) = 1 
\end{equation}
with $p^2 = \mu_p^2$. Here $\Gamma_{\mathrm {Born}} (p) $ is the 
Born term in the vertex function of $\cO$ computed on the lattice, and
$Z_q$ denotes the quark field renormalization constant. 
The latter is taken as 
\begin{equation} \label{defzq}
 Z_q (p) = \frac{ {\rm tr} \left( - {\rm i} \sum_\lambda \gamma_\lambda 
           \sin (a p_\lambda) a  S^{-1} (p) \right) }
           {12 \sum_\lambda \sin^2 (a p_\lambda) } \,, 
\end{equation}
again at $p^2 = \mu_p^2$. Aiming at a mass-independent renormalization 
scheme we finally have to extrapolate the resulting values of $Z$ to 
the chiral limit.

Note that there will be no $O(a)$ lattice artefacts in 
Eq.~(\ref{calcz}), because they come with operators of opposite chirality
in the vertex function, and these drop out when the trace is taken.
Still, matrix elements of the renormalized operators will in general
have $O(a)$ lattice artefacts because the operators are not improved.
Once improved operators are available one can evaluate their 
renormalization factors using the methods described in this paper.

Equations~(\ref{defz}) and (\ref{defzq}) (in the chiral limit)
together define
a renormalization scheme of the momentum subtraction type
which is called $\RI$ scheme~\cite{marti}. Here RI stands for 
``regularization independent''. This nomenclature refers to the fact 
that the definition of the $\RI$ scheme does not depend on a 
particular regularization -- here we have used a lattice
cutoff just for definiteness and because the lattice regularization
will be the basis of our numerical investigations. The $\MS$ scheme, 
on the other hand, can only be defined within dimensional regularization 
and is therefore restricted to perturbation theory.

The $\RI$ scheme differs from the RI-MOM scheme only in the definition
of the quark field renormalization constant, which in the $\RI$ scheme
is more suitable for the numerical evaluation.

In general, the $\RI$ scheme will not agree with any of the momentum
subtraction schemes used in continuum perturbation theory.
It is therefore desirable to convert our results to a more popular 
scheme like the $\MS$ scheme. Another reason for converting to 
the $\MS$ scheme lies in the fact that many of the operators 
discussed in this paper appear in the operator product expansion 
along with the corresponding Wilson coefficients, which are generally 
given in the $\MS$ scheme. Hence we have to perform a finite 
renormalization leading us from the $\RI$ scheme to the $\MS$ scheme
if we want to use our renormalized operators together with the
perturbative Wilson coefficients.
This finite renormalization factor can be computed in continuum
perturbation theory using, e.g., dimensional regularization.
The details needed for the evaluation of this factor will be
discussed in Sec.~\ref{sec.pert}.

If the operator under study belongs to an O(4) multiplet of dimension
greater than 1, i.e., if it carries at least one space-time index, the
trace in Eq.~(\ref{calcz}) will in general depend on the direction 
of $p$. This has the immediate consequence that the 
renormalization condition (\ref{defz}) violates 
O(4) covariance even in the continuum limit. In the continuum, one
can restore O(4) covariance by a summation over the members of 
the O(4) multiplet. On the lattice, each operator when renormalized
according to Eq.~(\ref{calcz}) has in general its own $Z$ factor, and
only after conversion to a covariant scheme all operators in an
irreducible H(4) multiplet will have the same renormalization factor.
However, it is also possible to define a common $Z$ factor for all 
members of such an H(4) multiplet already in the RI-MOM framework
by taking a suitable average. 
If $j=1,2,\ldots,N$ labels the members of the chosen basis 
of the multiplet we can average over this basis and calculate $Z$ from
\begin{equation} \label{calczav}
Z_q^{-1} Z \frac{1}{N} \sum_{j=1}^N \mbox{\small $\frac{1}{12}$}
{\rm tr} \left( \Gamma_j (p)
   \Gamma_j^{\mathrm {Born}} (p) ^{-1} \right) = 1 
\end{equation}
with $p^2 = \mu_p^2$. This procedure has two advantages. 
It is simpler than working with a different renormalization factor
for every single operator, and it leads to a smoother dependence
of the results on $p^2$, because it reduces the amount of O(4)
violation.

The bases actually used in our calculations are given in
Appendix~\ref{sec.bases}. 
Whenever we want to refer to this averaging procedure we shall write
the respective operator with a bar on top, i.e., $\cO_{v_{2,a}}$
means precisely the operator (\ref{opdefv2a}) while 
$\overline{\cO}_{v_{2,a}}$ refers to a result for the operator 
multiplet (\ref{avdefv2a}), and analogously for the currents.

Ideally, the scale $\mu_p$ at which our renormalization constants are 
defined should satisfy the conditions
\begin{equation} \label{window}
  1/L^2 \ll \Lambda^2_{\mathrm {QCD}} \ll \mu_p^2 \ll 1/a^2 
\end{equation}
on a lattice with linear extent $L$. The inequality 
$\Lambda^2_{\mathrm {QCD}} \ll \mu_p^2$ should ensure that we can safely
use (continuum) perturbation theory to transform our results from one 
scheme to another. The inequality $\mu_p^2 \ll 1/a^2$ is supposed to keep
discretization effects small. So we have to find a way between the
Scylla of difficult to control nonperturbative effects and the Charybdis
of lattice artefacts. Whether in a concrete calculation 
the conditions (\ref{window}) may be considered as fulfilled remains to
be seen. 

Let us finally comment on our notation for the renormalization scale. 
In the case of a general scheme $\mathcal S$ we use the letter $M$, 
in the case of the $\MS$ scheme we use $\mu$. We take $\mu_p$ when
dealing with the $\RI$ scheme and $\mu_M$ in the case of the MOM scheme
to be defined below. 

\section{Numerical implementation} \label{sec.numerics}

Let us sketch the main ingredients of our calculational 
procedure~\cite{reno}.
To simplify the notation we set the lattice spacing $a=1$ in this section.
Moreover we suppress Dirac and color indices.
In a first step the gauge field configurations are numerically
fixed to some convenient gauge, the Landau gauge in our case \cite{mandula}. 
Representing the operator under study in the form
\begin{equation}
\sum_z \cO (z) = \sum_{z,z^\prime} \bar{q}(z) J(z,z^\prime) q(z^\prime)
\end{equation}
we calculate the nonamputated Green function (\ref{Gdef}) as the
gauge field average of the quantity
\begin{equation} \label{greenf}
 \hat{G}(p) = \frac{1}{V}
   \sum_{x,y,z,z^\prime} {\rm e}^{- {\mathrm i} p \cdot (x-y) } 
    \hat{S}(x,z) J(z,z^\prime) \hat{S}(z^\prime,y) \,,
\end{equation}
constructed from the quark propagator $\hat{S}$ on the same gauge 
field configuration.
Working in the limit of exact isospin invariance we do not have 
to distinguish between $u$ and $d$ propagators. 
With the help of the relation
\begin{equation}
 \hat{S}(x,y) = \gamma_5 \hat{S}(y,x)^+ \gamma_5
\end{equation}
we rewrite $\hat{G}(p)$ as  
\begin{equation}
 \hat{G}(p) = \frac{1}{V} \sum_{z,z^\prime} \gamma_5 
  \left( \sum_x \hat{S}(z,x) {\rm e}^{{\mathrm i} p \cdot x} \right)^+ 
  \gamma_5 J(z,z^\prime)
  \left( \sum_y \hat{S}(z^\prime,y) {\rm e}^{{\mathrm i} p \cdot y} \right) \,.
\end{equation}
The quantities
\begin{equation}
   \sum_x \hat{S}(z,x) {\rm e}^{{\mathrm i} p \cdot x} 
\end{equation}
appearing in this expression can be calculated by solving the lattice 
Dirac equation with a momentum source:
\begin{equation}
 \sum_z M(y,z)   
   \left( \sum_x \hat{S}(z,x) {\rm e}^{{\mathrm i} p \cdot x} \right) 
   =  {\rm e}^{{\mathrm i} p \cdot y} \,.
\end{equation}
Here $M(x,y)$ represents the fermion matrix. So the number of required 
matrix inversions is proportional to the number of momenta
considered. But the quark propagators, which we need for the amputation
and the computation of the quark wave function renormalization, are 
immediately obtained from the quantities already calculated. 

Strictly speaking, one should evaluate the quark propagators 
going into the calculation of $S(p)$ in Eq.~(\ref{Sdef}) on 
configurations that are statistically independent of those used
for the computation of the Green functions (\ref{Gdef})
in order to avoid unwanted correlations \footnote{We thank F. Niedermayer 
for drawing our attention to this point.}. 
If we calculate these expectation values on two independent
ensembles, statistical fluctuations in the quark propagators
and the Green functions (\ref{Gdef}) are uncorrelated, and
(\ref{calczav}) gives a good estimate of the true $Z$.
If we calculate both expectation values on the same ensemble, the 
fluctuations will be correlated, particularly if the configuration 
number $N_{\mathrm {conf}}$ is small. It can be shown~\cite{efron} 
that this will give a bias proportional to $1/N_{\mathrm {conf}}$. 
In our case statistical fluctuations are very small, because of our 
use of momentum sources, so we do not expect a large problem. Indeed, 
estimating the bias introduced by our procedure (see, e.g., 
Refs.~\cite{bias,efron} for appropriate methods) we confirm this
expectation. 
Note that such correlations exist also in the calculation of hadronic
matrix elements from ratios of correlation functions. However, given
the typical number of configurations used in these studies, the bias
should be very small.

Another computational strategy would be to choose a particular 
location for the operator.
Translational invariance ensures that this will give the same 
expectation value after averaging over all gauge field 
configurations. For this method
we need to solve the Dirac equation with a point source at the
location of the operator and (in the case of extended operators)
for a small number of point sources in the immediate neighbourhood.
For operators with a small number of derivatives the point source
method would require fewer inversions, but it turns out that relying 
on translational invariance increases the statistical errors. 

The required gauge fixing necessarily raises the question of the influence of
Gribov copies. Fortunately, investigations of this problem 
indicate that the fluctuations induced by the Gribov
copies are not overwhelmingly large and may be less important than the
ordinary statistical fluctuations~\cite{gribov1,gribov2} 
(see also Ref.~\cite{zci}).

Since the numerical effort is proportional to the number of momenta,
the proper choice of the momenta considered is of particular importance.
In order to minimize cut-off effects we choose them close to the
diagonal of the Brillouin zone and achieve for most operators
an essentially smooth dependence on the renormalization scale $\mu_p^2$.
It goes without saying that in this way we cannot eliminate lattice 
artefacts completely. However, more sophisticated strategies for 
coping with the cut-off effects such as those suggested
in Refs.~\cite{roiesnel,boucaud} would require the use of far more 
momenta than we can afford when working with momentum sources. 
As the treatment of lattice artefacts is a subtle issue anyway we 
have decided to keep the advantage of small statistical errors 
provided by the above procedure and to deal with the discretization 
errors in a different manner.

A specific lattice artefact is caused by the $O(a)$ chiral symmetry 
breaking term of the quark propagator. In position space this term 
is concentrated at very short distances, for fermion actions obeying 
the Ginsparg-Wilson condition it is even exactly a delta function.
In momentum space it gives rise to the Wilson mass term, 
$\sim a p^2$ in the inverse propagator. The authors of 
Ref.~\cite{ferenc} discuss one method of suppressing this 
artefact (for an earlier treatment of the same effect see 
Ref.~\cite{offshell}). Here, our approach to this problem is
to define $Z_q$ from Eq.~(\ref{defzq}), in which the trace removes the
Wilson mass term, and to suppress the remaining $O(a^2)$
lattice artefacts by using the perturbative subtraction scheme
described in Sec.~\ref{sec.subtraction}.

\section{Monte Carlo ensembles} \label{sec.configs}

\begin{table}
\caption{Simulation parameters $\beta$, $\kappa = \kappa_{\mathrm {sea}}$,
clover coefficient $\csw$ and lattice volume along with the corresponding 
values of the pion mass in lattice units.}
\label{tab.param}
\begin{ruledtabular}
\begin{tabular}{lllll}
\multicolumn{1}{c}{$\beta$}   & \multicolumn{1}{c}{$\kappa$}  & $\csw$ &
\multicolumn{1}{c}{$V$}       & \multicolumn{1}{c}{$a m_\pi$} \\
\hline 
5.20 & 0.1342  & 2.0171 & $16^3 \times 32$ & 0.5847(12)   \\
5.20 & 0.1350  & 2.0171 & $16^3 \times 32$ & 0.4148(13)   \\
5.20 & 0.1355  & 2.0171 & $16^3 \times 32$ & 0.2907(15)   \\
5.25 & 0.1346  & 1.9603 & $16^3 \times 32$ & 0.4932(10)   \\
5.25 & 0.1352  & 1.9603 & $16^3 \times 32$ & 0.3821(13)   \\
5.25 & 0.13575 & 1.9603 & $24^3 \times 48$ & 0.25556(55)  \\
5.25 & 0.1360  & 1.9603 & $24^3 \times 48$ & 0.18396(56)  \\
5.29 & 0.1340  & 1.9192 & $16^3 \times 32$ & 0.5767(11)   \\
5.29 & 0.1350  & 1.9192 & $16^3 \times 32$ & 0.42057(92)  \\
5.29 & 0.1355  & 1.9192 & $24^3 \times 48$ & 0.32696(64)  \\
5.29 & 0.1359  & 1.9192 & $24^3 \times 48$ & 0.23997(47)  \\
5.29 & 0.1362  & 1.9192 & $24^3 \times 48$ & 0.15784(75)  \\
5.40 & 0.1350  & 1.8228 & $24^3 \times 48$ & 0.40301(43)  \\
5.40 & 0.1356  & 1.8228 & $24^3 \times 48$ & 0.31232(67)  \\
5.40 & 0.1361  & 1.8228 & $24^3 \times 48$ & 0.22081(72)  \\
5.40 & 0.13625 & 1.8228 & $24^3 \times 48$ & 0.19053(62)  \\
5.40 & 0.1364  & 1.8228 & $24^3 \times 48$ & 0.1535(13)   \\
\end{tabular}
\end{ruledtabular}
\end{table}

\begin{table}
\caption{Critical hopping parameters $\kappa_c$ along 
with chirally extrapolated values for the Sommer parameter $r_0/a$ 
and the average plaquette $P$.}
\label{tab.paramc}
\begin{ruledtabular}
\begin{tabular}{llll}
\multicolumn{1}{c}{$\beta$}   & \multicolumn{1}{c}{$\kappa_c$}  &
\multicolumn{1}{c}{$r_0/a$}   & \multicolumn{1}{c}{$P$} \\
\hline 
5.20 & 0.136008(15) & 5.454(59)  & 0.538608(49)   \\
5.25 & 0.136250(7)  & 5.880(26)  & 0.544780(89)   \\
5.29 & 0.136410(9)  & 6.201(25)  & 0.549877(109)  \\
5.40 & 0.136690(22) & 6.946(44)  & 0.562499(46)
\end{tabular}
\end{ruledtabular}
\end{table}

We work with two degenerate flavors of nonperturbatively 
improved Wilson fermions (clover fermions). For the explicit
form of the fermionic action see, e.g., Ref.~\cite{offshell}.
As our gauge field action we take Wilson's plaquette action. 
In Table~\ref{tab.param} we collect the parameters of our 
simulations, $\beta$, $\kappa = \kappa_{\mathrm {sea}}$,
the clover coefficient $\csw$ and the lattice volume along 
with $a m_\pi$, the pion mass in lattice units. 
Table~\ref{tab.paramc} contains the critical 
hopping parameters $\kappa_c$ as well as the values of the Sommer 
parameter $r_0/a$ and the average plaquette 
$P = \langle \third \mbox{tr} U_\Box \rangle$ in the
chiral limit~\cite{lambda} which are employed in this paper. 
Note that the results for the chiral extrapolation of $r_0/a$ 
given here are based on a larger set of data than that used in 
Ref.~\cite{lambda}.

The statistical errors will be calculated by means of the jackknife 
procedure. 

\section{Chiral extrapolation} \label{sec.chiex}

\begin{figure}
\begin{center}
\epsfig{file=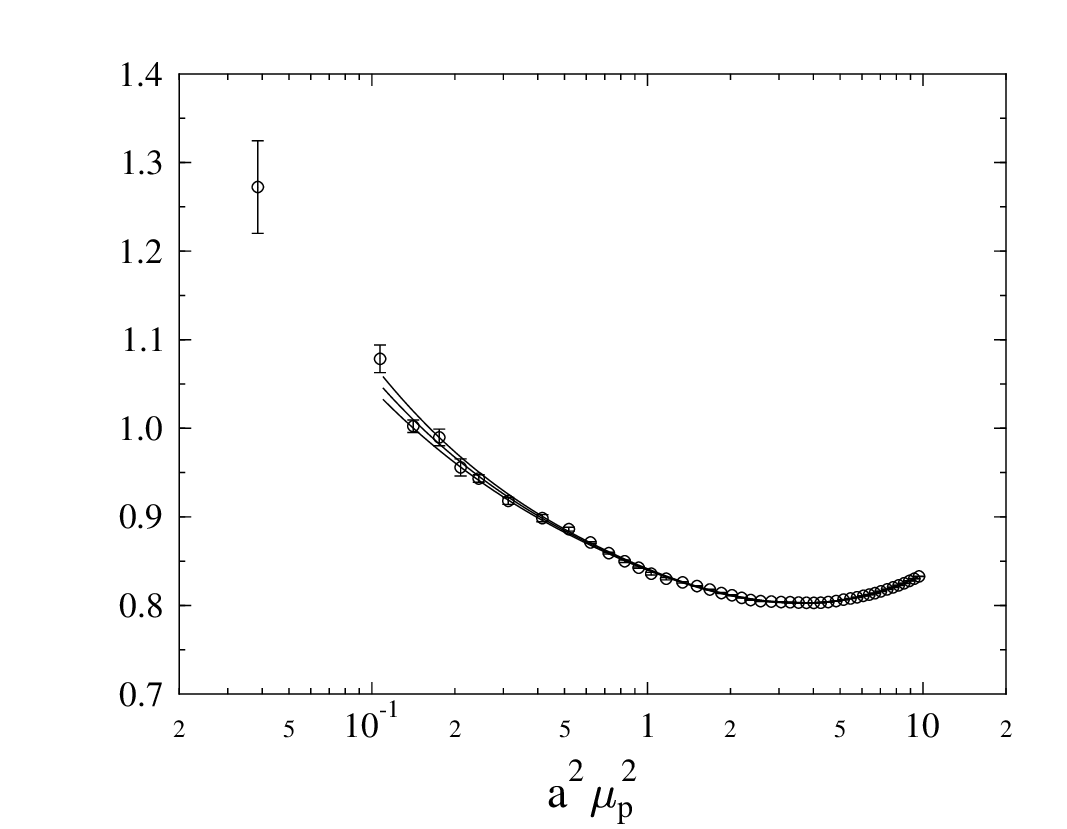,width=11cm}
\end{center}
\caption{$Z^{\RI}$ for the operator $\overline{\cO}_T$ at $\beta = 5.29$,
$\kappa = 0.1362$ on a $24^3 \times 48$ lattice. 
The curves represent splines with two interior knots fitted to the data 
and to the data $\pm$ the statistical error.}
\label{fig.splts}
\end{figure}

As already mentioned in Sec.~\ref{sec.method} we have to extrapolate
our results obtained at nonvanishing quark masses to the chiral limit.
This will be done for each $\beta$ at fixed values of $p^2$. In the cases
where the simulations for different values of $\kappa$ have been
performed on different volumes, i.e., for $\beta = 5.25$ and 
$\beta = 5.29$, the sets of momenta used depend on $\kappa$, and 
some kind of interpolation is required. For this purpose we 
fit the data on the larger lattices ($24^3 \times 48$) with 
cubic splines in $\ln (a^2 p^2)$. 
Except for very small momenta, which will not influence
the final results, these fits yield a very good description of our data.
An example is shown in Fig.~\ref{fig.splts}.

Of course, ``wiggles'' in the data (caused by lattice artefacts)
will be smoothed out by this interpolation. These wiggles are less
pronounced on the larger lattices than on the smaller ones. That is why
we have chosen to work with the momenta coming from the smaller 
lattices so that we can use the data on these lattices directly 
without any interpolation and have to interpolate only the results
obtained on the larger lattices.

Alternatively, one could use the interpolation for all $\kappa$ values.
This however leads to negligible differences in the final
results.

For the chosen momenta we can then extrapolate our data to the
chiral limit. This is done linearly in $(r_0 m_\pi)^2$, i.e., by
means of a fit of the form
\begin{equation}
Z = z_0 + z_1 (r_0 m_\pi)^2 \,,
\end{equation}
where the fit parameter $z_0$ is identified with the desired value of
the renormalization factor in the chiral limit. Note that this
is essentially a linear fit in the quark mass. The ansatz is motivated
by the fact that in perturbation theory the leading quark mass 
dependence is linear as the chiral limit is approached (see, e.g., 
Ref.~\cite{offshell}).

\begin{figure}[tbh]
\begin{center}
\epsfig{file=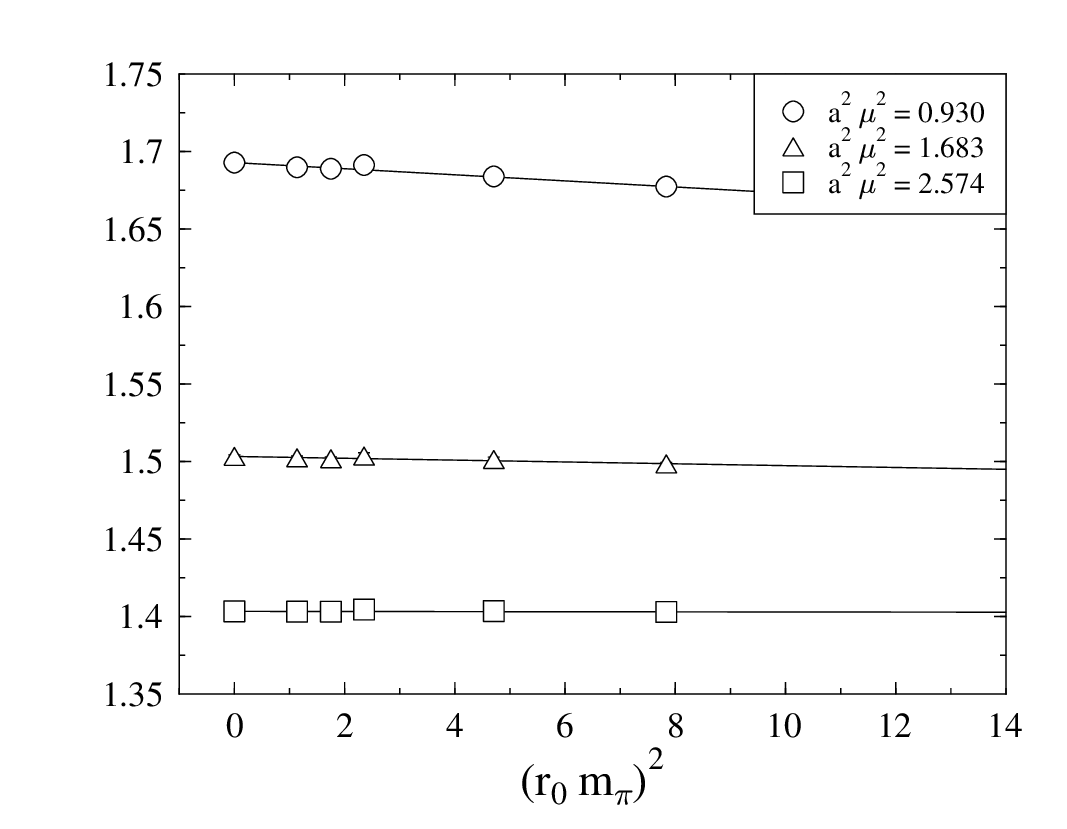,width=11cm}
\end{center}
\caption{Chiral extrapolation for $\cO_{a_2}$ at $\beta = 5.40$.}
\label{fig.chiex_a2}
\end{figure}

\begin{figure}[tbh]
\begin{center}
\epsfig{file=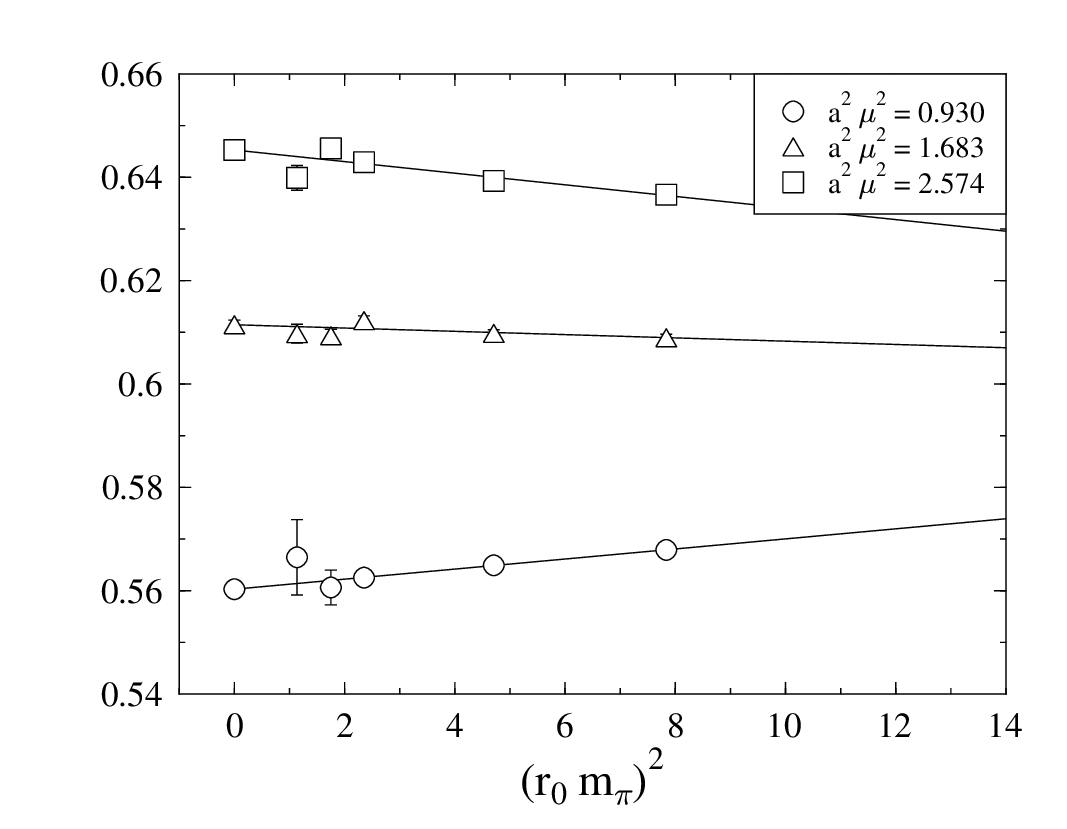,width=11cm}
\end{center}
\caption{Chiral extrapolation for $\cO^S$ (subtracted data, 
as explained in Sec.~\ref{sec.subtraction}) at $\beta = 5.40$.}
\label{fig.chiex_I}
\end{figure}

With the possible exception of the smallest momenta these fits work well.
Examples are shown in Figs.~\ref{fig.chiex_a2}, \ref{fig.chiex_I}.
Nevertheless, we have also performed quadratic extrapolations of the form
\begin{equation} \label{quex}
Z = z_0 + z_1 (r_0 m_\pi)^2 + z_2 (r_0 m_\pi)^4 
\end{equation}
in order to get an idea of the impact of the chiral extrapolation on the 
final results. (Note that this corresponds to a three-parameter fit at 
three data points for $\beta = 5.20$.) 

\begin{figure}[htb]
\begin{center}
\epsfig{file=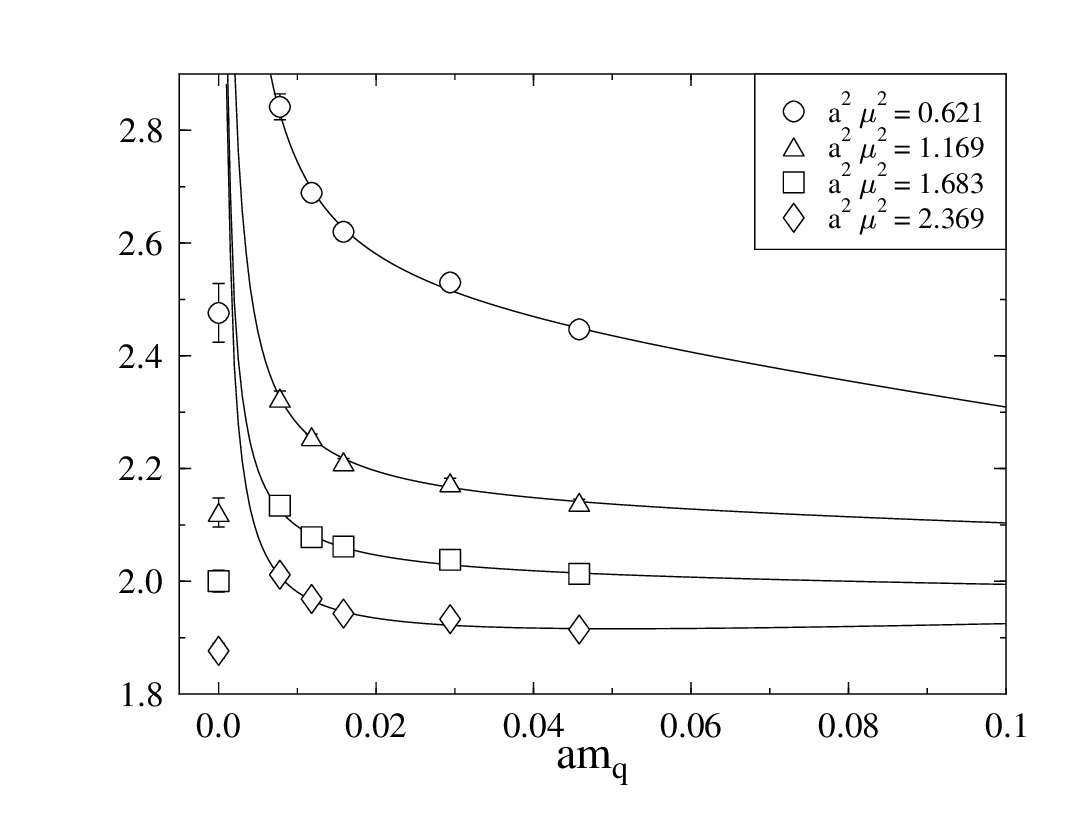,width=11cm}
\end{center}
\caption{Chiral extrapolation of $1/Z_P$ (subtracted data, 
as explained in Sec.~\ref{sec.subtraction}) at $\beta = 5.40$.
The symbols at $am = 0$ represent the chirally extrapolated values, 
i.e., the quantity $s_1$.}
\label{fig.g5chiex}
\end{figure}

However, there is an exceptional case where these simple extrapolations
are not trustworthy. This is the pseudoscalar density $\cO^P$. In
this case one expects that $Z = Z_P$ vanishes with the quark mass $m_q$,
because $Z_P^{-1}$ develops a pole in $m_q$. Therefore we follow
Ref.~\cite{cudell} and try to subtract the pole contribution using a fit of 
the form
\begin{equation} \label{modchiex}
\frac{1}{Z_P} = s_0 \, \frac{1}{a m_q} + s_1 + s_2 \, a m_q
\end{equation}
with 
\begin{equation}
a m_q = \frac{1}{2} \left( \frac{1}{\kappa} - \frac{1}{\kappa_c} \right) \,. 
\end{equation}
Here $\kappa_c$ is the critical hopping parameter defined for fixed
$\beta$ by the vanishing of the pseudoscalar mass $m_\pi$. The values 
used in this paper can be found in Table~\ref{tab.paramc}. The
fit parameter $s_1$ is then identified with the inverse of $Z_P$ in the 
chiral limit. Examples of such fits are shown in Fig.~\ref{fig.g5chiex}.
The curvature in the data is clearly visible establishing the existence
of the Goldstone pole.

\begin{figure}[htb]
\begin{center}
\epsfig{file=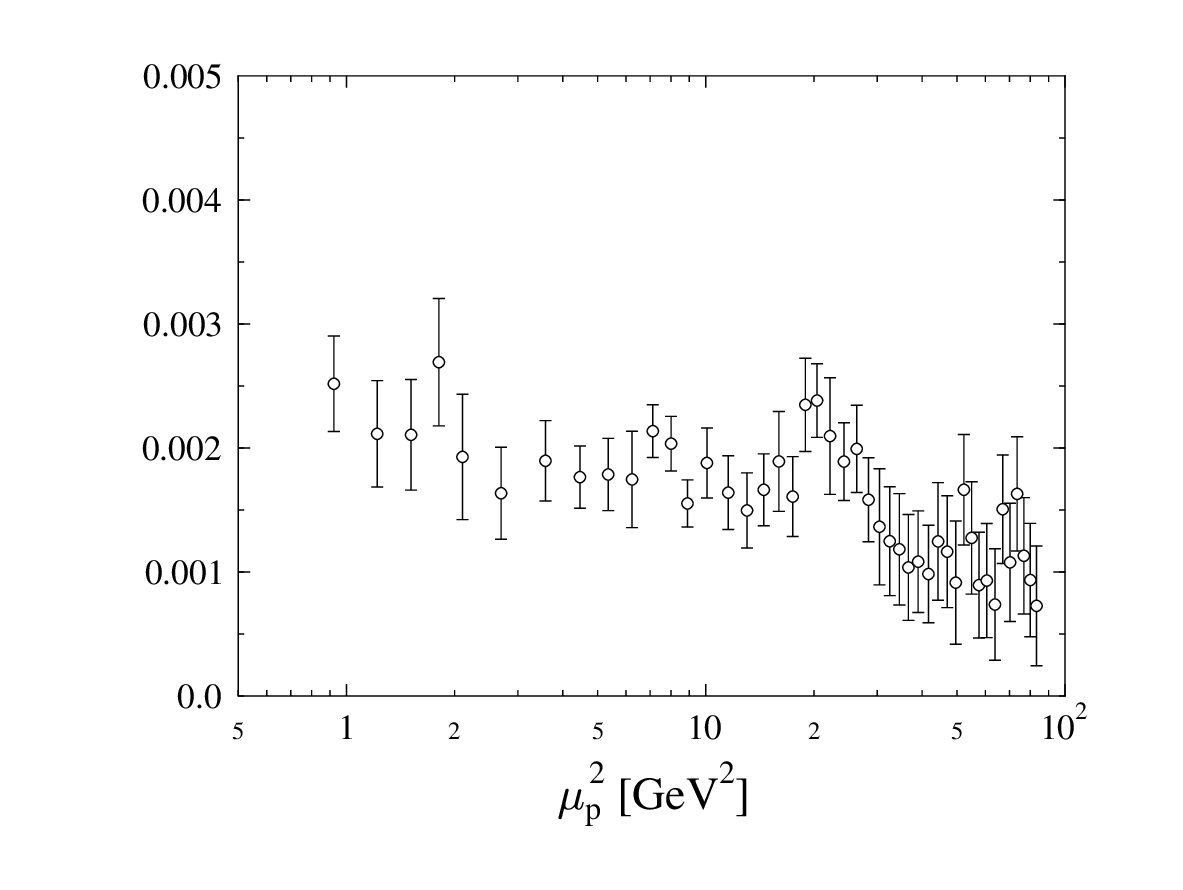,width=11cm}
\end{center}
\caption{$a^2 \mu_p^2 \, s_0$ (from a fit to subtracted data, as 
explained in Sec.~\ref{sec.subtraction})
as a function of $\mu_p^2$ for the pseudoscalar
density at $\beta = 5.40$.} 
\label{fig.g5coeff}
\end{figure}

How can we judge the reliability of the resulting numbers? 
From the operator product expansion~\cite{lane,pagels} we expect
that $s_0$ is inversely proportional to $\mu_p^2$, i.e., 
$\mu_p^2 \, s_0$ should become independent of $\mu_p^2$.
Therefore we plot $a^2 \mu_p^2 \, s_0$ versus $\mu_p^2$ in 
Fig.~\ref{fig.g5coeff}. The $\mu_p^2$ independence seems to be
satisfied with reasonable accuracy. Thus we are confident that 
our extrapolation for $Z_P$ works fairly well. Nevertheless, the 
results for $Z_P$ must be considered with some caution.

\section{Input from continuum perturbation theory} \label{sec.pert}

In Sec.~\ref{sec.method} we have explained how one can compute
nonperturbative renormalization factors leading us from the bare 
lattice operators (at lattice spacing $a$) to renormalized operators 
in the $\RI$ scheme (renormalization scale $\mu_p$). In this section
we collect results from continuum perturbation theory which will be
needed for the conversion to standard renormalization schemes such
as the $\MS$ scheme.

If the operator $\mathcal O$ under study is multiplicatively 
renormalizable the operator renormalized in some scheme 
$\mathcal S$ at the scale $M$ is related to the bare lattice 
operator $\mathcal O_{\mathrm {bare}}$ by
\begin{equation}
\mathcal O^{\mathcal S} (M) = 
Z_{\mathrm {bare}}^{\mathcal S} (M,a) \mathcal O_{\mathrm {bare}} \,.
\end{equation}
The scale dependence of the renormalized operator is determined by the
anomalous dimension
\begin{equation} \label{anodim}
\gamma^{\mathcal S} = - M \frac{\mathrm d}{\mathrm d M}
\ln Z_{\mathrm {bare}}^{\mathcal S} \,.
\end{equation}
Here the derivative is to be taken at fixed bare parameters, and it
is implicitly assumed that the cutoff has been removed in the end. In
perturbation theory $\gamma^{\mathcal S}$ is expanded in powers of some
renormalized coupling constant $g^{\mathcal S} (M)$:
\begin{equation} \label{defgamma}
\gamma^{\mathcal S} = \gamma_0 \frac{g^{\mathcal S} (M)^2}{16 \pi^2}
 + \gamma_1^{\mathcal S} 
       \left( \frac{g^{\mathcal S} (M)^2}{16 \pi^2} \right)^2
 + \gamma_2^{\mathcal S} 
       \left( \frac{g^{\mathcal S} (M)^2}{16 \pi^2} \right)^3
 + \gamma_3^{\mathcal S} 
       \left( \frac{g^{\mathcal S} (M)^2}{16 \pi^2} \right)^4 + \cdots 
\end{equation}
Note that the one-loop coefficient $\gamma_0$ is scheme independent.

Similarly we define the quark field renormalization constant 
$Z^{\mathcal S}_{q,\mathrm {bare}}(M,a)$ in the scheme $\mathcal S$
so that the renormalized quark propagator is given by
$Z^{\mathcal S}_{q,\mathrm {bare}}(M,a) S_{\mathrm {bare}}$. 
In the $\RI$ scheme, $Z^{\RI}_{q,\mathrm {bare}}$ is then specified
by Eq.~(\ref{defzq}) or its continuum analogue. For the
anomalous dimension of the quark field we adopt the definition
\begin{equation} 
\gamma^{\mathcal S}_q = - M \frac{\mathrm d}{\mathrm d M}
\ln Z_{q,\mathrm {bare}}^{\mathcal S} \,.
\end{equation}

The running of the coupling constant $g^{\mathcal S} (M)$ as the 
scale $M$ is varied is controlled by the $\beta$ function
\begin{equation} \label{betadef}
\beta^{\mathcal S} =  M \frac{\mathrm d}{\mathrm d M}
                      g^{\mathcal S} (M) \,.
\end{equation}
Again, the derivative is to be taken at fixed bare parameters and it
is implicitly assumed that the cutoff has been removed in the end. 
The perturbative expansion of the $\beta$ function can be written as
\begin{equation} \label{defbeta}
\beta^{\mathcal S} = - \beta_0 \frac{g^{\mathcal S} (M)^3}{16 \pi^2}
 - \beta_1 \frac{g^{\mathcal S} (M)^5}{(16 \pi^2)^2}
 - \beta_2^{\mathcal S} \frac{g^{\mathcal S} (M)^7}{(16 \pi^2)^3}
 - \beta_3^{\mathcal S} \frac{g^{\mathcal S} (M)^9}{(16 \pi^2)^4}
 + \cdots
\end{equation}
In this case the first two coefficients $\beta_0$ and $\beta_1$
are scheme independent.

Integrating Eq.~(\ref{betadef}) we obtain
\begin{eqnarray}
\frac{M}{\Lambda^{\mathcal S}} & = &  
 \left( \frac{\beta_0}{16 \pi^2} {g^{\mathcal S} (M)^2} \right)
                             ^{\frac{\beta_1}{2 \beta_0^2}}
 \exp \left(\frac{1}{2 \beta_0} \cdot 
              \frac{16 \pi^2}{g^{\mathcal S} (M)^2} \right)
\nonumber \\ & & {} \times  \label{alphacalc}
 \exp \left \{ \int_0^{g^{\mathcal S} (M)} \! \mathrm d g'
  \left( \frac{1}{\beta^{\mathcal S} (g')} 
     + \frac{1}{\beta_0} \frac{16 \pi^2}{g'^3}
       - \frac{\beta_1}{\beta_0^2} \frac{1}{g'} \right) \right \} 
\end{eqnarray}
with the $\Lambda$ parameter $\Lambda^{\mathcal S}$ appearing as an
integration constant.

In the same spirit we define the so-called RGI (renormalization 
group invariant) operator, which is independent of scale and scheme, by 
\begin{equation} 
\mathcal O^{\mathrm {RGI}} 
       = \Delta Z ^{\mathcal S} (M) \mathcal O^{\mathcal S} (M)
       = Z^{\mathrm {RGI}} (a) \mathcal O_{\mathrm {bare}}
\end{equation}
with
\begin{equation} \label{deltaz}
\Delta Z ^{\mathcal S} (M) = 
  \left( 2 \beta_0 \frac {g^{\mathcal S} (M)^2}{16 \pi^2}\right)
                             ^{-\frac{\gamma_0}{2 \beta_0}}
 \exp \left \{ \int_0^{g^{\mathcal S} (M)} \! \mathrm d g'
  \left( \frac{\gamma^{\mathcal S}(g')}{\beta^{\mathcal S} (g')} 
   + \frac{\gamma_0}{\beta_0 g'} \right) \right \} 
\end{equation}
and 
\begin{equation} 
Z^{\mathrm {RGI}} (a) = 
  \Delta Z ^{\mathcal S} (M) Z_{\mathrm {bare}}^{\mathcal S} (M,a) \,,
\end{equation}
where $Z^{\mathrm {RGI}}$ depends only on $a$ (or on the bare 
coupling parameter $\beta$).
Once we know $\mathcal O^{\mathrm {RGI}}$ (or equivalently 
$Z^{\mathrm {RGI}}$), multiplication with $\Delta Z ^{\mathcal S} (M)^{-1}$
will allow us to evaluate $Z_{\mathrm {bare}}^{\mathcal S} (M,a)$
and hence the operator $\mathcal O$ (or rather its matrix elements) 
in any scheme and at any scale we like, provided we know the 
$\beta$ and $\gamma$ functions sufficiently well.

In the two-loop approximation, i.e., setting 
$\beta_n^{\mathcal S} = \gamma_n^{\mathcal S} = 0$ for $n \geq 2$,
one can evaluate the integral in Eq.~(\ref{deltaz}) easily:
\begin{equation} \label{deltaz2l}
\Delta Z ^{\mathcal S} (M) = 
  \left( 2 \beta_0 \frac {g^{\mathcal S} (M)^2}{16 \pi^2}\right)
                             ^{-\frac{\gamma_0}{2 \beta_0}}
\left( 1 + \frac{\beta_1}{\beta_0} \frac {g^{\mathcal S} (M)^2}{16 \pi^2}
     \right)^{\frac{\gamma_0 \beta_1 - \gamma_1^{\mathcal S} \beta_0}
             {2 \beta_0 \beta_1}} \,.
\end{equation}
Whenever we need them we evaluate the integrals in Eqs.~(\ref{alphacalc})
and (\ref{deltaz}) exactly (by numerical methods) and do not re-expand
them in $g^{\mathcal S} (M)$.

With the help of the methods described in Secs.~\ref{sec.method}
and \ref{sec.numerics} we can compute 
$Z_{\mathrm {bare}}^{\RI} (\mu_p)$ numerically 
for some range of scales $\mu_p$. 
Knowledge of $\Delta Z^\RI (\mu_p)$ would then permit us to 
compute $Z^{\mathrm {RGI}}$. However, being not covariant for most
operators, the $\RI$ scheme is not very suitable for evaluating 
anomalous dimensions. Therefore we will adopt a two-step procedure
for computing $Z^{\mathrm {RGI}}$. In the first step we transform
the numerical results for $Z_{\mathrm {bare}}^{\RI}$ to a covariant 
``intermediate'' scheme $\mathcal S$, e.g., the $\MS$ scheme, 
and in the second step we use the anomalous dimension and the 
$\beta$ function in this scheme to compute $\Delta Z ^{\mathcal S}$
and hence $Z^{\mathrm {RGI}}$. Thus we could in principle
compute $Z^{\mathrm {RGI}} (a)$ as
\begin{equation} \label{compzrgi}
Z^{\mathrm {RGI}} (a) = 
  \Delta Z ^{\mathcal S} (M = \mu_p) Z_{\RI}^{\mathcal S} (M = \mu_p) 
Z_{\mathrm {bare}}^{\RI} (\mu_p,a) \,,
\end{equation}
where $Z_{\RI}^{\mathcal S}$ denotes the finite renormalization factor
leading from the $\RI$ scheme to the scheme $\mathcal S$ and all the 
scales have been identified with the scale $\mu_p$ initially set 
in the $\RI$ scheme. 

The most obvious choice for $\mathcal S$ is of course the $\MS$ scheme. 
However, it will turn out to be advantageous to consider also a kind of 
(perturbative) momentum subtraction scheme, which we call MOM scheme. 
This is defined by requiring 
that in the renormalized vertex function the coefficient of the tree-level 
(or Born) term equals one at the renormalization scale $\mu_M$.
To make this definition unambiguous one has to specify in each case
the basis used for the other contributions.
The quark field renormalization constant
is taken to be the same as in the $\RI$ scheme:
\begin{equation} 
Z^{\MOM}_{q,\mathrm {bare}}(\mu_M,a) = Z^{\RI}_{q,\mathrm {bare}}(\mu_M,a) \,. 
\end{equation}
The MOM scheme is covariant and rather ``close'' to the $\RI$ scheme 
so that the conversion factor $Z_{\RI}^{\MOM}$ from $\RI$ to MOM
usually differs less from one than the factor $Z_{\RI}^{\MS}$ leading from
$\RI$ to $\MS$. 

We can expand the conversion factor leading from $\RI$ to $\MS$
in powers of the $\MS$ coupling constant:
\begin{equation} \label{convri}
Z_{\RI}^{\MS} (\mu) =
1 + c_1 \frac{g^{\MS} (\mu)^2}{16 \pi^2} 
  + c_2 \left(  \frac{g^{\MS} (\mu)^2}{16 \pi^2} \right)^2
  + c_3 \left(  \frac{g^{\MS} (\mu)^2}{16 \pi^2} \right)^3 + \cdots
\end{equation}
Using
\begin{equation} \label{convmom}
Z_{\MOM}^{\MS} (\mu) =
1 + c'_1 \frac{g^{\MS} (\mu)^2}{16 \pi^2} 
  + c'_2 \left(  \frac{g^{\MS} (\mu)^2}{16 \pi^2} \right)^2
  + c'_3 \left(  \frac{g^{\MS} (\mu)^2}{16 \pi^2} \right)^3 + \cdots
\end{equation}
we obtain for the conversion factor from $\RI$ to MOM:
\begin{eqnarray}
Z_{\RI}^{\MOM} (\mu) & = & \frac{Z_{\RI}^{\MS} (\mu)}{Z_{\MOM}^{\MS} (\mu)} 
\nonumber \\
& = & 
1 + (c_1 - c'_1)  \frac{g^{\MS} (\mu)^2}{16 \pi^2}
  + (c_2 - c_1 c'_1 + {c'_1}^2 - c'_2) 
                       \left(  \frac{g^{\MS} (\mu)^2}{16 \pi^2} \right)^2
\nonumber \\
& & \hphantom{1}{}  
+ (c_3 - c_2 c'_1 + c_1 {c'_1}^2 - c_1 c'_2 - {c'_1}^3 + 2 c'_1 c'_2 - c'_3)
       \left(  \frac{g^{\MS} (\mu)^2}{16 \pi^2} \right)^3 + \cdots
\nonumber \\ {} \label{convmom2}
\end{eqnarray}

If the vertex function $\Gamma (p)$ is proportional to the Born term
$\Gamma_{\mathrm {Born}} (p)$ (as it happens in simple cases), the 
$\RI$ scheme and the MOM scheme do not differ. So we have
\begin{equation} 
Z_{\RI}^{\MS} (\mu) = Z_{\MOM}^{\MS} (\mu)
\end{equation}
and consequently $c_i = c'_i$. However, in the generic case the
matrix $\Gamma (p)$ will contain also contributions that are not a 
multiple of $\Gamma_{\mathrm {Born}} (p)$. As we consider only
operators which are multiplicatively renormalizable in the continuum
these additional contributions
are finite, but they make $Z_{\RI}^{\MS}$ different from $Z_{\MOM}^{\MS}$
and are responsible for the dependence of $Z_{\RI}^{\MS}$ on the direction
of the momentum $p$. An explicit example is discussed in 
Appendix~\ref{sec.rexa}.

Working with the MOM scheme it is quite natural to expand in a 
coupling constant which is similarly defined through a momentum subtraction
procedure. Therefore we have also considered the $\momt$ scheme as defined 
in Ref.~\cite{cheret2}. The corresponding coupling constant 
$g^{\momt}$ is related to the $\MS$ coupling constant $g^{\MS}$ by 
\begin{equation}
\frac{g^{\momt} (\mu)^2}{16 \pi^2} = \frac{g^{\MS} (\mu)^2}{16 \pi^2}
 + d_1 \left(  \frac{g^{\MS} (\mu)^2}{16 \pi^2} \right)^2
 + d_2 \left(  \frac{g^{\MS} (\mu)^2}{16 \pi^2} \right)^3
 + d_3 \left(  \frac{g^{\MS} (\mu)^2}{16 \pi^2} \right)^4 + \cdots \,,
\end{equation}
where in the Landau gauge
\begin{eqnarray}
d_1 & = & \frac{70}{3} - \frac{22}{9} n_f \,, \\ 
d_2 & = & \frac{516217}{576} - \frac{153}{4} \zeta_3 - \left( \frac{8125}{54} 
           + \frac{4}{3} \zeta_3 \right) n_f + \frac{376}{81} n_f^2 \,, \\
d_3 & = & \frac{304676635}{6912} - \frac{299961}{64}\zeta_3 
          - \frac{81825}{64} \zeta_5
              + \left(- \frac{13203725}{1296} + \frac{13339}{27} \zeta_3 
                      + \frac{1885}{9} \zeta_5 \right) n_f
              \nonumber \\ & & {}
           + \left( \frac{580495}{972} + \frac{40}{9} \zeta_3 \right) n_f^2
       - \frac{5680}{729} n_f^3 \,.
\end{eqnarray}
As usual, $\zeta_n$ denotes the value of Riemann's $\zeta$ function 
at the argument $n$. Choosing $\mathcal S = \MOM$ in Eq.~(\ref{deltaz})
we shall always work with the $\momt$ coupling.

Using the above expressions, all expansions in the $\MS$ coupling may be 
rewritten as expansions in powers of the $\momt$ coupling. For example,
the coefficients $\gamma_i^{\MOM}$ of the anomalous dimension 
in the MOM scheme expanded in powers of $g^{\momt}$ are related 
to the coefficients $\gamma_i^{\MS}$ of the anomalous dimension 
in the $\MS$ scheme expanded in powers of $g^{\MS}$ by
\begin{eqnarray}
\gamma_1^{\MOM} & = & \gamma_1^{\MS}  - 2 \beta_0 c'_1 
                                       - d_1 \gamma_0  \,, 
\label{conv1} \\
\gamma_2^{\MOM} & = & \gamma_2^{\MS}  - 2 \beta_1 c'_1
         -  2 \beta_0 \left( 2 c'_2 - c_1^{\prime 2} \right)    
         - 2 d_1 \left( \gamma_1^{\MS} - 2 c'_1 \beta_0 \right)
         + \gamma_0 \left( 2 d_1^2 - d_2 \right) \,, 
\label{conv2} \\
\gamma_3^{\MOM} & = & \gamma_3^{\MS}  - 2 \beta_2^{\MS} c'_1
         - 3 d_1 \left[ \gamma_2^{\MS} - 2 c'_1 \beta_1 
                 - 2 \beta_0 \left( 2 c'_2 - c_1^{\prime 2} \right) \right]
         + \left( \gamma_1^{\MS} - 2 \beta_0 c'_1 \right)
                     \left( 5 d_1^2 - 2 d_2 \right)
    \nonumber \\ & & {}
         - 2 \beta_1 \left( 2 c'_2 - c_1^{\prime 2} \right)
         - 2 \beta_0 \left( 3 c'_3 - 3c'_1 c'_2 + c_1^{\prime 3} \right)
         + \gamma_0 \left( 5 d_1 d_2 - 5 d_1^3 - d_3 \right) \,.  
\label{conv3} 
\end{eqnarray}

For the actual evaluation of the expansion coefficients $c_i$ and
$c'_i$ one starts from the bare vertex function for the operator
under consideration computed in dimensional
regularization and imposes the respective renormalization conditions 
yielding the renormalized vertex functions in the different schemes.
These differ only by (scale dependent) factors, from which the
desired conversion factors can be derived immediately, once the
quark field renormalization factor has been extracted from
the quark propagator. The results for the coefficients $c_1$, 
$c_2$, $c_3$ and $c_1^\prime$, $c_2^\prime$, $c_3^\prime$  
as well as the sources from which we have taken the required 
perturbative vertex functions are given in Appendix~\ref{sec.coeff}. 

\section{Lattice perturbation theory} \label{sec.latpert}

As long as no mixing with operators of lower dimension is involved
it is possible to compute renormalization factors in lattice 
perturbation theory. Although straightforward in principle, the actual
calculations tend to become rather cumbersome in practice. Hence
they rarely extend beyond one-loop order (see, however, 
Refs.~\cite{mason,panagopoulos1,panagopoulos2}). This is a severe 
limitation since lattice perturbation theory converges rather slowly
in most cases of interest. Therefore various improvement schemes
have been devised, such as boosted perturbation theory and 
tadpole improvement~\cite{lepmac}.

In spite of these problems we want to compare our nonperturbative 
results with the corresponding values obtained in (improved) 
lattice perturbation theory. For the renormalization factor 
$Z_{\mathrm {bare}}^{\MS} (\mu,a)$ a straightforward application 
of one-loop lattice perturbation theory yields results of the form
\begin{equation} \label{zpert}
Z_{\mathrm {bare}}^{\MS} (\mu,a)_{\mathrm {pert}}
 = 1 - \frac{g^2}{16 \pi^2} ( \gamma_0 \ln (a \mu) + C_F \Delta ) \,,
\end{equation}
where $\Delta = \Delta (\csw)$ is a finite constant depending on the 
details of the lattice action and we have $C_F = 4/3$ for the 
gauge group SU(3). 
If mixing occurs, the single renormalization factor of a 
multiplicatively renormalizable operator is replaced by a 
matrix of $Z$ factors. However, we shall neglect this complication
and restrict ourselves to the matrix element on the diagonal
corresponding to the operator under consideration.
Working with an anticommuting $\gamma_5$ 
also in the continuum part of the calculation one obtains 
the values given in Table~\ref{tab.finco1}. They do not depend 
on the particular operator but only on the H(4) multiplet to which the
operator belongs. In the case $\csw=0$ results for $\cO_{v_4}$, 
$\cO_{h_{1,b}}$, $\cO_{h_{2,b}}$ and $\cO^T_{\mu \nu}$ have 
already been obtained in Ref.~\cite{stefano2}. Note that $\Delta$ 
is gauge invariant for our quark-antiquark operators, however, in 
the case of the quark wave function renormalization constant $Z_q$ 
it is not. The result given in Table~\ref{tab.finco1} corresponds 
to the Landau gauge. 

\begin{table}
\caption{Finite contributions to the renormalization factors 
         $Z_{\mathrm {bare}}^{\MS}$ in lattice perturbation theory.
         The result for the quark field is given in the Landau gauge.}
\label{tab.finco1}
\begin{ruledtabular}
\begin{tabular}{ccc}
Op. & $\Delta (\csw)$ & Ref.\\
\hline
$\cO^S$          &  $ 12.95241 + 7.73792 \csw - 1.38038 \csw^2 $  &
                                                 \cite{offshell} \\
$\cO^P$          &  $ 22.59544 - 2.24887 \csw + 2.03602 \csw^2 $  &
                                                 \cite{offshell} \\
$\cO^V_\mu$      &  $ 20.61780 - 4.74556 \csw - 0.54317 \csw^2 $  &
                                                 \cite{offshell} \\
$\cO^A_\mu$      &  $ 15.79628 + 0.24783 \csw - 2.25137 \csw^2 $  &
                                                 \cite{offshell} \\
$\cO^T_{\mu \nu}$&  $ 17.01808 - 3.91333 \csw - 1.97230 \csw^2 $  &
                                                 \cite{offshell} \\
$Z_q$            &  $ 16.64441 - 2.24887 \csw - 1.39727 \csw^2 $ &
                                                 \cite{offshell} \\
$\cO_{v_{2,a}} $ &  $ 1.27959 - 3.87297 \csw - 0.67826 \csw^2 $  &
                                                 \cite{offshell} \\
$\cO_{v_{2,b}} $ &  $ 2.56184 - 3.96980 \csw - 1.03973 \csw^2 $  &
                                                 \cite{offshell} \\
$\cO_{r_{2,a}} $ &  $ 0.34512 - 1.35931 \csw  - 1.89255 \csw^2 $  &
                                                 \cite{offshell} \\
$\cO_{r_{2,b}} $ &  $ 0.16738 - 1.24953 \csw - 1.99804 \csw^2 $  &
                                                 \cite{offshell} \\
$\cO_{h_{1,a}} $ &  $ 1.25245 - 3.10180 \csw - 1.59023 \csw^2 $  &
                                                 \cite{clovernf} \\
$\cO_{h_{1,b}} $ &  $ 0.52246 - 2.99849 \csw - 1.46224 \csw^2 $  &
                                                 \cite{clovernf} \\
$\cO_{v_3} $     &  $ -12.12740 - 2.92169 \csw  - 0.98166 \csw^2 $ &
                                       \cite{scprivate,timid,clovernf} \\
$\cO_{v_{3,a}} $ &  $ -11.56318 - 2.89800 \csw - 0.98387 \csw^2 $  &
                                                 \cite{clovernf} \\
$\cO_{r_3} $     &  $ -12.86094 - 1.49316 \csw - 1.68673 \csw^2 $ &
                                                 \cite{clovernf} \\
$\cO_{a_2} $     &  $ -12.11715 - 1.51925 \csw - 1.71846 \csw^2 $  &
                                                 \cite{clovernf} \\
$\cO_{h_{2,a}} $ &  $ -11.54826 - 2.41077 \csw - 1.51175 \csw^2 $  &
                                                 \cite{clovernf} \\
$\cO_{h_{2,b}} $ &  $ -11.86877 - 2.30651 \csw - 1.34908 \csw^2 $  &
                                                 \cite{clovernf} \\
$\cO_{h_{2,c}} $ &  $ -11.74773 - 2.36201 \csw - 1.45084 \csw^2 $  &
                                                 \cite{clovernf} \\
$\cO_{h_{2,d}} $ &  $ -12.9268 - 2.38849 \csw - 1.3900 \csw^2 $  &
                                                 this work \\
$\cO_{v_4} $     &  $ -25.50303 - 2.41788 \csw - 1.12826 \csw^2 $ &
                                               \cite{scprivate,timid} 
\end{tabular}
\end{ruledtabular}
\end{table}

In order to obtain the corresponding results in tadpole improved
perturbation theory we write (with $\mu = 1/a$)
for an operator with $n_D$ covariant derivatives
\begin{equation}
  1 - \frac{g^2}{16 \pi^2} C_F \Delta =
  \frac{u_0}{u_0^{n_D}} u_0^{n_D-1}  
           \left( 1 - \frac{g^2}{16 \pi^2} C_F \Delta \right) =
 \frac{u_0}{u_0^{n_D}} 
 \left( 1 - \frac{g_{\mathrm {LAT}}^2}{16 \pi^2} C_F \overline{\Delta} \right) 
       + O(g_{\mathrm {LAT}}^4) \,,
\end{equation}
where
\begin{equation}
  u_0 = \langle \third \mbox{tr} U_\Box \rangle ^{\frac{1}{4}} =
     1 - \frac{g^2}{16 \pi^2} C_F \pi^2 + O(g^4) 
\end{equation}
and
\begin{equation} 
  \overline{\Delta} = \Delta + (n_D - 1) \pi^2 \,. 
\end{equation}
This reflects the fact that one has $n_D$ operator tadpole diagrams
and one leg tadpole diagram, which are of the same magnitude but
contribute with opposite sign. It remains to make a physically 
reasonable choice for the expansion parameter $g_{\mathrm {LAT}}$. 
Here we identify $g_{\mathrm {LAT}}$ with the boosted coupling 
\begin{equation} \label{boosted}
g_\Box = \frac{g}{u_0^2} \,.
\end{equation}

Now we have two options. Either we stay with the expression
(\ref{zpert}) and its tadpole improved analogue  
\begin{equation} \label{zpertti}
Z_{\mathrm {bare}}^{\MS} (\mu,a)_{\mathrm {ti}}
= u_0^{1-n_D} \left[
   1 - \frac{g_{\mathrm {LAT}}^2}{16 \pi^2} ( \gamma_0 \ln (a \mu) + 
          C_F \overline{\Delta} ) \right] 
\end{equation}
or we apply these formulae only at a fixed scale $\mu = \mu_0$ 
(e.g., $\mu_0 = 1/a$) using the
renormalization group to change $\mu$:
\begin{eqnarray} \label{zpert.rg}
Z_{\mathrm {bare}}^{\MS} (\mu,a)_{\mathrm {pert}}^{\mathrm {RG}}
& = & \Delta Z ^{\MS} (\mu)^{-1} \Delta Z ^{\MS} (\mu_0) 
               Z_{\mathrm {bare}}^{\MS} (\mu_0,a)_{\mathrm {pert}} \,, \\
\label{zpertti.rg}
Z_{\mathrm {bare}}^{\MS} (\mu,a)_{\mathrm {ti}}^{\mathrm {RG}}
& = & \Delta Z ^{\MS} (\mu)^{-1} \Delta Z ^{\MS} (\mu_0) 
               Z_{\mathrm {bare}}^{\MS} (\mu_0,a)_{\mathrm {ti}} \,.
\end{eqnarray}
The latter option seems preferable leading to the estimates
$\Delta Z ^{\MS} (\mu_0) Z_{\mathrm {bare}}^{\MS} (\mu_0,a)_{\mathrm {pert}}$
and 
$\Delta Z ^{\MS} (\mu_0) Z_{\mathrm {bare}}^{\MS} (\mu_0,a)_{\mathrm {ti}}$
for $Z^{\mathrm {RGI}}$.
Working in the chiral limit we compute $u_0$ from the chirally 
extrapolated values for $P = u_0^4$ given in Table~\ref{tab.paramc}.
To be consistent with lowest order perturbation theory we
set $\csw = 1$. 

Further improvement can be attempted by tadpole-improved, 
renormalization-group-improved boosted perturbation theory or TRB
perturbation theory~\cite{berlin,timid}. This works as follows.
In Eq.~(\ref{anodim}) we have defined the anomalous dimension
$\gamma^{\mathcal S}$ by differentiating 
$Z_{\mathrm {bare}}^{\mathcal S} $ with respect to the renormalization 
scale $M$ at fixed cutoff and bare parameters. Alternatively one
can keep the renormalized quantities fixed and take the derivative 
with respect to the cutoff, the lattice spacing $a$ in our case. 
Then one obtains
\begin{equation} 
\gamma^{\mathrm {LAT}} = - a \frac{\mathrm d}{\mathrm d a}
\ln Z_{\mathrm {bare}}^{\mathcal S} \,.
\end{equation}
Note that the derivative with respect to $a$ also acts on $\csw$,
unless only the tree-level value $\csw = 1$ is used. The
anomalous dimension $\gamma^{\mathrm {LAT}}$ is to be considered as 
a function of some bare coupling constant 
$g_{\mathrm {LAT}} = g_{\mathrm {LAT}} (a)$.
This could be the usual bare coupling $g$, but for our purposes it
will be more advantageous to work with the boosted coupling 
$g_\Box$. Expanding in $g_{\mathrm {LAT}}$
and recalling that the one-loop coefficient $\gamma_0$ is universal 
we can write
\begin{equation} 
\gamma^{\mathrm {LAT}}(g_{\mathrm {LAT}}) 
= \gamma_0 \frac{g_{\mathrm {LAT}}^2}{16 \pi^2} +
\gamma_1^{\mathrm {LAT}} \left( \frac{g_{\mathrm {LAT}}^2}{16 \pi^2} \right)^2 
                                                              + \cdots
\end{equation}
Similarly we define 
\begin{equation} 
\beta^{\mathrm {LAT}} (g_{\mathrm {LAT}})  
= - a \frac{\mathrm d g_{\mathrm {LAT}}}{\mathrm d a}
=  - \beta_0 \frac{g_{\mathrm {LAT}}^3}{16 \pi^2} 
   - \beta_1 \frac{g_{\mathrm {LAT}}^5}{(16 \pi^2)^2}
   + O(g_{\mathrm {LAT}}^7) \,,
\end{equation}
where $\beta_0$ and $\beta_1$ have the same values as in the $\beta$
function (\ref{betadef}). Expressing $Z^{\mathrm {RGI}} (a) = 
\Delta Z ^{\mathcal S} (M) Z_{\mathrm {bare}}^{\mathcal S} (M,a)$
in terms of $\gamma^{\mathrm {LAT}}$ and $\beta^{\mathrm {LAT}}$ we find
\begin{equation} \label{zrgilat}
Z^{\mathrm {RGI}} = 
  \left( 2 \beta_0 \frac {g_{\mathrm {LAT}}^2}{16 \pi^2}\right)
                             ^{-\frac{\gamma_0}{2 \beta_0}}
 \exp \left \{ \int_0^{g_{\mathrm {LAT}}} \! \mathrm d g_0
  \left( \frac{\gamma^{\mathrm {LAT}}(g_0)}{\beta^{\mathrm {LAT}} (g_0)} 
   + \frac{\gamma_0}{\beta_0 g_0} \right) \right \} \,.
\end{equation}
In the two-loop approximation we obtain
\begin{equation} \label{zrgilat2}
Z^{\mathrm {RGI}} = 
  \left( 2 \beta_0 \frac {g_{\mathrm {LAT}}^2}{16 \pi^2}\right)
                             ^{- \frac{\gamma_0}{2 \beta_0}}
\left( 1 + \frac{\beta_1}{\beta_0} \frac {g_{\mathrm {LAT}}^2}{16 \pi^2}
     \right)^{\frac{\gamma_0 \beta_1 - \gamma_1^{\mathrm {LAT}} \beta_0}
                                             {2 \beta_0 \beta_1}} \,.
\end{equation}
Choosing $g_{\mathrm {LAT}} = g_\Box$ one has 
\begin{equation} \label{g1lat} 
\gamma_1^{\mathrm {LAT}} = \gamma_1^\Box 
= \gamma_1^{\MS} + 2 \beta_0 C_F \Delta(\csw)
                  + 16 \pi^2 \gamma_0 \left( t_1 - \quarter C_F \right)  \,,
\end{equation}
where~\cite{luwe,bode,christou,bode2}
\begin{equation} 
t_1 = 0.4682013 - 
    \left( 0.0066960 - 0.0050467 \csw + 0.0298435 \csw^2 \right) n_f  \,.
\end{equation}
Tadpole improvement finally yields the result in TRB perturbation theory:
\begin{equation} \label{zrgitrb}
Z^{\mathrm {RGI}}_{\mathrm {TRB}} =
u_0^{1-n_D}   \left( 2 \beta_0 \frac {g_\Box^2}{16 \pi^2}\right)
                             ^{-\frac{\gamma_0}{2 \beta_0}}
\left( 1 + \frac{\beta_1}{\beta_0} \frac {g_\Box^2}{16 \pi^2} \right)
^{\frac{\gamma_0 \beta_1 - \gamma_1^\Box \beta_0}
{2 \beta_0 \beta_1} + \pi^2 (1-n_D ) C_F \frac{\beta_0}{\beta_1} } \,.
\end{equation}
Applying this formula we shall set again $\csw = 1$ to be consistent 
with lowest order perturbation theory.

For the operators $\cO^S$, $\cO^P$, $\cO^V_\mu$, $\cO^A_\mu$, and 
$\cO^T_{\mu \nu}$ without derivatives two-loop calculations of the
renormalization factors in lattice perturbation theory have recently 
appeared~\cite{panagopoulos1,panagopoulos2}. The various improvement
schemes can be applied also to these two-loop expressions. However,
the resulting formulae become considerably more complicated. Therefore 
we defer the corresponding discussion to Appendix~\ref{sec.2loops}. 

\section{Perturbative subtraction of lattice artefacts} \label{sec.subtraction}

In the perturbative form (\ref{zpert}) of the renormalization 
factors the lattice spacing $a$ only appears in the logarithm 
(and implicitly in the bare gauge coupling $g$). In the remaining 
contributions the limit $a \to 0$ has been performed at fixed $\mu$
leading to the finite constant $\Delta$. In this way all lattice
artefacts vanishing like powers of $a$ have been eliminated.
However, there is no need to do so. In fact, $a \mu$ is not 
necessarily small in our Monte Carlo results. Hence it is worthwhile 
to keep $a$ finite and to compare the lattice artefacts in the 
perturbative expressions with their nonperturbative counterparts.

To do this we simply write down the one-loop integrals for general 
external momentum $p$ and perform the integrations numerically. 
The integrals can no longer be reduced to a small number of standard 
integrals, they have to be done independently at each value of $p$ and
can only be obtained in numerical form. They will in general not only
depend on $p^2$ but also on the direction of the momentum. 

\begin{figure}[htb]
\begin{center}
\epsfig{file=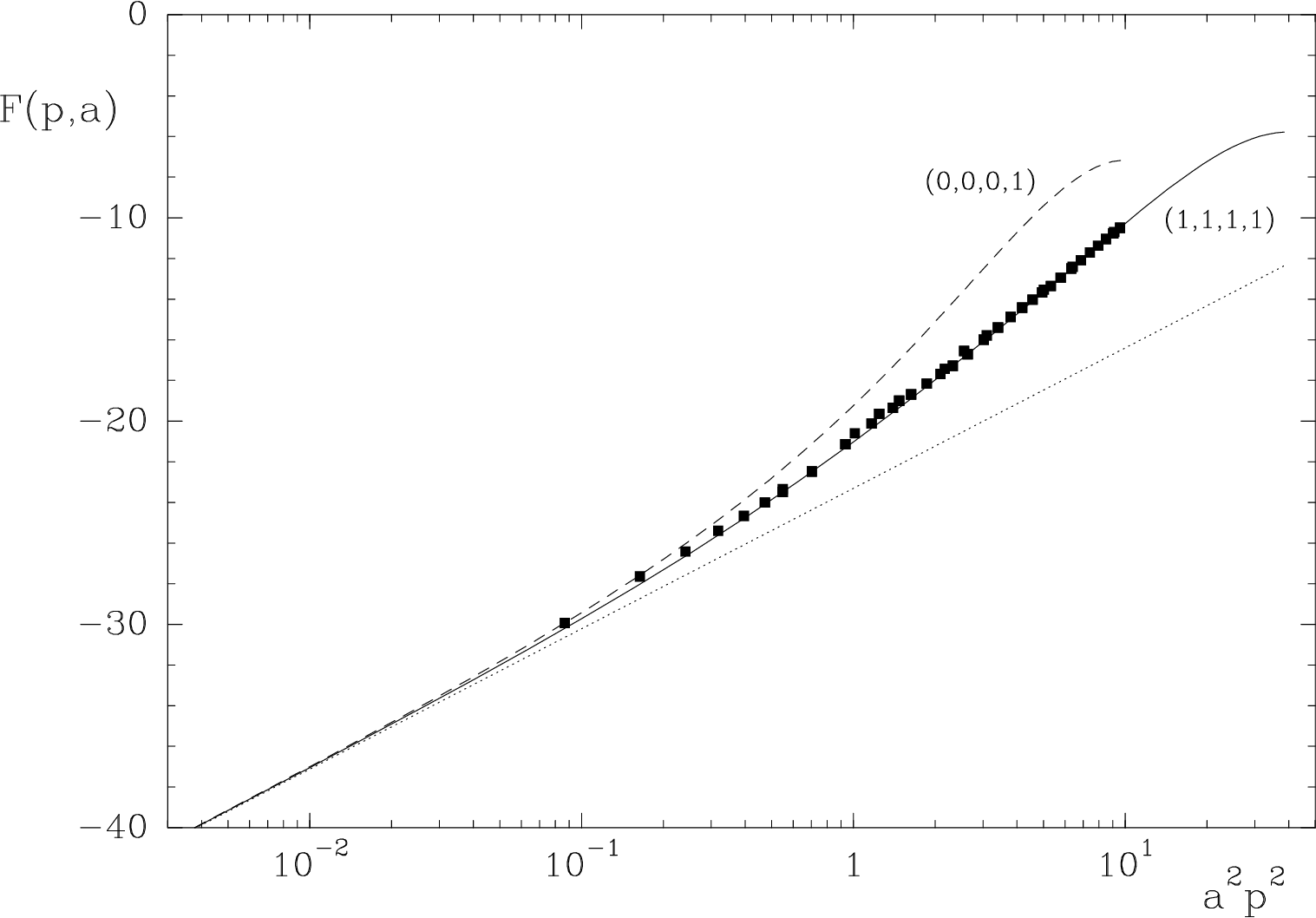,angle=270,width=11cm}
\end{center}
\caption{Lattice artefacts  for the scalar density. The dotted straight
line shows $\tilde{F}(p,a)$, while the two other curves represent
$F(p,a)$ for the momentum directions indicated in the plot.
The black squares denote the values of $F(p,a)$ at the momenta 
used on the $16^3 \times 32$ lattices, which have been chosen close
to the diagonal of the Brillouin zone.}
\label{fig.artefacts}
\end{figure}

For general $p$ we can write the one-loop expression for 
$Z^{\RI}_{\mathrm {bare}}$ in the form
\begin{equation} 
Z^{\RI}_{\mathrm {bare}} (p,a) = 1 + \frac{g^2}{16 \pi^2} C_F F(p,a) 
                                   + O(g^4) \,,
\end{equation}
where the quark mass has been set equal to zero. Neglecting all 
contributions which vanish as $a \to 0$ we get from $F(p,a)$ the expression
$\tilde{F}(p,a)$; e.g., for the scalar density $\cO^S$ it is given by
\begin{equation} 
\tilde{F}(p,a) = 3 \ln (a^2 p^2) - 16.9524 - 7.73792 \csw + 1.38038 \csw^2
\end{equation}
in the Landau gauge. The difference between $F$ and $\tilde{F}$ 
represents the lattice artefacts in one-loop perturbation theory.
Though being $O(a^2)$, $F - \tilde{F}$ can be fairly large for the 
momenta in the actual simulations. An example for the case of the 
scalar density $\cO^S$ is shown in Fig.~\ref{fig.artefacts}. 

We can (and will) use this calculated difference to correct for 
the discretization errors in our lattice data (see 
also Ref.~\cite{damir}). We take
\begin{equation} 
D(p,a) = F(p,a) - \tilde{F}(p,a) 
\end{equation}
as an estimate of the perturbative discretization errors in our 
Monte Carlo renormalization constants 
$Z^{\RI}_{\mathrm {bare}} (p,a)_{\mathrm {MC}}$
and define subtracted renormalization constants by
\begin{equation} 
Z^{\RI}_{\mathrm {bare}} (p,a)_{\mathrm {MC,sub}} 
= Z^{\RI}_{\mathrm {bare}} (p,a)_{\mathrm {MC}} 
   - \frac{g_\Box^2}{16 \pi^2} C_F D(p,a)
\end{equation}
employing boosted perturbation theory with $u_0$ in $g_\Box$
taken at the respective $\kappa$ value, i.e., before the chiral 
extrapolation. Working consistently with one-loop perturbation 
theory we set $\csw = 1$. This procedure removes all the $g^2 a^2$ 
discretization terms in $Z^{\RI}_{\mathrm {bare}} (p,a)_{\mathrm {MC}}$, 
leaving lattice artefacts $O(g^4 a^2)$. As we shall see, the use of the 
boosted coupling $g_\Box$ seems to do a reasonable job of 
estimating the higher-order discretization effects.

Unfortunately, this procedure becomes rather cumbersome for operators
with more than one derivative. So we can use it only for the quark wave
function renormalization, the currents and the operators with one derivative.

\section{Extracting the renormalization factors} \label{sec.fit}

\begin{figure}
\begin{center}
\epsfig{file=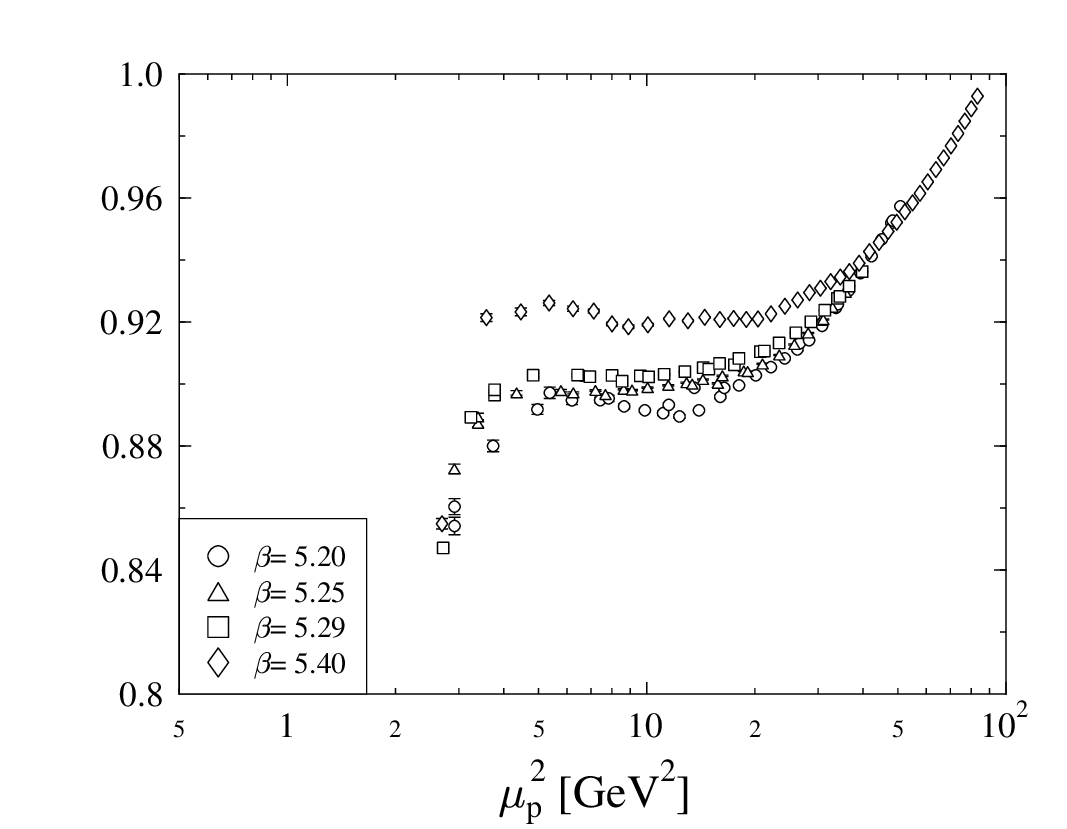,width=11cm} \\
\epsfig{file=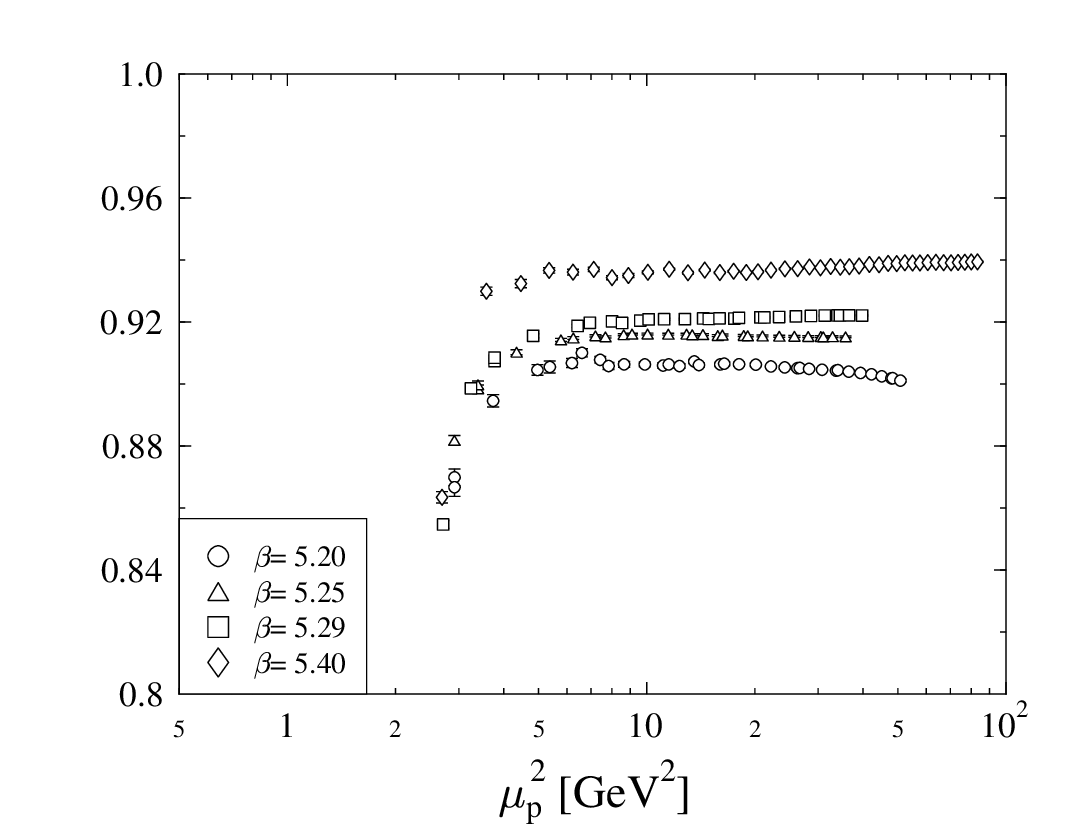,width=11cm}
\end{center}
\caption{$Z^{\mathrm {RGI}}$ for the operator $\overline{\cO}_T$ before 
(upper plot) and after (lower plot) the perturbative subtraction of 
lattice artefacts.}
\label{fig.rgits}
\end{figure}

\begin{figure}
\begin{center}
\epsfig{file=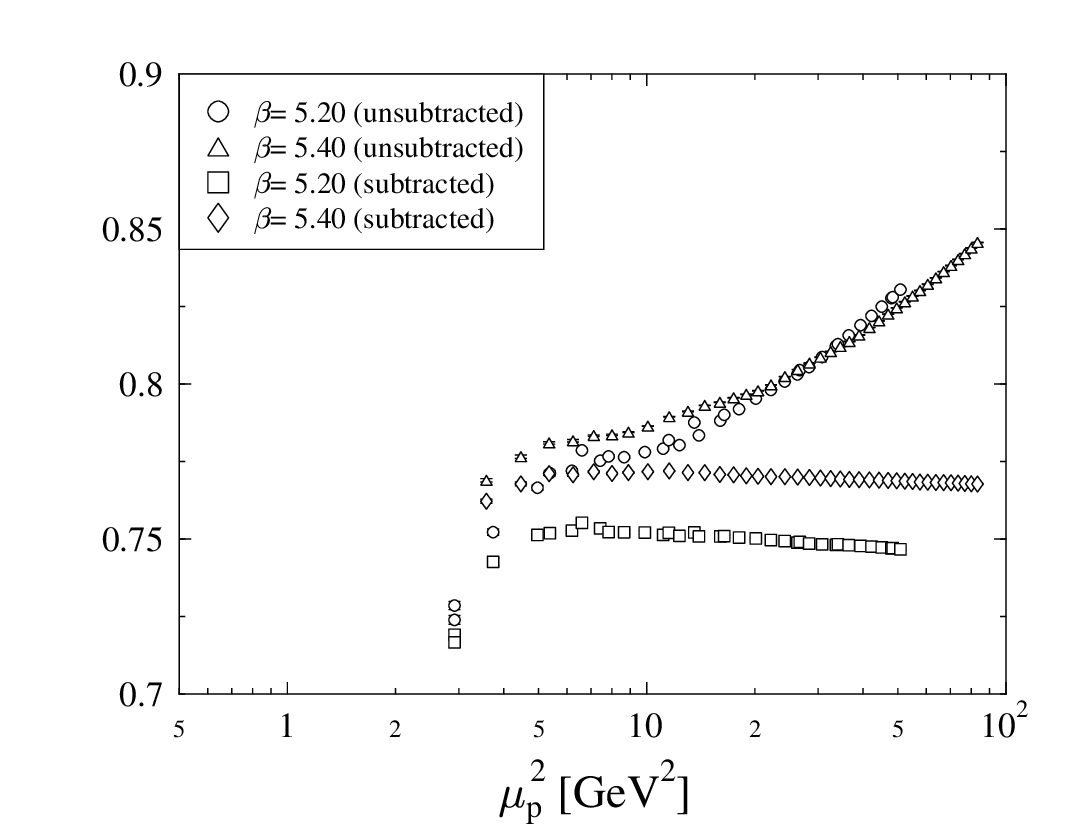,width=11cm}
\end{center}
\caption{$Z^{\mathrm {RGI}}$ for the quark field renormalization 
constant $Z_q$ before and after the perturbative subtraction of 
lattice artefacts.}
\label{fig.rgizq}
\end{figure}

The simplest procedure for obtaining a value of 
$Z^{\mathrm {RGI}} (a)$ would be to plot the right-hand side 
of Eq.~(\ref{compzrgi}), i.e., of the relation
\begin{equation} 
Z^{\mathrm {RGI}} (a) = 
  \Delta Z ^{\mathcal S} (M = \mu_p) Z_{\RI}^{\mathcal S} (M = \mu_p) 
Z_{\mathrm {bare}}^{\RI} (\mu_p,a) \,,
\end{equation}
versus $\mu_p$ and to read off 
$Z^{\mathrm {RGI}} (a)$ in an interval of $\mu_p$ where the 
inequalities (\ref{window}) are satisfied. In this region the
value of $Z^{\mathrm {RGI}} (a)$ would be independent of $\mu_p$,
i.e., one would observe a plateau, and one could determine the
final result by fitting a constant to the data for $Z^{\mathrm {RGI}} (a)$.
Examples of such plots before and after the perturbative subtraction
of lattice artefacts are shown in Fig.~\ref{fig.rgits}, and in
Fig.~\ref{fig.rgizq} subtracted and unsubtracted results for the quark 
field renormalization constant $Z_q$ are directly compared at 
$\beta=5.20$ and $\beta=5.40$.
Equivalently one could fit the values obtained for
$Z_{\mathrm {bare}}^{\mathcal S} (\mu_p,a) 
= Z_{\RI}^{\mathcal S} (\mu_p) Z_{\mathrm {bare}}^{\RI} (\mu_p,a)$ 
in the plateau region by 
$\Delta Z ^{\mathcal S} (\mu_p)^{-1} Z^{\mathrm {RGI}}(a)$ 
(with $Z^{\mathrm {RGI}}(a)$ as fit parameter) using the 
most accurate perturbative expressions for $Z_{\RI}^{\mathcal S}$
and $\Delta Z ^{\mathcal S}$. 

However, in our actual simulations it is not so clear how well the 
inequalities (\ref{window}) are fulfilled, and there are
two effects to be considered that jeopardize the reliability 
of this approach. Firstly, there will be lattice artefacts which 
vanish like powers (up to logarithms) of $a$ for 
$a \to 0$~\cite{roiesnel,boucaud}.
In order to reduce the corresponding contamination one would like to
perform the fit at small values of $\mu_p$.
Secondly, the truncation of the perturbative expansions in 
$\Delta Z ^{\mathcal S}$ and $Z_{\RI}^{\mathcal S}$ will produce
noticeable effects in the region of small $\mu_p$ leading, in
particular, to a dependence of the results on the intermediate
scheme $\mathcal S$. In order to minimize the related uncertainties
one would like to move the fit interval to large values of $\mu_p$.
Because of these conflicting requirements it is a nontrivial matter 
to extract a final value for $Z^{\mathrm {RGI}}$ from the data.

\begin{figure}
\begin{center}
\epsfig{file=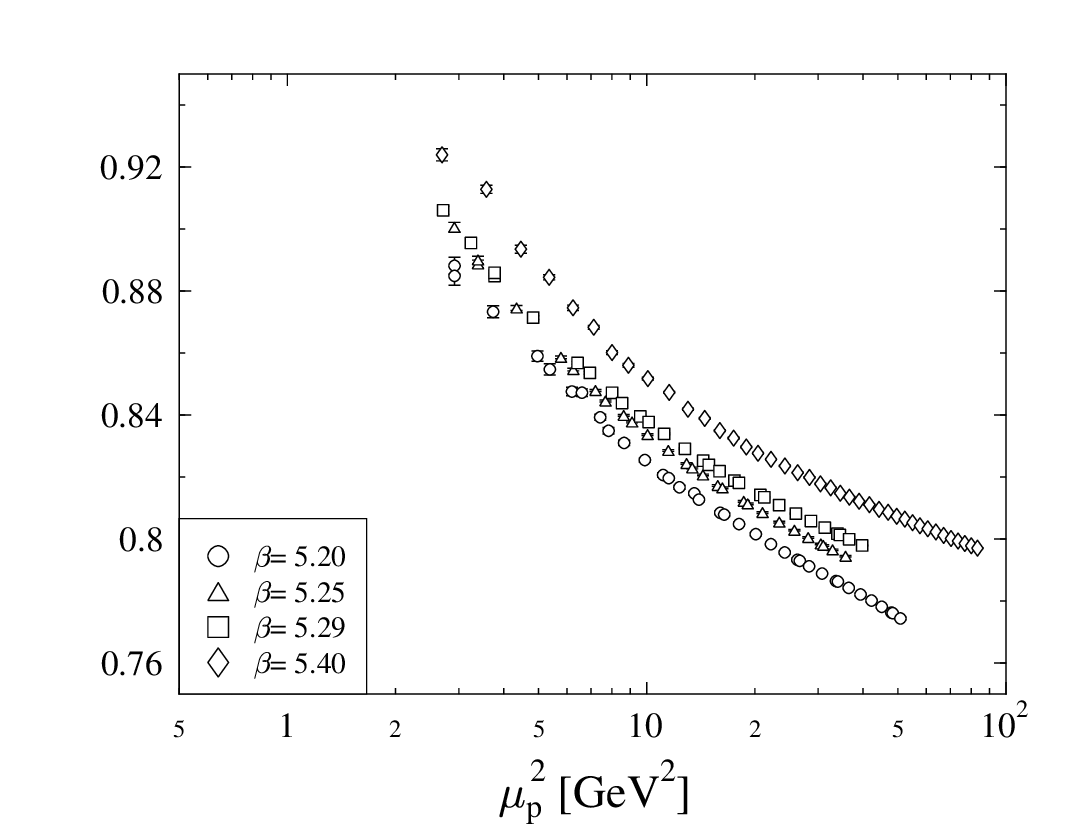,width=11cm} \\
\epsfig{file=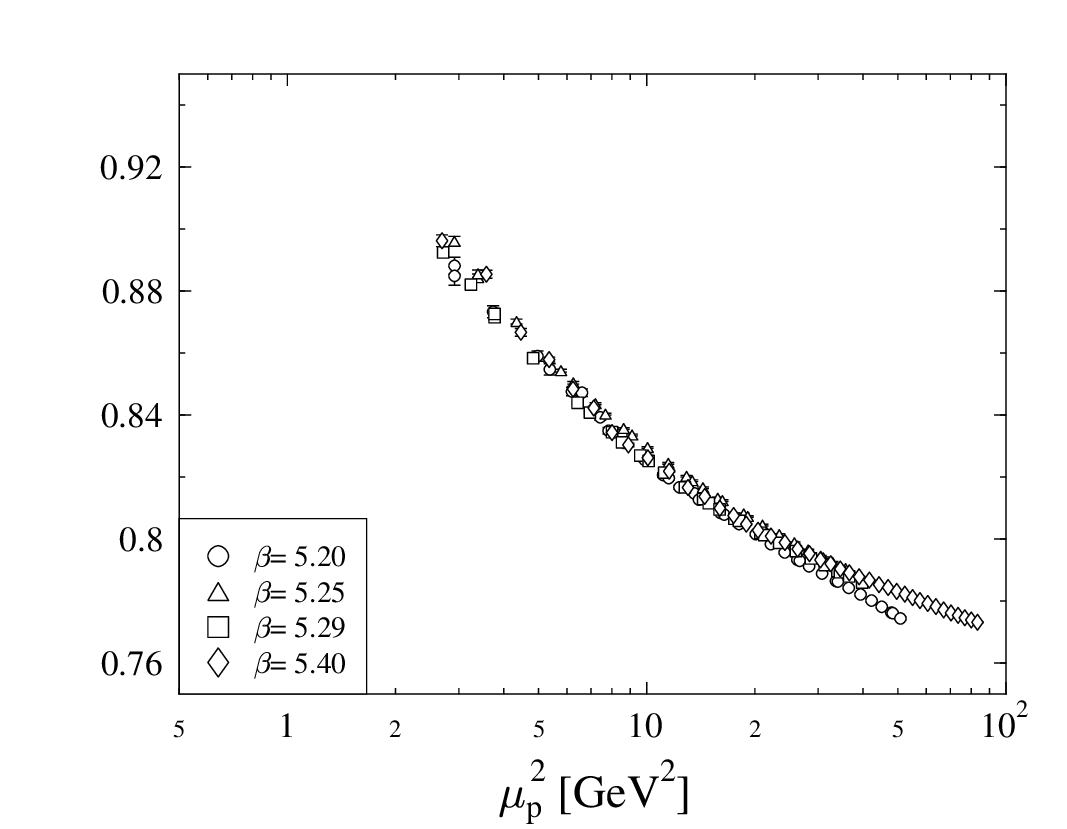,width=11cm}
\end{center}
\caption{$Z_{\mathrm {bare}}^{\MOM}$ (perturbatively subtracted) 
for the operator $\overline{\cO}_T$
as a function of the renormalization scale $\mu_p$. The upper plot shows the 
actual results, in the lower plot they have been multiplied by suitable 
($\mu_p$ independent) scaling factors.} 
\label{fig.zscalets}
\end{figure}

Let us first investigate to which extent we can separate truncation
effects from lattice artefacts. According to Eq.~(\ref{compzrgi}),  
$Z_{\mathrm {bare}}^{\mathcal S} (\mu_p,a) =
Z_{\RI}^{\mathcal S} (\mu_p) Z_{\mathrm {bare}}^{\RI} (\mu_p,a)$
can be written as
\begin{equation} \label{factor}
Z_{\mathrm {bare}}^{\mathcal S} (\mu_p,a)
= \Delta Z ^{\mathcal S} (\mu_p)^{-1} Z^{\mathrm {RGI}}(a) \,.
\end{equation}
In this way, we have factorized the dependence
of $Z_{\mathrm {bare}}^{\mathcal S} (\mu_p,a)$ on the renormalization scale
$\mu_p$ and on the cutoff $a$. Consequently we can write
\begin{equation} \label{scal}
Z_{\mathrm {bare}}^{\mathcal S} (\mu_p,a)
= \frac{Z^{\mathrm {RGI}} (a)}{Z^{\mathrm {RGI}} (a^\prime)} 
Z_{\mathrm {bare}}^{\mathcal S} (\mu_p,a^\prime) \,.
\end{equation}

Hence multiplication by an appropriate ($\mu_p$ independent) scaling 
factor should bring the values of $Z_{\mathrm {bare}}^{\mathcal S} (\mu_p,a)$ 
obtained for different values of $a$ (or $\beta$) onto a single
curve representing a function $f^{\mathcal S}(\mu_p)$ of $\mu_p$ only, provided 
$\mu_p$ is small enough so that lattice artefacts can be neglected.
Note that the ratio
$Z_{\mathrm {bare}}^{\mathcal S} (s \mu_p,a) / 
Z_{\mathrm {bare}}^{\mathcal S} (\mu_p,a)$ for some fixed value of $s$,
the so-called step scaling function, has a decent continuum limit.
This coincides with $f^{\mathcal S}(s \mu_p)/f^{\mathcal S}(\mu_p)$ in the 
region where $f^{\mathcal S}(\mu_p)$ is well defined.

In most cases, this collapse onto a single function works quite well 
for a reasonable range of renormalization scales, even if mixing is
allowed. For an example see Fig.~\ref{fig.zscalets}.
So the factorization of $\mu_p$ dependence and $a$ dependence seems 
to be possible (except for the highest values of $\mu_p$).

However, the available perturbative results cannot describe
the $\mu_p$ dependence below $\mu_p^2 \approx 5 \, \mbox{GeV}^2$,
as exemplified by Fig.~\ref{fig.rgits}.
It would be interesting to investigate 
whether this observation is related to the claim~\cite{landshoff} that 
DGLAP evolution cannot be used below $Q^2 = 5 \, \mbox{GeV}^2$ 
(at least in the region of larger values of Bjorken's variable $x$
\footnote{We thank M. Stratmann for a discussion of this issue.}).
Note also that an even later onset of (three-loop) perturbative behavior
has been found for the quenched gluon propagator~\cite{gluonprop1,gluonprop2}.

\begin{figure}
\begin{center}
\epsfig{file=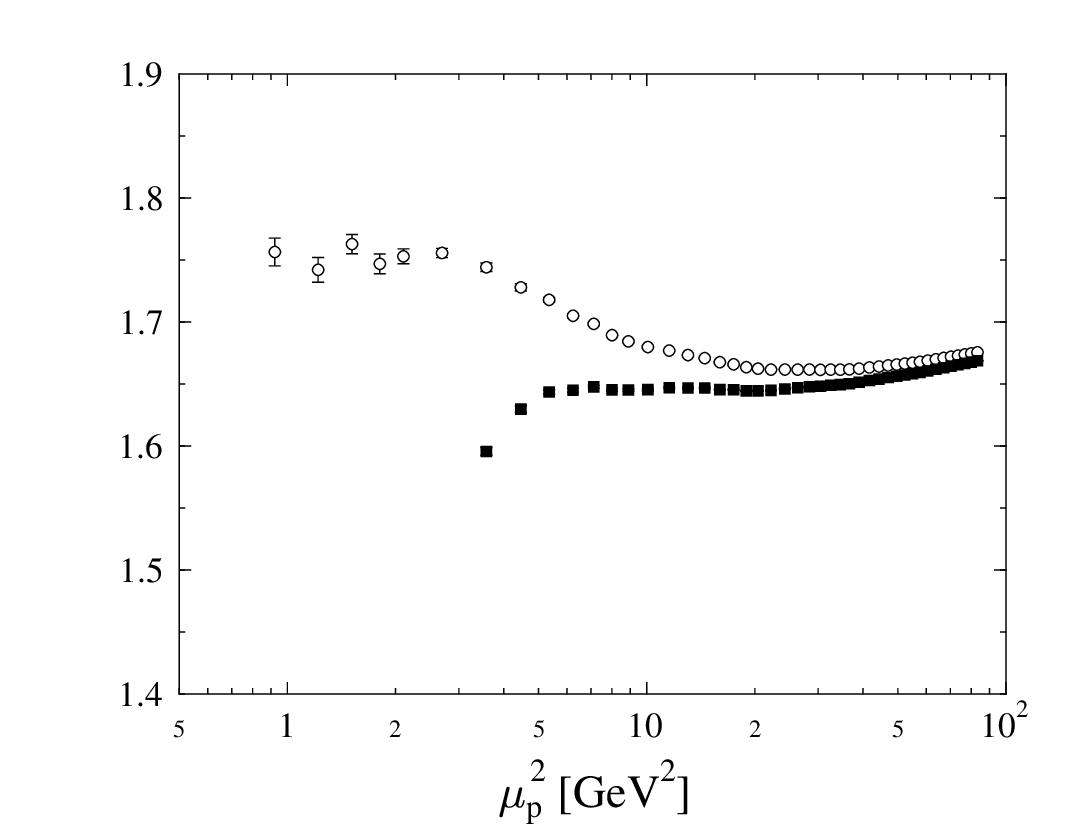,width=11cm}
\end{center}
\caption{$Z^{\mathrm {RGI}}$ (perturbatively subtracted) 
for the operator $\overline{\cO}_{h_{1,a}}$
at $\beta = 5.40$
as a function of the renormalization scale $\mu_p$. The open circles
(filled squares) have been obtained with $\mathcal S = \mathcal S' = \MS$
($\mathcal S = \MOM$, $\mathcal S' = \momt$).}
\label{fig.2scheme}
\end{figure}

In Fig.~\ref{fig.rgits} we have taken $\mathcal S = \MOM$ and we
have exploited the freedom to select the scheme $\mathcal S'$ 
for the coupling used in the perturbative expansion of 
$Z_{\RI}^{\mathcal S}$ choosing $\mathcal S' = \momt$.
Generally, the plateaus in $Z^{\mathrm {RGI}}$ look better for 
$\mathcal S = \MOM$ than for $\mathcal S = \MS$. 
This may be due to the fact that the perturbative expansion
of $Z_{\RI}^{\mathcal S}$ seems to be better behaved for $\mathcal S = \MOM$. 
For $\mathcal S'$, the choice $\mathcal S' = \momt$ turns out
to be preferable. An example comparing results obtained with 
$\mathcal S = \MS$, $\mathcal S' = \MS$ to results obtained with 
$\mathcal S = \MOM$, $\mathcal S' = \momt$ is shown in 
Fig.~\ref{fig.2scheme}.
Note that the difference between the two sets of results is caused
by the different truncation errors of the respective perturbative 
expansions.

\begin{figure}[t]
\begin{center}
\epsfig{file=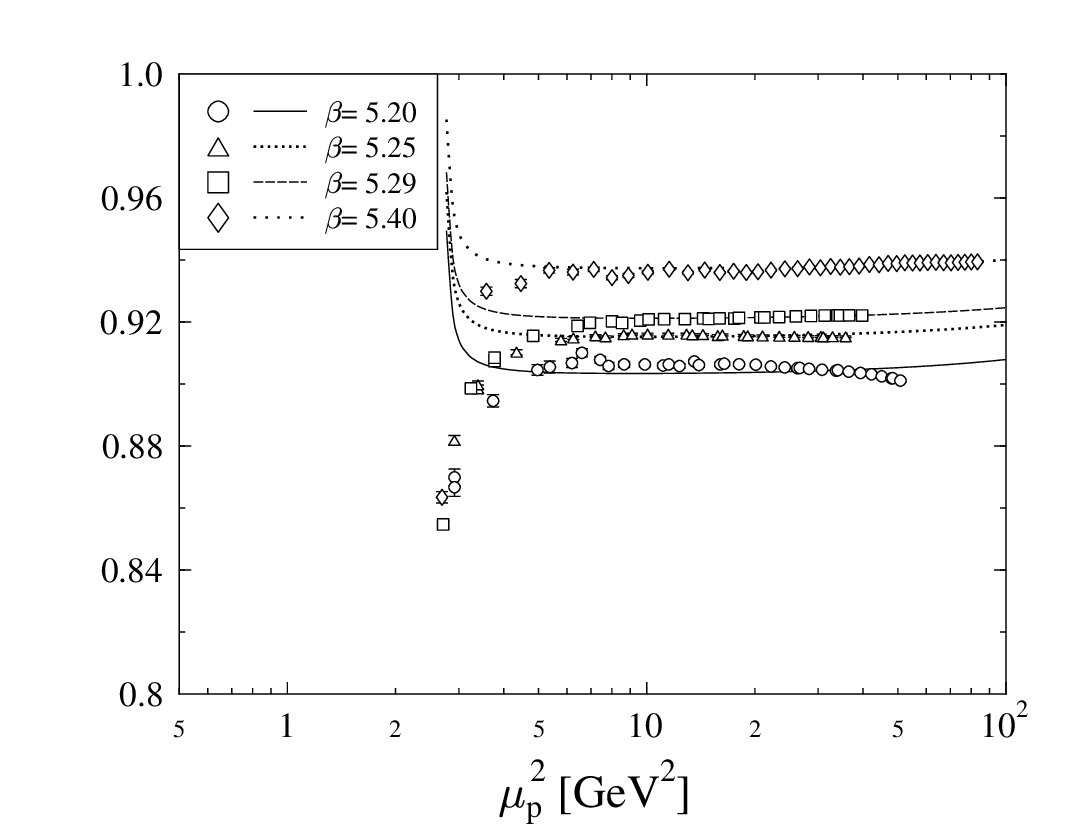,width=11cm} \\
\epsfig{file=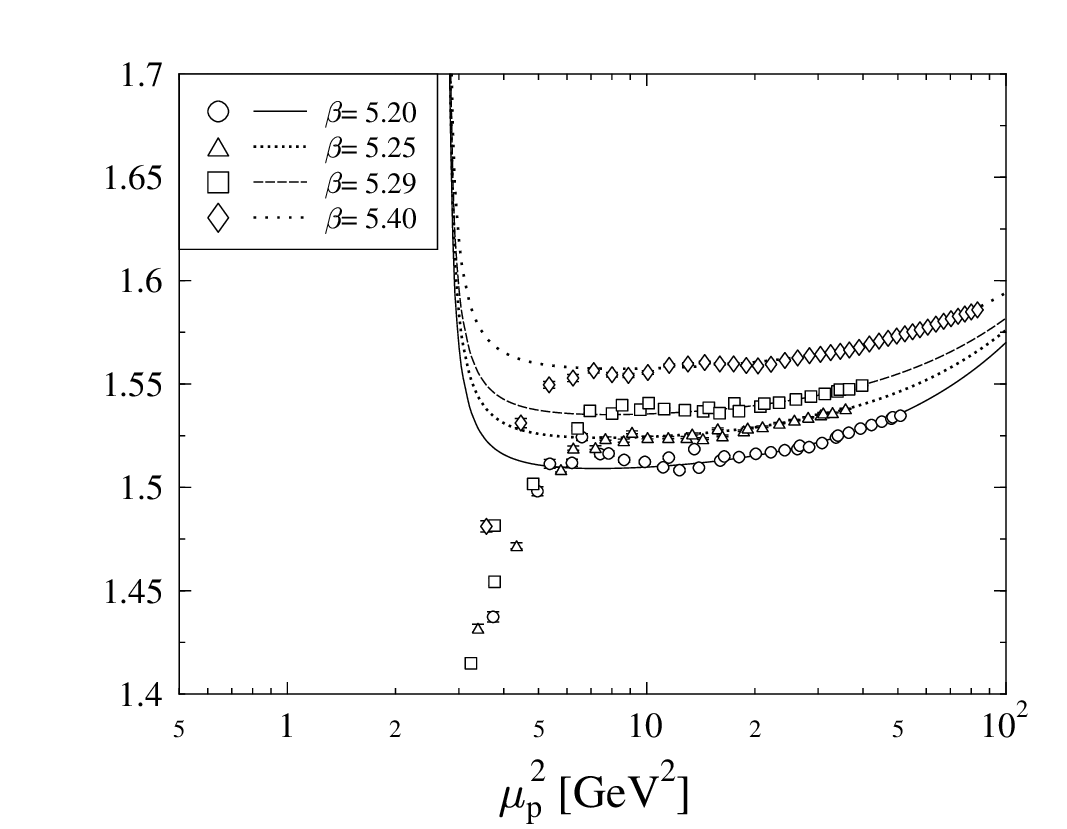,width=11cm}
\end{center}
\caption{$Z^{\mathrm {RGI}}$ (perturbatively subtracted) 
for the operators $\overline{\cO}_T$ (upper plot) and 
$\overline{\cO}_{v_{2,a}}$ (lower plot) as a function of the 
renormalization scale. Also shown are the fit curves used for the 
determination of $Z^{\mathrm {RGI}}$.}
\label{fig.fitexamples}
\end{figure}

In order to account for the deviations of the data from a perfect
plateau, e.g., due to lattice artefacts and the truncation of the 
perturbative expansions we have then applied a more complicated procedure 
than a simple fit with a constant. 
In particular, one should take
into account that residual lattice artefacts and truncation errors 
might conspire to produce a fake plateau. Therefore we consider
it important to include correction terms for both types of errors.
We have tried to incorporate 
higher terms in the perturbative expansions of 
$\Delta Z ^{\mathcal S} (\mu_p)$ and $Z_{\RI}^{\mathcal S} (\mu_p)$
treating the corresponding coefficients as additional fit parameters.
Similarly we have attempted to correct for discretization effects by 
including a simple ansatz for lattice artefacts. Again, the parameters
in this ansatz have to be fitted. Nevertheless, the number of fit
parameters will not get too large because we fit the data for all
four $\beta$ values simultaneously. Only the quantities 
$Z^{\mathrm {RGI}} (a)$, our final results, depend on $\beta$, 
the other parameters do not.

When we perform the fits we make the 
following choices. In the expansions for $Z_{\RI}^{\mathcal S} (\mu_p)$ 
and $\Delta Z ^{\mathcal S} (\mu_p)$ originating from continuum 
perturbation theory we use as many terms as are available. The same applies 
to the $\beta$ function used when computing the running coupling 
$g^{\mathcal S} (\mu_p)$. We choose the $\MOM$ scheme as the intermediate 
scheme $\mathcal S$ and expand
$Z_{\RI}^{\mathcal S}$ in the $\momt$ coupling. All data for
$\mu_p^2 \geq 10 \, \mbox{GeV}^2$ are included in the fit.
The correlations between the data at different momenta but the same $\beta$
are not taken into account. Two examples of such fits are shown in 
Fig.~\ref{fig.fitexamples}. More details concerning the fit procedure
can be found in Appendix~\ref{sec.fits}. 

In some cases, the data on our coarsest lattice ($\beta = 5.20$) 
are not very well reproduced
by the fit, see, e.g., the upper plot in Fig.\ref{fig.fitexamples}. 
Excluding these data would, however, lead only 
to tiny changes in the results. Therefore we have kept $\beta = 5.20$
in the fit for all operators.
Note that the ``divergence'' of the fit curves (and the data) in 
the vicinity of $\mu_p^2 = 3 \, \mbox{GeV}^2$ is mainly caused by 
the Landau pole in the renormalized coupling constant in the 
$\momt$ scheme. 

\begin{figure}[htb]
\begin{center}
\epsfig{file=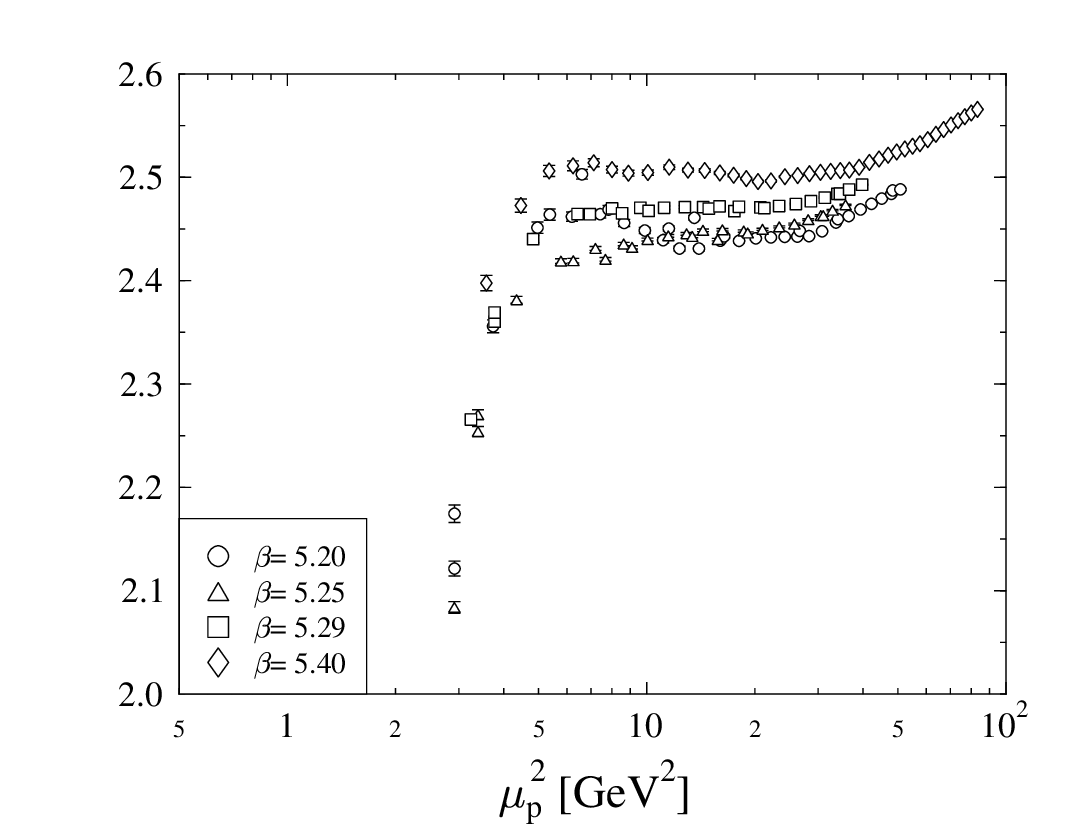,width=11cm}
\end{center}
\caption{$Z^{\mathrm {RGI}}$ for the operator $\overline{\cO}_{h_{2,b}}$
as a function of the renormalization scale.}
\label{fig.zrgi}
\end{figure}

While the fits for the subtracted data are reasonable it
was hardly possible to obtain a satisfactory fit for the unsubtracted 
numbers. Although plots of unsubtracted data do not differ dramatically
from plots of subtracted data (see Fig.~\ref{fig.zrgi} for results 
for an operator where no subtracted data are available), the fit curves
look quite strange. Therefore we have to conclude that our fit 
procedure is applicable only to perturbatively subtracted data and 
we must apply a different procedure to unsubtracted data. So we 
choose the following method. We read off $Z^{\mathrm {RGI}}$
at a reasonable value of $\mu_p^2$ and take as the error the
maximum of the differences with the results at one lower and one higher
value of the scale. The choice of these three scales is
to some extent dictated by the necessity to avoid large lattice
artefacts as well as large truncation errors in the perturbative 
expansions. We take the values 
$\mu_p^2 = 10 \, \mbox{GeV}^2$, $20 \, \mbox{GeV}^2$ and 
$30 \, \mbox{GeV}^2$.

\section{Results} \label{sec.results}

Before we present and discuss our results we have to 
consider the influence of the two parameters that enter
our analysis: the physical value of $r_0$ and
$r_0 \Lambda_\MS$. The perturbative expressions which are needed
in the evaluation of $Z^{\mathrm {RGI}}$ are functions of 
$\mu_p^2 / \Lambda_\MS^2$, where $\mu_p^2$ is related to the momenta
in lattice units $q^2$ by $a^2 \mu_p^2 = q^2$. Since we use $r_0$
to set the scale we write
\begin{equation} 
\mu_p^2 = \frac{\left( r_0/a \right)^2 q^2}{r_0^2} 
\end{equation}
so that
\begin{equation} 
\frac{\mu_p^2}{\Lambda_\MS^2} 
 = \frac{\left( r_0/a \right)^2 q^2}{\left( r_0 \Lambda_\MS \right)^2}  \,.
\end{equation}
This shows that the physical value of $r_0$ has an influence only on 
the scale $\mu_p$ to which a particular $Z$ value is 
associated. As the question of the scale at which perturbation theory
becomes applicable is not completely immaterial we shall set $r_0$
to a reasonable value, for which we take 
$0.467 \, \mbox{fm}$~\cite{scale1,scale2} (see also Ref.~\cite{scale3}). 
However, the precise number does not matter too much because $r_0$
enters only logarithmically.

The value chosen for $r_0 \Lambda_\MS$, on the other hand, has an 
impact on the results for the renormalization factors. In particular,
varying $r_0 \Lambda_\MS$ modifies the scale dependence of the
right-hand side of Eq.~(\ref{compzrgi}) and improves or deteriorates
the appearance of the plateau. 
We take $r_0 \Lambda_\MS = 0.617$ from Ref.~\cite{lambda}, which is 
consistent with the value found in Ref.~\cite{alpha}.

In order to estimate the systematic errors due to the uncertainties
in the values of $r_0$ and $r_0 \Lambda_\MS $ we have repeated
our analysis using $r_0 = 0.467 \, \mbox{fm}$, $r_0 \Lambda_\MS = 0.662$
and $r_0 = 0.5 \, \mbox{fm}$, $r_0 \Lambda_\MS = 0.617$ instead
of our standard values $r_0 = 0.467 \, \mbox{fm}$, $r_0 \Lambda_\MS = 0.617$.
Note that $r_0 \Lambda_\MS = 0.662$ results from 
$r_0 \Lambda_\MS = 0.617(40)(21)$ as given in Ref.~\cite{lambda}
by adding the errors in quadrature. 
The ensuing differences will be shown in the same format
as the (statistical) errors, however with the sign information included.
For reasons of better readability they are given below the 
results themselves. 
The first (second) number corresponds to the difference caused by the 
variation of $r_0 \Lambda_\MS$ ($r_0$).
For example, the entry
\begin{displaymath}  \begin{array}{c}
0.45155(80) \\[-0.2cm]
(568)(15)
\end{array}
\end{displaymath}
means that the analysis with $r_0 = 0.467 \, \mbox{fm}$ and 
$r_0 \Lambda_\MS = 0.617$ produced the result $0.45155 \pm 0.00080$
while using $r_0 = 0.467 \, \mbox{fm}$ along with $r_0 \Lambda_\MS = 0.662$
led to 0.45723 and working with $r_0 = 0.5 \, \mbox{fm}$,
$r_0 \Lambda_\MS = 0.617$ gave 0.45170. 
Note that the first error ((80) in the example) is determined from
the deviation of the $Z^{\mathrm {RGI}}$ data from a perfect plateau
as explained in more detail below, where also further sources of 
systematic errors will be discussed.

\begin{table}
\caption{Final nonperturbative results for operators with at most 
one derivative obtained with $r_0 = 0.467 \, \mbox{fm}$ and 
$r_0 \Lambda_\MS = 0.617$ from perturbatively subtracted data. 
Estimates of systematic errors have been included.}
\label{tab.final1}
\begin{ruledtabular}
\begin{tabular}{ccccccc}
Op. & $Z^{\mathrm {RGI}}|_{\beta = 5.20}$
 & $Z^{\mathrm {RGI}}|_{\beta = 5.25}$ & $Z^{\mathrm {RGI}}|_{\beta = 5.29}$ 
 & $Z^{\mathrm {RGI}}|_{\beta = 5.40}$ \\
\hline
$\cO^S$  
   & $0.45061(28)$ & $0.44990(65)$ & $0.44880(72)$ & $0.45155(80)$ \\
   & $(608)(21)$   & $(593)(18)$   & $(584)(17)$   & $(568)(15)$   \\
$\cO^P$  
   & $0.3376(95)$  & $0.3422(95)$  & $0.347(11)$   & $0.367(11)$   \\
   & $(43)(-5)$    & $(42)(-6)$    & $(4)(-1)$     & $(4)(-1)$     \\
$\overline{\cO}_V$ 
   & $0.7228(35)$  & $0.7323(28)$  & $0.7373(37)$  & $0.7513(25)$  \\
   & $(-3)(1)$     & $(-3)(1)$     & $(-3)(2)$     & $(-3)(1)$     \\
$\overline{\cO}_A$ 
   & $0.7527(21)$  & $0.76024(78)$ & $0.76439(30)$ & $0.77682(54)$ \\
   & $(1)(-8)$     & $(6)(-70)$    & $(6)(-62)$    & $(6)(-60)$    \\
$\overline{\cO}_T$ 
   & $0.9027(16)$  & $0.91453(58)$ & $0.92055(68)$ & $0.9368(14)$  \\
   & $(-40)(-1)$   & $(-390)(-6)$  & $(-389)(1)$   & $(-39)(0)$    \\
$Z_q$                                   
   & $0.7501(18)$  & $0.7557(12)$  & $0.75958(80)$ & $0.7703(14)$  \\
   & $(-7)(-6)$    & $(-7)(-6)$    & $(-66)(-49)$  & $(-7)(-5)$    \\
$\overline{\cO}_{v_{2,a}}$              
   & $1.5028(47)$  & $1.5182(58)$  & $1.5298(61)$  & $1.5526(54)$  \\
   & $(-164)(8)$   & $(-162)(7)$   & $(-161)(6)$   & $(-159)(6)$   \\
$\overline{\cO}_{v_{2,b}}$              
   & $1.5089(56)$  & $1.5233(81)$  & $1.5336(96)$  & $1.5555(28)$  \\
   & $(-159)(-7)$  & $(-156)(-5)$  & $(-157)(-5)$  & $(-155)(-6)$  \\
$\overline{\cO}_{r_{2,a}}$              
   & $1.4920(17)$  & $1.5071(38)$  & $1.5194(55)$  & $1.5430(34)$  \\
   & $(-159)(0)$   & $(-156)(1)$   & $(-156)(-1)$  & $(-154)(0)$   \\
$\overline{\cO}_{r_{2,b}}$              
   & $1.5382(23)$  & $1.5514(67)$  & $1.5614(92)$  & $1.5822(15)$  \\
   & $(-159)(-15)$ & $(-158)(-14)$ & $(-158)(-13)$ & $(-155)(-12)$ \\
$\overline{\cO}_{h_{1,a}}$              
   & $1.5791(39)$  & $1.5963(58)$  & $1.6096(41)$  & $1.6363(37)$  \\
   & $(-187)(0)$   & $(-185)(1)$   & $(-185)(0)$   & $(-184)(-1)$  \\
$\overline{\cO}_{h_{1,b}}$              
   & $1.5989(43)$  & $1.6155(63)$  & $1.6282(47)$  & $1.6541(44)$  \\
   & $(-191)(-1)$  & $(-189)(0)$   & $(-188)(0)$   & $(-186)(-1)$
\end{tabular}
\end{ruledtabular}
\end{table}

\begin{table}
\caption{Final nonperturbative results for operators with two and 
three derivatives obtained with $r_0 = 0.467 \, \mbox{fm}$ and
$r_0 \Lambda_\MS = 0.617$ from unsubtracted data. 
Estimates of systematic errors have been included.}
\label{tab.final2}
\begin{ruledtabular}
\begin{tabular}{ccccccc}
Op. & $Z^{\mathrm {RGI}}|_{\beta = 5.20}$
 & $Z^{\mathrm {RGI}}|_{\beta = 5.25}$ & $Z^{\mathrm {RGI}}|_{\beta = 5.29}$ 
 & $Z^{\mathrm {RGI}}|_{\beta = 5.40}$ \\
\hline
$\overline{\cO}_{v_3}$                  
   & $2.3796(97)$  & $2.385(23)$  & $2.410(30)$  & $2.4337(69)$  \\
   & $(-481)(17)$  & $(-48)(5)$   & $(-49)(-2)$  & $(-489)(18)$  \\
$\overline{\cO}_{v_{3,a}}$              
   & $2.3586(70)$  & $2.365(13)$  & $2.385(18)$  & $2.4084(98)$  \\
   & $(-501)(0)$   & $(-50)(4)$   & $(-51)(3)$   & $(-515)(13)$  \\
$\overline{\cO}_{r_3}$                  
   & $2.3979(67)$  & $2.401(25)$  & $2.426(32)$  & $2.4488(59)$  \\
   & $(-485)(7)$   & $(-49)(4)$   & $(-49)(-3)$  & $(-492)(14)$  \\
$\overline{\cO}_{a_2}$                  
   & $2.357(14)$   & $2.360(11)$  & $2.383(24)$  & $2.4084(94)$  \\
   & $(-50)(-2)$   & $(-50)(3)$   & $(-51)(2)$   & $(-515)(4)$   \\
$\overline{\cO}_{h_{2,a}}$              
   & $2.4265(77)$  & $2.435(14)$  & $2.4588(69)$ & $2.486(10)$   \\
   & $(-518)(10)$  & $(-52)(4)$   & $(-525)(6)$  & $(-53)(1)$    \\
$\overline{\cO}_{h_{2,b}}$              
   & $2.4408(68)$  & $2.448(14)$  & $2.4710(76)$ & $2.4967(79)$  \\
   & $(-521)(13)$  & $(-52)(4)$   & $(-528)(8)$  & $(-533)(13)$  \\
$\overline{\cO}_{h_{2,c}}$              
   & $2.4363(72)$  & $2.444(14)$  & $2.4680(68)$ & $2.494(10)$   \\
   & $(-520)(4)$   & $(-52)(4)$   & $(-527)(4)$  & $(-53)(1)$    \\
$\overline{\cO}_{h_{2,d}}$              
   & $2.420(10)$   & $2.427(19)$  & $2.450(12)$  & $2.475(11)$   \\
   & $(-52)(2)$    & $(-52)(5)$   & $(-52)(2)$   & $(-53)(2)$    \\
$\overline{\cO}_{v_4}$                  
   & $3.59(25)$    & $3.60(12)$   & $3.631(47)$  & $3.720(87)$   \\
   & $(-9)(-1)$    & $(-9)(-5)$   & $(-93)(-11)$ & $(-95)(-141)$
\end{tabular}
\end{ruledtabular}
\end{table}

In the cases where lattice artefacts have been subtracted 
perturbatively as explained in Sec.~\ref{sec.subtraction} we have
determined values for $Z^{\mathrm {RGI}}$ by means of the fit
procedure described in the previous section. 
The corresponding results will be called the fit results.

For unsubtracted data we apply the alternative method 
mentioned at the end of Sec.~\ref{sec.fit}.
We evaluate $Z^{\mathrm {RGI}}$ at the
scales $\mu_p^2 = 10 \, \mbox{GeV}^2$, $20 \, \mbox{GeV}^2$ and 
$30 \, \mbox{GeV}^2$, interpolating linearly in $\mu_p^2$ between adjacent
data points. We take the value at $20 \, \mbox{GeV}^2$ as our
central value and estimate the error from the maximum of the deviations
of the values at the other two scales. This error is always
larger (in most cases considerably larger) than the statistical error.
These results will be called the interpolation results.

Of course, the same method can also be applied to the subtracted data.
Except for $\cO^S$ at $\beta=5.20$, the error of the interpolation 
results is again larger than the statistical error.
So, for the operators for which perturbatively subtracted data exist
we have interpolation results and fit results, both based on
the subtracted numbers, as well as interpolation results extracted from
the unsubtracted data. For the operators for which no subtracted 
data are available we have only the interpolation results.
Note that the $v_4$ operators are particularly difficult: 
$\Gamma_{\mathrm {Born}} (p)$ vanishes on the diagonal of the Brillouin
zone close to which all our momenta lie, and large lattice artefacts
are obvious in the Monte Carlo data. So the corresponding nonperturbative 
results should be considered with caution.

Finally, we have to decide which numbers we want to consider as the
most reliable results to be used in the applications. It is clear that
we make use of the perturbatively subtracted data whenever they are 
available. In these cases our fits seem to exploit the Monte Carlo 
data in an optimal manner, and therefore we take the fit results 
as our final numbers. However, the errors computed by the MINUIT 
program~\cite{minuit} appear to be seriously underestimated as 
they are mainly determined by the statistical uncertainties.
Hence we adopt a kind of hybrid approach taking the errors from 
the interpolation results (based on the subtracted numbers) because 
they take into account the deviation of our $Z^{\mathrm {RGI}}$ data 
from a perfect plateau. The uncertainties due to the scale setting 
and the value of $\Lambda_\MS$ are again taken from the fit results.

For the operators with two or more derivatives we do not have much 
choice. So we take the interpolation results (including the systematic
uncertainties) as our final numbers. 
All our final results are
collected in Tables~\ref{tab.final1} (operators with at most
one derivative, based on perturbatively subtracted data) and 
\ref{tab.final2} (operators with more than one derivative, based
on unsubtracted data).

The impact of an uncertainty in $r_0$ or $r_0 \Lambda_\MS$ 
is easy to quantify and therefore given in Tables~\ref{tab.final1}
and \ref{tab.final2}. Further systematic errors are more difficult
to control. In particular, the error caused by gauge fixing is hard to
estimate reliably. However, as already remarked above, the existing 
investigations indicate that the ``Gribov noise'' does not exceed 
the present statistical errors~\cite{gribov1,gribov2,zci}.
Since they are not the dominating uncertainty it seems justified to
neglect the influence of Gribov copies although a more detailed
study would clearly be desirable.

The necessity of gauge fixing in the Rome-Southampton approach has also
another consequence: The operators of interest, though of course
gauge invariant, can mix with non gauge-invariant (NGI) operators. 
However, in perturbation theory this effect shows up only in two-loop order
and can thus reasonably be expected to be small.
NGI improvement terms for the quark propagator were discussed at 
some length in Refs.~\cite{impnp,sharpe}. While $O(g^2)$ perturbation 
theory cannot distinguish between the gauge-invariant and NGI
improvement terms, in an $O(g^3)$ calculation of the $qqg$ vertex
\cite{cswpert}, we could calculate the NGI improvement coefficient, 
and found it indeed to be numerically small.
Mixing with gauge invariant operators, on the other hand, can in most
cases be excluded by means of symmetry arguments (see 
Sec.~\ref{sec.operators}).

A few further systematic uncertainties can be estimated more easily.
We have tested the sensitivity to the chiral extrapolation by
repeating the analysis employing a quadratic chiral 
extrapolation (see Eq.(\ref{quex})). This changed the results by less
than 1\%, except for the case of $\overline{\cO}_{v_4}$ at 
$\beta = 5.20$, where a change of 1.8\% was observed.
In order to estimate the error caused by the truncation of the 
perturbative series we have reduced the order of all perturbative
expressions involved by one compared to the maximal value available. 
This led to changes of at most 1\%. Finally, we have considered the 
uncertainty related to the chiral extrapolation of $r_0/a$. Since 
little is known about the quark mass dependence of $r_0$, we had 
to rely on some phenomenological ansatz~\cite{lambda} leading to 
chirally extrapolated values of $r_0/a$ with errors of the order 
of 1\% (see Table~\ref{tab.paramc}). Varying the values of $r_0/a$ 
used in the analysis by 1\% produced changes of at most 0.5\% in 
the results for $Z^{\mathrm {RGI}}$. Thus it seems justified to
assign an additional uncertainty of about 2\% to our results.

\begin{table}
\caption{Perturbative estimates for $Z^{\mathrm {RGI}}$ based on
one-loop lattice perturbation theory. The intermediate 
scheme $\mathcal S$ is taken to be the $\MS$ scheme and 
$r_0 \Lambda_\MS = 0.617$. In all cases $\csw = 1$ is used.}
\label{tab.pertres3}
\begin{ruledtabular}
\begin{tabular}{cccccc}
Op. & {} 
& $Z^{\mathrm {RGI}}|_{\beta = 5.20}$ & $Z^{\mathrm {RGI}}|_{\beta = 5.25}$ 
& $Z^{\mathrm {RGI}}|_{\beta = 5.29}$ & $Z^{\mathrm {RGI}}|_{\beta = 5.40}$ \\
\hline
$\cO^S$             & bare PT &  0.5602 &  0.5532 &  0.5486 &  0.5399 \\
                    & TI PT   &  0.4902 &  0.4865 &  0.4843 &  0.4811 \\
                    & TRB PT  &  0.4637 &  0.4615 &  0.4598 &  0.4553 \\
$\cO^P$             & bare PT &  0.5396 &  0.5331 &  0.5288 &  0.5209 \\
                    & TI PT   &  0.4573 &  0.4547 &  0.4533 &  0.4519 \\
                    & TRB PT  &  0.4398 &  0.4382 &  0.4370 &  0.4336 \\
$\cO^V_\mu$         & bare PT &  0.8507 &  0.8521 &  0.8532 &  0.8562 \\
                    & TI PT   &  0.7721 &  0.7760 &  0.7792 &  0.7872 \\
                    & TRB PT  &  0.7800 &  0.7837 &  0.7866 &  0.7940 \\
$\cO^A_\mu$         & bare PT &  0.8656 &  0.8669 &  0.8679 &  0.8706 \\
                    & TI PT   &  0.7959 &  0.7994 &  0.8023 &  0.8094 \\
                    & TRB PT  &  0.8008 &  0.8042 &  0.8069 &  0.8136 \\
$\cO^T_{\mu \nu}$   & bare PT &  0.9811 &  0.9878 &  0.9926 &  1.0029 \\
                    & TI PT   &  0.9212 &  0.9296 &  0.9357 &  0.9494 \\
                    & TRB PT  &  0.9547 &  0.9617 &  0.9673 &  0.9815 \\
\end{tabular}
\end{ruledtabular}
\end{table}

\begin{table}
\caption{Perturbative estimates for $Z^{\mathrm {RGI}}$ based on 
two-loop lattice perturbation theory. The intermediate 
scheme $\mathcal S$ is taken to be the $\MS$ scheme and 
$r_0 \Lambda_\MS = 0.617$. In all cases the one-loop value for
$\csw$ is used.}
\label{tab.pertres2loop}
\begin{ruledtabular}
\begin{tabular}{cccccc}
Op. & {} 
& $Z^{\mathrm {RGI}}|_{\beta = 5.20}$ & $Z^{\mathrm {RGI}}|_{\beta = 5.25}$ 
& $Z^{\mathrm {RGI}}|_{\beta = 5.29}$ & $Z^{\mathrm {RGI}}|_{\beta = 5.40}$ \\
\hline
$\cO^S$                 & bare PT &  0.5165 &  0.5110 &  0.5074 &  0.5012 \\
                        & TI PT   &  0.4618 &  0.4596 &  0.4585 &  0.4578 \\
                        & TRB PT  &  0.4577 &  0.4558 &  0.4543 &  0.4503 \\
$\cO^P$                 & bare PT &  0.4860 &  0.4813 &  0.4784 &  0.4735 \\
                        & TI PT   &  0.4175 &  0.4170 &  0.4171 &  0.4192 \\
                        & TRB PT  &  0.4201 &  0.4194 &  0.4188 &  0.4170 \\
$\cO^V_\mu$             & bare PT &  0.7861 &  0.7887 &  0.7908 &  0.7963 \\
                        & TI PT   &  0.6943 &  0.7013 &  0.7068 &  0.7204 \\
                        & TRB PT  &  0.7067 &  0.7129 &  0.7179 &  0.7301 \\
$\cO^A_\mu$             & bare PT &  0.8101 &  0.8124 &  0.8142 &  0.8191 \\
                        & TI PT   &  0.7320 &  0.7380 &  0.7428 &  0.7545 \\
                        & TRB PT  &  0.7401 &  0.7456 &  0.7500 &  0.7608 \\
$\cO^T_{\mu \nu}$       & bare PT &  0.9789 &  0.9857 &  0.9905 &  1.0009 \\
                        & TI PT   &  1.0007 &  1.0064 &  1.0105 &  1.0190 \\
                        & TRB PT  &  1.0085 &  1.0136 &  1.0178 &  1.0285 \\
\end{tabular}
\end{ruledtabular}
\end{table}

In the case of the perturbative estimates we consider the choices
``bare PT'', ``TI PT'', and ``TRB PT''. Here ``bare PT'' and 
``TI PT'' refer to the expressions 
$\Delta Z ^{\MS} (\mu_0) Z_{\mathrm {bare}}^{\MS} (\mu_0,a)_{\mathrm {pert}}$
and 
$\Delta Z ^{\MS} (\mu_0) Z_{\mathrm {bare}}^{\MS} (\mu_0,a)_{\mathrm {ti}}$,
respectively, both evaluated at $\mu_0 = 1/a$ (see Eqs.(\ref{zpert.rg})
and (\ref{zpertti.rg})); ``TRB PT'' corresponds to the estimate by 
tadpole-improved, renormalization-group-improved boosted perturbation theory
in Eq.~(\ref{zrgitrb}). 
Because of our choice $\mu_0 = 1/a$ the perturbative estimates do not
depend on $r_0$. The bare PT and TI PT values do however depend on
$r_0 \Lambda_\MS $ as well as on the intermediate scheme $\mathcal S$,
which was taken to be the $\MS$ scheme in Eqs.(\ref{zpert.rg}) 
and (\ref{zpertti.rg}). We stick to this choice and set 
$r_0 \Lambda_\MS = 0.617$. The TRB PT value, on the other hand, is 
independent of these choices. Note, however, that all three 
perturbative estimates depend on the chosen value of $\csw$. 
As remarked above, we have set $\csw = 1$. For the operators
without derivatives we give the resulting numbers in 
Table~\ref{tab.pertres3}. The analogous results obtained from the two-loop
calculations as described in Appendix~\ref{sec.2loops} are given in 
Table~\ref{tab.pertres2loop}. All these perturbative numbers apply to
any member of the H(4) multiplet to which the operator listed belongs.

Let us now compare the results obtained by the various methods, i.e.,
by the different procedures of extracting $Z^{\mathrm {RGI}}$ from
the Monte Carlo data and by the different versions of lattice
perturbation theory. In particular, for the operators
for which perturbatively subtracted data exist we can compare
the results extracted from the perturbatively subtracted data,
both by interpolation and by means of the fit procedure, and
the interpolation results based on the unsubtracted numbers.
Of course, ideally they should agree within the errors. In reality, 
this is not always true. Note, however, that
the errors of the fit results only account for the (rather small) 
statistical uncertainties of the raw data while the errors of
the interpolation results are dominated by systematic effects.

\begin{figure}
\begin{center}
\epsfig{file=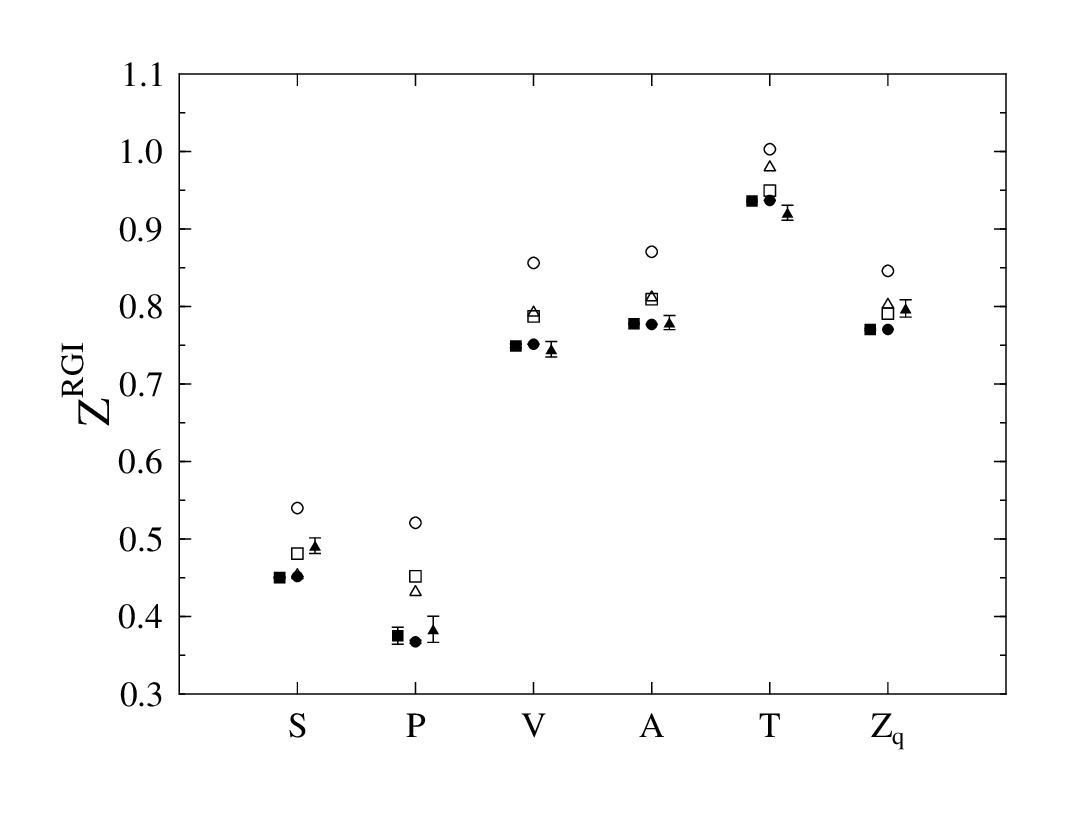,width=11cm}
\end{center}
\caption{Results for operators without derivatives at $\beta = 5.40$. 
The filled symbols correspond to our fit results (circles), 
interpolation results based on subtracted (squares) and unsubtracted
(triangles) data. Our final numbers are the fit results with the
errors taken from the interpolation results based on the subtracted data.
The open symbols represent estimates from bare perturbation theory (circles),
tadpole-improved perturbation theory (squares) and TRB perturbation theory
(triangles) based on one-loop calculations. }
\label{fig.res0}
\end{figure}

\begin{figure}
\begin{center}
\epsfig{file=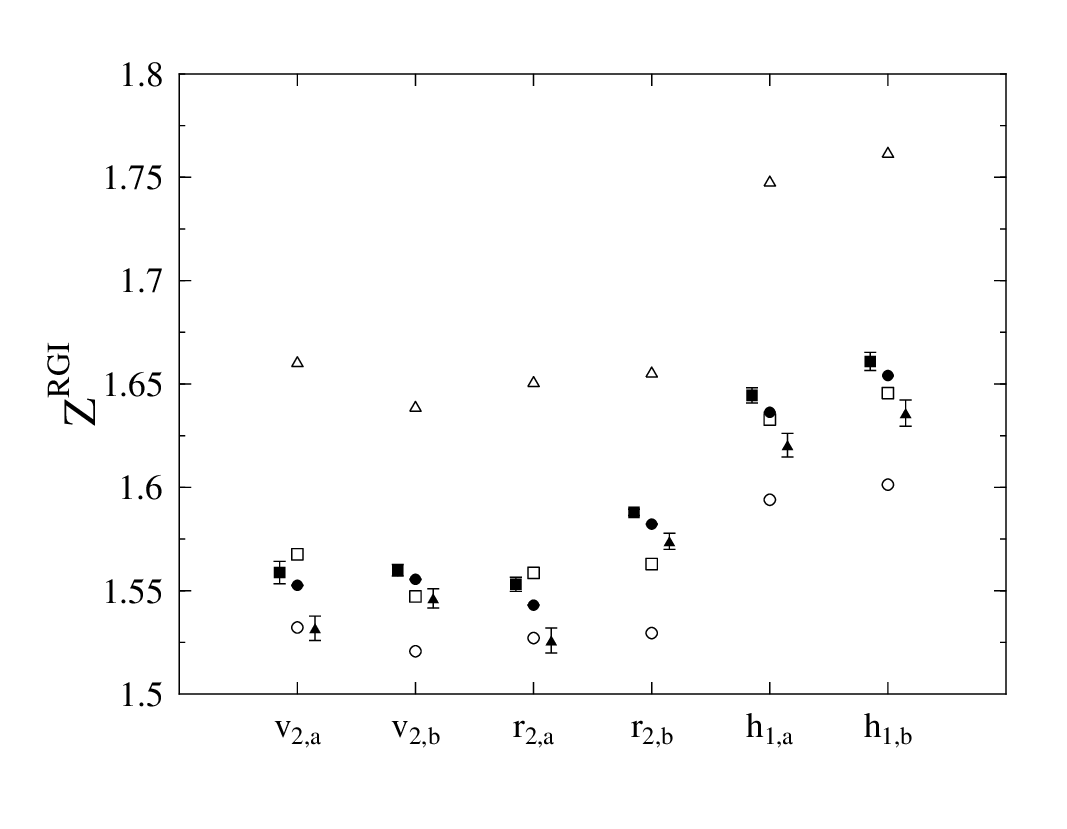,width=11cm}
\end{center}
\caption{Results for operators with one derivative at $\beta = 5.40$. 
The filled symbols correspond to our fit results (circles), 
interpolation results based on subtracted (squares) and unsubtracted
(triangles) data. Our final numbers are the fit results with the
errors taken from the interpolation results based on the subtracted data.
The open symbols represent estimates from bare perturbation theory (circles),
tadpole-improved perturbation theory (squares) and TRB perturbation theory
(triangles) based on one-loop calculations.}
\label{fig.res1}
\end{figure}

\begin{figure}
\begin{center}
\epsfig{file=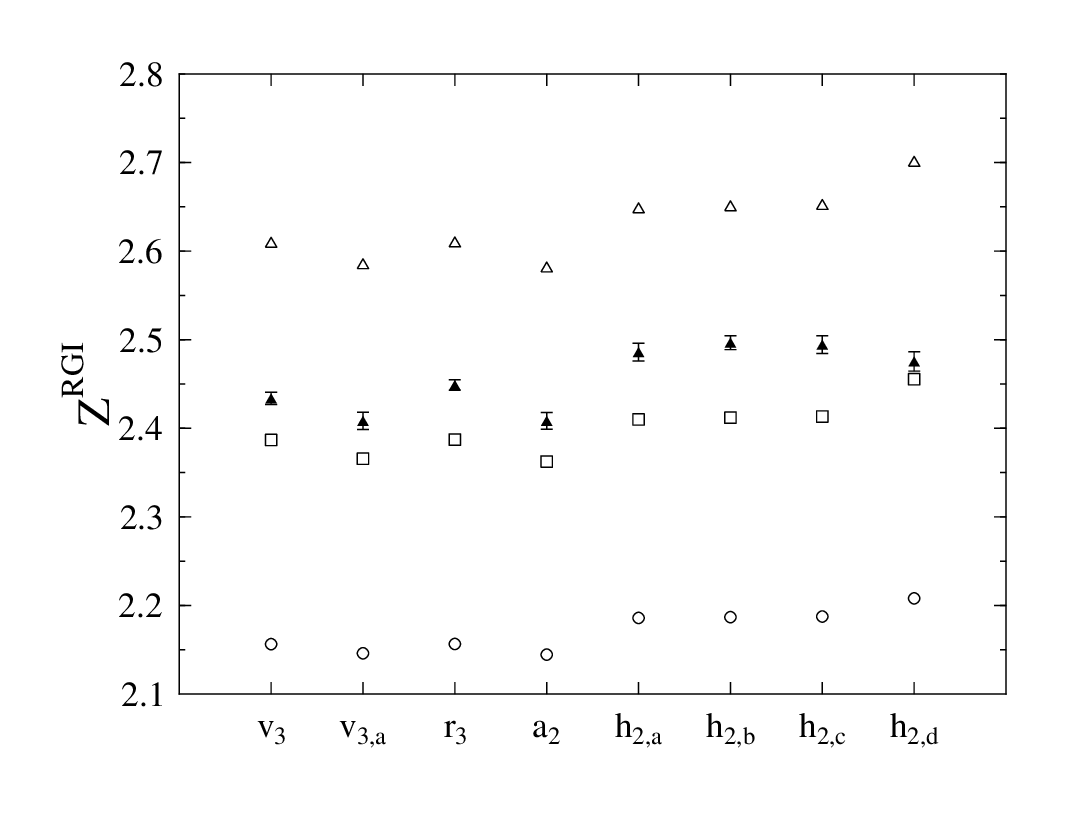,width=11cm}
\end{center}
\caption{Results for operators with two derivatives at $\beta = 5.40$. 
The filled triangles correspond to our nonperturbative results 
obtained by the interpolation method.
The open symbols represent estimates from bare perturbation theory (circles),
tadpole-improved perturbation theory (squares) and TRB perturbation theory
(triangles) based on one-loop calculations.}
\label{fig.res2}
\end{figure}

Figures~\ref{fig.res0}, \ref{fig.res1} and \ref{fig.res2} give an 
overview of our results for $\beta = 5.40$. The corresponding plots
for the other $\beta$ values look similar. For the operators without
derivatives (see Fig.~\ref{fig.res0}) the nonperturbative results
obtained with the different methods (with and without perturbative 
subtraction of lattice artefacts) are well consistent in most cases.
The one-loop perturbative estimates are larger, but tadpole improvement works. 
TRB perturbation theory, on the other hand, leads to
further improvement only in a few cases, for some operators it is even
worse than ordinary tadpole improved perturbation theory. 

\begin{figure}
\begin{center}
\epsfig{file=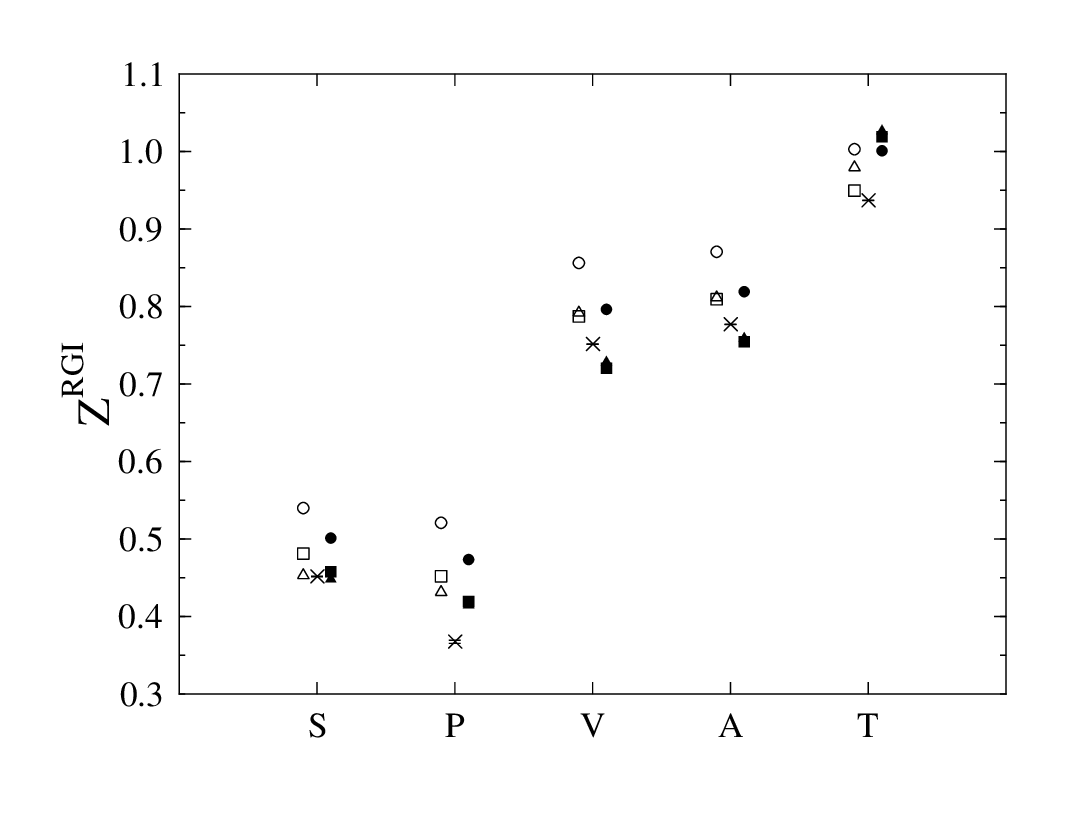,width=11cm}
\end{center}
\caption{Results for operators without derivatives at $\beta = 5.40$. 
The crosses correspond to our nonperturbative results obtained
by fits of the subtracted data. The open symbols represent estimates 
from bare perturbation theory (circles), tadpole-improved perturbation 
theory (squares) and TRB perturbation theory (triangles) in the 
one-loop approximation. The corresponding estimates based on two-loop
calculations are shown by the filled symbols.}
\label{fig.res2loops}
\end{figure}

\begin{figure}
\begin{center}
\epsfig{file=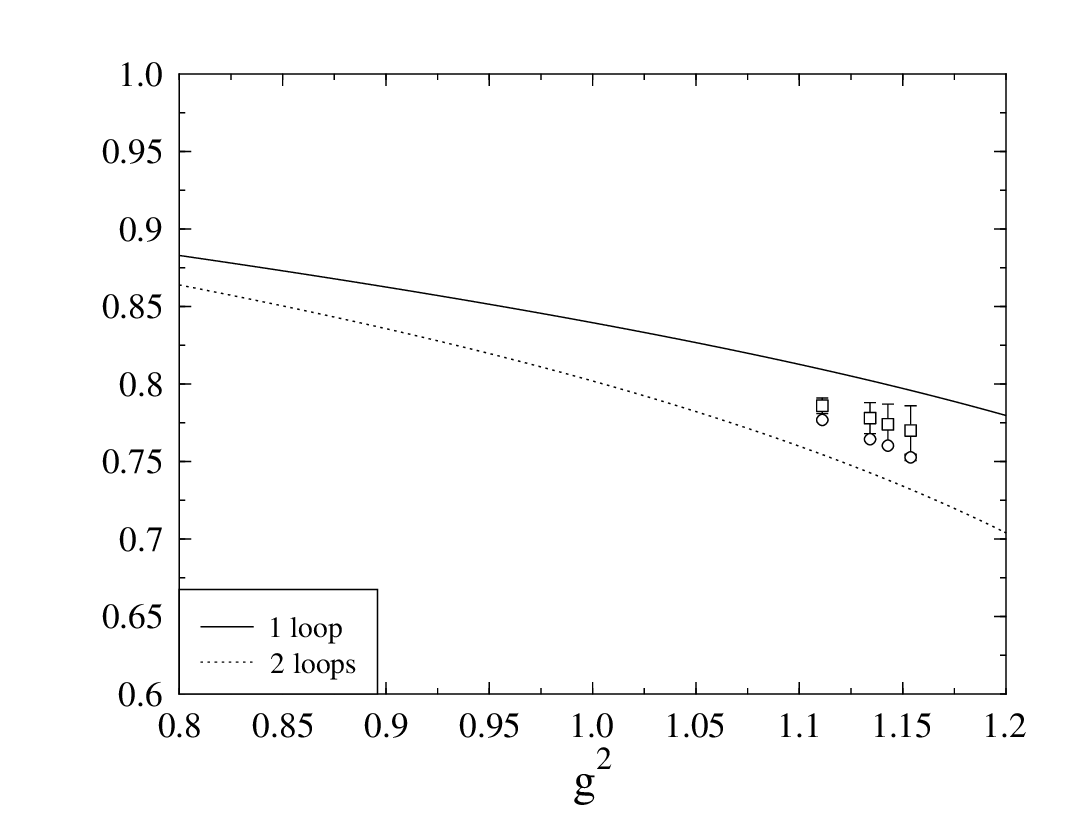,width=11cm}
\end{center}
\caption{Renormalization factor of the local axial current
as a function of $g^2$. The curves represent one- and two-loop
tadpole-improved perturbation theory. The circles are our 
nonperturbative results from Table~\ref{tab.final1}, the squares
are numbers obtained by the ALPHA collaboration~\cite{alpha2}.}
\label{fig.za}
\end{figure}

In the case of the operators with one derivative (see 
Fig.~\ref{fig.res1}) the agreement between the different methods 
for the nonperturbative results is less convincing,
in particular, the interpolation results obtained without perturbative
subtraction of lattice artefacts lie about 1 - 2\% lower.
Also the numbers from bare perturbation theory are smaller than
the nonperturbative results. Again, tadpole improvement 
moves the perturbative estimates in the right direction, though
too far in some cases. However, TRB perturbation theory leads to a 
significant overestimation.

For operators with two derivatives (see Fig.~\ref{fig.res2}) the 
only nonperturbative numbers we have at our disposal are those 
obtained without perturbative subtraction of lattice artefacts. 
Since the corresponding results for operators with one derivative
lie consistently below those coming from the perturbatively 
subtracted data, it is tempting to guess that this 
is also the case for operators with two derivatives though we cannot
quantify the difference. The behavior of the perturbative estimates 
is similar to that observed for operators with one derivative. Bare 
perturbation theory underestimates the results considerably, while 
tadpole improved perturbation theory comes much closer to the 
nonperturbative numbers. The results from TRB perturbation theory 
lie too high again. It should however be noted that the relative
positions of the nonperturbative renormalization factors for almost 
all operators considered are surprisingly well reproduced by any 
of the perturbative estimates. 

In Fig.~\ref{fig.res2loops} we plot our fit results
for the operators without derivatives
together with the one-loop and two-loop perturbative estimates,
again for $\beta = 5.40$. Let us comment on the numbers from
bare lattice perturbation theory first, represented by circles in the 
figure. They exhibit the expected behavior: The two-loop results come 
closer to the nonperturbative numbers than the one-loop estimates,
though only slightly in the case of the tensor current 
$\cO^T_{\mu \nu}$. Except for the tensor current, tadpole improvement
works also in the two-loop approximation moving the perturbative
values, indicated by squares, closer to the nonperturbative numbers. 
However, the results from TRB perturbation theory, shown by triangles,
do not differ much from the values found by tadpole improved
two-loop perturbation theory. At the moment it is unclear why 
the tensor current shows such a peculiar behavior.

The perturbative estimates can easily be calculated
at arbitrary values of the bare coupling constant $g$. However, for
tadpole improvement one also needs nonperturbative values for
$u_0$ (or for the average plaquette $P = u_0^4$) at these couplings.
Such values can (approximately) be obtained from the results for 
$P$ given in Table~\ref{tab.paramc} by a simple Pad\'e fit
taking into account the known two-loop expression for $P$~\cite{athen}.
In Fig.~\ref{fig.za} we plot the tadpole-improved perturbative results
for the renormalization factor of the local axial current $\cO^A_\mu$ 
in the one- and two-loop aproximation along with our nonperturbative
numbers and those from the ALPHA collaboration~\cite{alpha2}.

In a few cases we can compare our nonperturbative renormalization
factors with results obtained by other methods.
The renormalization factor of the local vector current $\cO^V_\mu$,
usually called $Z_V$, can also be extracted from hadron three-point
functions by considering the time component of the current and imposing
charge conservation. Some time ago we have employed this approach in the
case of the nucleon~\cite{zv} on a subset of the gauge field ensembles
used in the present work. The results are given in Table~\ref{tab.resother}
along with the renormalization factor of the local axial current 
$\cO^A_\mu$, usually called $Z_A$, obtained by the ALPHA collaboration 
by means of the Schr\"odinger functional method~\cite{alpha2}.
We also include the results for $Z_V$ from the ALPHA 
collaboration~\cite{alpha3}. Note that in these determinations no 
gauge fixing is required. Hence the reasonable
agreement of the numbers in Table~\ref{tab.resother} with those found
in the present work indicates that the Gribov noise is small.

In addition, we give in this table values for 
$Z_m^{\mathrm {RGI}} = (Z_A/Z_P)^{\mathrm {RGI}}$, where $Z_P$ is
the renormalization factor of the pseudoscalar density $\cO^P$.
The factor $Z_m^{\mathrm {RGI}}$ renormalizes the quark mass as determined
from the lattice axial Ward identity. In Ref.~\cite{strange} 
we have calculated it by a method similar to that used in the present 
paper and applied the results in an evaluation of the strange quark
mass. For easier comparison we also give the numbers following from
Table~\ref{tab.final1}. They differ from the older results by at most 2\%.

\begin{table}
\caption{Results from alternative approaches. See Table 3 in 
Ref.~\cite{zv} for $Z_V$, Eq.(3.7) in Ref.~\cite{alpha3} for $Z_V$(ALPHA),
and Eq.~(4.10) in Ref.~\cite{alpha2} for $Z_A$(ALPHA). The numbers 
for $Z_m$ in Ref.~\cite{strange} are taken
from Table 3 in that reference.}
\label{tab.resother}
\begin{ruledtabular}
\begin{tabular}{cllll}
{ } & $Z^{\mathrm {RGI}}|_{\beta = 5.20}$ & $Z^{\mathrm {RGI}}|_{\beta = 5.25}$ 
& $Z^{\mathrm {RGI}}|_{\beta = 5.29}$  & $Z^{\mathrm {RGI}}|_{\beta = 5.40}$ \\
\hline
$Z_V$        & $0.7304(18)$   & $0.7357(13)$  & $0.7420(7)$    &  {}    \\
$Z_V$(ALPHA) & $0.739(5)$     & $0.744(5)$    & $0.749(5)$     & $0.759(5)$ \\
$Z_A$(ALPHA) & $0.770(16)$    & $0.774(13)$   & $0.778(11)$    & $0.786(5)$ \\
$Z_m$~\cite{strange}  
             & $2.270(12)$    & $2.191(24)$   & $2.177(14)$    & $2.124(6)$ \\
$Z_m$ (this work)
             & $2.230(63)$     & $2.222(62)$    & $2.203(70)$     & $2.117(63)$
\end{tabular}
\end{ruledtabular}
\end{table}

Factors for converting $Z^{\mathrm {RGI}}$ to the $\MS$ scheme,
i.e., $\Delta Z ^{\MS} (\mu)^{-1}$
evaluated for our standard values of $r_0$ and $r_0 \Lambda_\MS $
can be found in Appendix~\ref{sec.conv}.  

\section{Conclusions} \label{sec.concl}

As more and more detailed questions about hadron structure are
treated in lattice QCD the renormalization of composite operators
has become an important issue and perturbative as well as 
nonperturbative methods have been developed. In this paper we
have presented results of a nonperturbative study in the 
RI-MOM scheme for a large variety of quark-antiquark operators, 
based on simulations with $n_f=2$ dynamical clover fermions. 
The results for the renormalization constants will be applied 
in the evaluation of phenomenologically relevant hadron 
matrix elements. Apart from these numbers, there are also a 
few lessons of a more general nature to be learned from our investigation.

The renormalization factors connecting the bare operators on the 
lattice with their renormalized counterparts in some renormalization 
scheme, e.g., the $\MS$ scheme, depend on the cutoff used, the lattice 
spacing $a$ in our case, and the renormalization scale $M$. The 
dependence on these two quantities should factorize, and this is
indeed observed in a broad range of $M$. However, the available 
results from continuum perturbation theory for the anomalous 
dimensions and the $\beta$ function can describe the $M$ 
dependence only for relatively large values of the scale, 
above $M^2 \approx 5 \, \mbox{GeV}^2$.

Only in the region where the scale dependence is well described 
by continuum perturbation theory is it possible to extract
reliable values for the renormalization factors. On the other 
hand, for large values of the renormalization scale lattice artefacts
may jeopardize the whole approach. It is therefore important
to keep discretization effects under control, and we have seen that
this purpose can be achieved (at least approximately) by subtracting 
lattice artefacts with the help of lattice perturbation theory.
We did this at the one-loop level, but to all orders in $a$.
Unfortunately, our procedure turned out to be too complicated for
operators with more than one covariant derivative. 
Alternatively, one can calculate the lattice artefacts proportional
to $a^2$ in one-loop lattice perturbation theory. This has recently been
done for operators without derivatives~\cite{a2corr}. Since it
should be possible to extend such calculations to more complicated
operators, it would be interesting to see if subtraction of the 
$a^2$ contributions is already sufficient for our purposes.

With the help of lattice perturbation theory one can not only 
calculate lattice artefacts, but also the renormalization factors
themselves. However, due to the notoriously poor convergence properties
of bare lattice perturbation theory some kind of improvement
is mandatory, at least if only one-loop calculations are available.
Indeed, we have found that tadpole improvement does quite a good job,
although it is hard to predict how good the results actually are.
For operators without derivatives, there are now even two-loop 
results. In bare perturbation theory they lead to a reduction of
the difference with our nonperturbative renormalization factors,
but the situation is less clear when tadpole improvement is included.
Perhaps the ideal perturbative scheme is still to be found.

Let us finally mention a few possible directions for future research.
The RI-MOM scheme has the disadvantage that it requires gauge fixing.
In principle, this problem could be overcome by working with 
correlation functions in coordinate space, and a first implementation
of this idea has been published~\cite{coord}. It seems, however,
that very fine lattices are necessary for this method.

Another possible modification concerns the choice
of the momenta. In our application we have followed the original 
RI-MOM scheme, where the momentum transfer at the operator insertion 
vanishes. However, a generalization to nonexceptional momenta is 
possible~\cite{nonex1,nonex2,nonex3}.

A third variant of nonperturbative renormalization
is motivated by the fact that the renormalization
condition (\ref{defz}) involves only a particular trace of the
vertex function $\Gamma (p)$. On the other hand, we have the
complete vertex functions (as $4 \times 4$-matrices after averaging 
over color) at our disposal, the bare ones computed nonperturbatively 
on the lattice as well as the renormalized ones calculated perturbatively
in the $\MS$ scheme. So instead of introducing the intermediate 
$\RI$ scheme by imposing (\ref{defz}) one could directly 
compare the bare nonperturbative vertex function $\Gamma (p)$ with 
the renormalized perturbative vertex function $\Gamma^{\MS} (p)$ in the 
$\MS$ scheme. Up to lattice artefacts we should have
\begin{equation} \label{zdirect}
\Gamma^{\MS} (p) = \left( Z^{\MS}_{q,\mathrm {bare}} \right)^{-1} 
       Z^{\MS}_{\mathrm {bare}} \, \Gamma (p) \,,
\end{equation}
where $\Gamma^{\MS} (p)$ as well as the renormalization factors
$Z^{\MS}_{q,\mathrm {bare}}$ and $Z^{\MS}_{\mathrm {bare}}$ also depend
on the renormalization scale $\mu$. An analogous relation should
hold for the quark propagator and $Z^{\MS}_{q,\mathrm {bare}}$.

Of course, it is not to be expected that (\ref{zdirect}) is satisfied 
exactly: Not only lattice artefacts would spoil the identity, but also 
the truncation of the perturbative expansion. So one would have to 
develop some kind of fit procedure for extracting $Z^{\MS}_{\mathrm {bare}}$ 
from (\ref{zdirect}). In any case, it might be an interesting exercise
to see how well (\ref{zdirect}) is fulfilled for our data.

\section*{Acknowledgements}
The numerical calculations have been performed on the 
apeNEXT and APEmille at NIC/DESY (Zeuthen). 
This work has been supported in part by the EU Integrated Infrastructure
Initiative HadronPhysics2 and by the DFG under contract SFB/TR55
(Hadron Physics from Lattice QCD).

\clearpage

\appendix

\section{Operator bases}
\label{sec.bases}

In this Appendix we list the operator bases we used when calculating 
the renormalization factors with the help of Eq.~(\ref{calczav}).

For $v_{2,a}$ (representation $\tau^{(6)}_3$, $C=+1$):
\begin{equation} \label{avdefv2a}
\cO _{\{\mu \nu \}} \,, \quad 1 \leq \mu < \nu \leq 4 \,.
\end{equation}

For $v_{2,b}$ (representation $\tau^{(3)}_1$, $C=+1$):
\begin{equation}
\cO _{11} +  \cO _{22} -  \cO _{33} -  \cO _{44} \,, \quad
\cO _{33} -  \cO _{44} \,, \quad
\cO _{11} -  \cO _{22} \,.
\end{equation}

For $r_{2,a}$ (representation $\tau^{(6)}_4$, $C=-1$):
\begin{equation}
\cO  ^5_{\{\mu \nu \}} \,, \quad 1 \leq \mu < \nu \leq 4 \,.
\end{equation}

For $r_{2,b}$ (representation $\tau^{(3)}_4$, $C=-1$):
\begin{equation}
\cO ^5_{11} +  \cO ^5_{22} -  \cO ^5_{33} 
                                   -  \cO ^5_{44} \,, \quad
\cO ^5_{33} -  \cO ^5_{44} \,, \quad
\cO ^5_{11} -  \cO ^5_{22} \,.
\end{equation}

For $h_{1,a}$ (representation $\tau^{(8)}_2$, $C=+1$):
\begin{equation} 
 2 {\cO}^T_{\nu_1 \{ \nu_2 \nu_3 \}}
 + {\cO}^T_{\nu_2 \{ \nu_1 \nu_3 \}} \,, \quad
 {\cO}^T_{\nu_2 \{ \nu_1 \nu_3 \}} \,,
   \quad 1 \leq \nu_1 < \nu_2 < \nu_3  \leq 4 \,.
\end{equation}

For $h_{1,b}$ (representation $\tau^{(8)}_1$, $C=+1$):
\begin{equation} \begin{array}{ll} \displaystyle
\cO^T_{1 2 2} - \cO^T_{1 3 3}  \,, &
\displaystyle
\cO^T_{1 2 2} + \cO^T_{1 3 3} - 2 \cO^T_{1 4 4} \,,
\\ \displaystyle
\cO^T_{2 1 1} - \cO^T_{2 3 3} \,, &
\displaystyle
\cO^T_{2 1 1} + \cO^T_{2 3 3} - 2 \cO^T_{2 4 4} \,,
\\ \displaystyle
\cO^T_{3 1 1} - \cO^T_{3 2 2} \,, &
\displaystyle
\cO^T_{3 1 1} + \cO^T_{3 2 2} - 2 \cO^T_{3 4 4} \,,
\\ \displaystyle
\cO^T_{4 1 1} - \cO^T_{4 2 2} \,, &
\displaystyle
\cO^T_{4 1 1} + \cO^T_{4 2 2} - 2 \cO^T_{4 3 3}  \,.
\end{array}
\end{equation}

For $v_3$ (representation $\tau^{(8)}_1$, $C=-1$):
\begin{equation}
\begin{array}{ll} \displaystyle
\cO_{\{122\}} - \cO_{\{133\}} \,, &
\displaystyle
\cO_{\{122\}} + \cO_{\{133\}} - 2 \cO_{\{144\}} \,,
\\  \displaystyle
\cO_{\{211\}} - \cO_{\{233\}} \,, &
\displaystyle
\cO_{\{211\}} + \cO_{\{233\}} - 2 \cO_{\{244\}} \,,
\\  \displaystyle
\cO_{\{311\}} - \cO_{\{322\}} \,, &
\displaystyle
\cO_{\{311\}} + \cO_{\{322\}} - 2 \cO_{\{344\}} \,,
\\  \displaystyle
\cO_{\{411\}} - \cO_{\{422\}} \,, &
\displaystyle
\cO_{\{411\}} + \cO_{\{422\}} - 2 \cO_{\{433\}} \,.
\end{array}
\end{equation}

For $v_{3,a}$ (representation $\tau^{(4)}_2$, $C=-1$):
\begin{equation}
\cO_{\{234\}} \,, \cO_{\{134\}} \,,
\cO_{\{124\}} \,, \cO_{\{123\}} \,.
\end{equation}

For $r_3$ (representation $\tau^{(8)}_2$, $C=+1$):
\begin{equation}
\begin{array}{ll} \displaystyle
\cO^5_{\{122\}} - \cO^5_{\{133\}} \,, &
\displaystyle
\cO^5_{\{122\}} + \cO^5_{\{133\}} - 2 \cO^5_{\{144\}} \,,
\\  \displaystyle
\cO^5_{\{211\}} - \cO^5_{\{233\}} \,, &
\displaystyle
\cO^5_{\{211\}} + \cO^5_{\{233\}} - 2 \cO^5_{\{244\}} \,,
\\  \displaystyle
\cO^5_{\{311\}} - \cO^5_{\{322\}} \,, &
\displaystyle
\cO^5_{\{311\}} + \cO^5_{\{322\}} - 2 \cO^5_{\{344\}} \,,
\\  \displaystyle
\cO^5_{\{411\}} - \cO^5_{\{422\}} \,, &
\displaystyle
\cO^5_{\{411\}} + \cO^5_{\{422\}} - 2 \cO^5_{\{433\}} \,.
\end{array}
\end{equation}

For $a_2$ (representation $\tau^{(4)}_3$, $C=+1$):
\begin{equation}
\cO^5_{\{234\}} \,, \cO^5_{\{134\}} \,,
\cO^5_{\{124\}} \,, \cO^5_{\{123\}} \,.
\end{equation}

For $v_4$ (representation $\tau^{(2)}_1$, $C=+1$):
\begin{equation} \begin{array}{c} \displaystyle
\cO_{\{1122\}} + \cO_{\{3344\}} - \cO_{\{1133\}} - \cO_{\{2244\}} \,,
\\ \displaystyle
\cO_{\{1122\}} + \cO_{\{3344\}} +\cO_{\{1133\}} + \cO_{\{2244\}} -
2 \cO_{\{1144\}} - 2 \cO_{\{2233\}}  \,.
\end{array}
\end{equation}

For $h_{2,a}$ (representation $\tau^{(3)}_2$, $C=-1$):
\begin{equation} \begin{array}{c} \displaystyle
\cO^T_{14 \{23\}} 
  + \cO^T_{24 \{13\}} + \cO^T_{34 \{12\}}  = - 3 \cO^T_{4 \{123\}} \,,
\\ \displaystyle
3 \cO^T_{13 \{24\}}
  + \cO^T_{14 \{23\}} + 3 \cO^T_{23 \{14\}}  
  + \cO^T_{24 \{13\}} - 2 \cO^T_{34 \{12\}} 
  =  - 9 \cO^T_{3 \{124\}}  -  3 \cO^T_{4 \{123\}} \,, 
\\ \displaystyle
2 \cO^T_{12 \{34\}}
  + \cO^T_{13 \{24\}} + \cO^T_{14 \{23\}} 
  - \cO^T_{23 \{14\}} -  \cO^T_{24 \{13\}} 
  = 3 \cO^T_{1 \{234\}} -  3 \cO^T_{2 \{134\}} \,. 
\end{array}
\end{equation}

For $h_{2,b}$ (representation $\tau^{(3)}_3$, $C=-1$):
\begin{equation} \begin{array}{c} \displaystyle
\cO^T_{1 \{122\}} - \cO^T_{1 \{133\}} + \cO^T_{2 \{233\}} \,,
\\ \displaystyle
 -2 \cO^T_{1 \{122\}}  - \cO^T_{1 \{133\}} + 3 \cO^T_{1 \{144\}}
  + \cO^T_{2 \{233\}} - 3 \cO^T_{2 \{244\}} \,,
\\ \displaystyle
 - \cO^T_{1 \{133\}} + \cO^T_{1 \{144\}} - \cO^T_{2 \{233\}}
  + \cO^T_{2 \{244\}} - 2 \cO^T_{3 \{344\}} \,.
\end{array}
\end{equation}

For $h_{2,c}$ (representation $\tau^{(6)}_2$, $C=-1$):
\begin{equation} \begin{array}{c} \displaystyle
    \cO^T_{13 \{32\}} + \cO^T_{23 \{31\}} 
  - \cO^T_{14 \{42\}} - \cO^T_{24 \{41\}} \,, 
\\ \displaystyle
    \cO^T_{12 \{23\}} + \cO^T_{32 \{21\}} 
  - \cO^T_{14 \{43\}} - \cO^T_{34 \{41\}} \,, 
\\ \displaystyle
    \cO^T_{12 \{24\}} + \cO^T_{42 \{21\}} 
  - \cO^T_{13 \{34\}} - \cO^T_{43 \{31\}} \,, 
\\ \displaystyle
    \cO^T_{21 \{13\}} + \cO^T_{31 \{12\}} 
  - \cO^T_{24 \{43\}} - \cO^T_{34 \{42\}} \,, 
\\ \displaystyle
    \cO^T_{21 \{14\}} + \cO^T_{41 \{12\}} 
  - \cO^T_{23 \{34\}} - \cO^T_{43 \{32\}} \,, 
\\ \displaystyle
    \cO^T_{31 \{14\}} + \cO^T_{41 \{13\}} 
  - \cO^T_{32 \{24\}} - \cO^T_{42 \{23\}} \,. 
\end{array}
\end{equation}

For $h_{2,d}$ (representation $\tau^{(6)}_3$, $C=-1$):
\begin{equation} \begin{array}{c} \displaystyle
\cO^T_{1211} - \cO^T_{1222}
  + \cO^T_{13 \{32\}} + \cO^T_{23 \{31\}} 
  + \cO^T_{14 \{42\}} + \cO^T_{24 \{41\}} \,, 
\\ \displaystyle
\cO^T_{1311} - \cO^T_{1333}
  + \cO^T_{12 \{23\}} + \cO^T_{32 \{21\}} 
  + \cO^T_{14 \{43\}} + \cO^T_{34 \{41\}} \,, 
\\ \displaystyle
\cO^T_{1411} - \cO^T_{1444}
  + \cO^T_{12 \{24\}} + \cO^T_{42 \{21\}} 
  + \cO^T_{13 \{34\}} + \cO^T_{43 \{31\}} \,, 
\\ \displaystyle
\cO^T_{2322} - \cO^T_{2333}
  + \cO^T_{21 \{13\}} + \cO^T_{31 \{12\}} 
  + \cO^T_{24 \{43\}} + \cO^T_{34 \{42\}} \,, 
\\ \displaystyle
\cO^T_{2422} - \cO^T_{2444}
  + \cO^T_{21 \{14\}} + \cO^T_{41 \{12\}} 
  + \cO^T_{23 \{34\}} + \cO^T_{43 \{32\}} \,, 
\\ \displaystyle
\cO^T_{3433} - \cO^T_{3444}
  + \cO^T_{31 \{14\}} + \cO^T_{41 \{13\}} 
  + \cO^T_{32 \{24\}} + \cO^T_{42 \{23\}} \,. 
\end{array}
\end{equation}

For the vector current (representation $\tau^{(4)}_1$, $C=-1$):
\begin{equation} 
 \cO^V_\mu \,, \quad 1 \leq \mu \leq 4 \,.
\end{equation}

For the axial vector current (representation $\tau^{(4)}_4$, $C=+1$):
\begin{equation} 
 \cO^A_\mu \,, \quad 1 \leq \mu \leq 4 \,.
 \end{equation}

For the tensor current (representation $\tau^{(6)}_1$, $C=-1$):
\begin{equation} 
 \cO^T_{\mu \nu} \,, \quad 1 \leq \mu < \nu \leq 4 \,.
\end{equation}

\section{Example for conversion factors}
\label{sec.rexa}

We want to explain in detail, how the difference between $Z_{\RI}^{\MS}$
and $Z_{\MOM}^{\MS}$ arises, considering the operator 
\begin{equation} \label{exop}
\cO _{\{ \mu \nu \}} = 
\half \bar{u} ( \gamma_{\mu} \Dd{\nu} + \gamma_{\nu} \Dd{\mu} ) d 
\end{equation}
for $\mu \neq \nu$ as an instructive example. The operator $\cO_{v_{2,a}}$
belongs to this multiplet of operators. 

Using dimensional regularization, straightforward perturbation theory
in $4 - \epsilon$ dimensions yields in the Landau gauge
\begin{eqnarray} 
& & \Gamma (p) = 
  \mathrm i \left( \gamma_\mu p_\nu + \gamma_\nu p_\mu \right) \nonumber \\
& & {} + \frac{g^2}{16 \pi^2} C_F \left \{ \mathrm i 
  \left( \gamma_\mu p_\nu + \gamma_\nu p_\mu \right) 
   \left[ - \frac{8}{3} \left( \frac{2}{\epsilon} + \ln (4 \pi) 
         - \gamma_{\mathrm E} - \ln (p^2/\mu^2) \right) - \frac{31}{9} \right]
    - \frac{2}{3} p_\mu p_\nu \frac{\mathrm i \slashed{p}}{p^2} \right \} 
                                                                    + O(g^4)
\nonumber \\ {}
\end{eqnarray}
up to terms which vanish for $\epsilon \to 0$. For QCD we have $C_F = 4/3$. 
In this Appendix we restrict ourselves to one-loop order. Hence
the coupling constant $g$ can be identified with the bare coupling 
or with some renormalized coupling. Moreover, to this order and in the
Landau gauge the quark wave function renormalization constant $Z_q$ 
is equal to one in all schemes of interest to us ($\RI$, $\MS$ and $\MOM$), 
see also Eq.~(\ref{zqexp}). So we can ignore it in the following.

For the operator (\ref{exop}) we have
\begin{equation} 
\Gamma_{\mathrm {Born}} (p) = 
  \mathrm i \left( \gamma_\mu p_\nu + \gamma_\nu p_\mu \right)
\end{equation}
and the term proportional to $\slashed{p}$ represents an additional structure
which is not a multiple of the Born term. The renormalized
vertex function in the $\MS$ scheme reads
\begin{equation} 
\Gamma_{\MS} (p) = \Gamma_{\mathrm {Born}} (p) 
   + \frac{g^2}{16 \pi^2} C_F \left \{ \Gamma_{\mathrm {Born}} (p)
   \left[ \frac{8}{3}  \ln (p^2/\mu^2) - \frac{31}{9} \right]
    - \frac{2}{3} p_\mu p_\nu \frac{\mathrm i \slashed{p}}{p^2} \right \} 
                                                                 + O(g^4) 
\end{equation}
and we get
\begin{equation} 
Z^{\MS}_{\mathrm {dimreg}} = 1 + \frac{g^2}{16 \pi^2} C_F 
   \cdot \frac{8}{3} \left( \frac{2}{\epsilon} + \ln (4 \pi) 
         - \gamma_{\mathrm E} \right) + O(g^4) \,.
\end{equation}
On the other hand, we can represent $\Gamma (p)$ as a linear combination
of $\Gamma_{\mathrm {Born}} (p)$ and 
$p_\mu p_\nu (\mathrm i \slashed{p} / p^2)$. 
Requiring that in the $\MOM$ scheme the coefficient of 
$\Gamma_{\mathrm {Born}} (p)$ be unity for $p^2 = \mu^2$ we find
\begin{equation} 
Z^{\MOM}_{\mathrm {dimreg}} = 1 + \frac{g^2}{16 \pi^2} C_F 
   \left[ \frac{8}{3} \left( \frac{2}{\epsilon} + \ln (4 \pi) 
         - \gamma_{\mathrm E} \right) + \frac{31}{9} \right] + O(g^4)
\end{equation}
such that 
\begin{equation} 
\Gamma_{\MOM} (p) = \Gamma_{\mathrm {Born}} (p) 
   + \frac{g^2}{16 \pi^2} C_F \left \{ \Gamma_{\mathrm {Born}} (p)
                                 \cdot \frac{8}{3}  \ln (p^2/\mu^2) 
    - \frac{2}{3} p_\mu p_\nu \frac{\mathrm i \slashed{p}}{p^2} \right \} 
                                                               + O(g^4) \,.
\end{equation}
Finally we obtain from Eq.~(\ref{calcz})
\begin{equation} 
Z^{\RI}_{\mathrm {dimreg}} = 1 + \frac{g^2}{16 \pi^2} C_F 
   \left[ \frac{8}{3} \left( \frac{2}{\epsilon} + \ln (4 \pi) 
         - \gamma_{\mathrm E} \right) + \frac{31}{9} 
    + \frac{4}{3} \frac{p_\mu^2 p_\nu^2}{p^2 (p_\mu^2 + p_\nu^2)} \right] 
                                                                + O(g^4)\,.
\end{equation}
Now we can calculate 
\begin{equation} 
Z_{\RI}^{\MS} (\mu) = \frac{Z^{\MS}_{\mathrm {dimreg}}}
                                 {Z^{\RI}_{\mathrm {dimreg}}} =
  1 + \frac{g^2}{16 \pi^2} C_F \left( - \frac{31}{9} 
      - \frac{4}{3} \frac{p_\mu^2 p_\nu^2}{p^2 (p_\mu^2 + p_\nu^2)} \right) 
                                                                   + O(g^4)
\end{equation}
and 
\begin{equation} 
Z_{\MOM}^{\MS} (\mu) = \frac{Z^{\MS}_{\mathrm {dimreg}}}
                                 {Z^{\MOM}_{\mathrm {dimreg}}} =
  1 + \frac{g^2}{16 \pi^2} C_F \left( - \frac{31}{9} \right) + O(g^4) \,,
\end{equation}
in agreement with the results given in Appendix~\ref{sec.coeff}.

\section{Results from continuum perturbation theory}
\label{sec.coeff}

In this Appendix we collect the results from continuum perturbation 
theory that go into our computations. They all refer to $n_c = 3$ colors
and Landau gauge, but the number of flavors $n_f$ is left free. 
Note that we quote only the papers which give the results with the
largest number of loops.

We begin with the coefficients of the $\beta$ function 
(see Eq.~(\ref{defbeta})). In the $\MS$ 
scheme they are given by (see Ref.~\cite{ritbergen})
\begin{eqnarray}
\beta_0 & = & 11 - \frac{2}{3} n_f \,, \\
\beta_1 & = & 102 - \frac{38}{3} n_f \,, \\
\beta_2 & = & \frac{2857}{2} - \frac{5033}{18} n_f 
              + \frac{325}{54} n_f^2 \,, \\
\beta_3 & = & \frac{149753}{6} + 3564 \zeta_3 
          - \left( \frac{1078361}{162} + \frac{6508}{27} \zeta_3  \right) n_f
        \nonumber \\ & & {}
           + \left( \frac{50065}{162} + \frac{6472}{81} \zeta_3 \right) n_f^2
           + \frac{1093}{729} n_f^3 \,.
\end{eqnarray}
In the $\momt$ scheme one finds~\cite{cheret2}
\begin{eqnarray}
\beta_2 & = & \frac{186747}{64} - \frac{1683}{4} \zeta_3
           - \left( \frac{35473}{96} - \frac{65}{6} \zeta_3  \right) n_f
           - \left( \frac{829}{54} - \frac{8}{9} \zeta_3 \right) n_f^2
           + \frac{8}{9} n_f^3 \,, \\
\beta_3 & = & \frac{20783939}{128} - \frac{1300563}{32} \zeta_3
                                      - \frac{900075}{32} \zeta_5
          - \left( \frac{2410799}{64} - \frac{1323259}{144} \zeta_3  
                     - \frac{908995}{144} \zeta_5 \right) n_f
         \nonumber \\ & & {}
           + \left( \frac{1464379}{648} - \frac{12058}{27} \zeta_3 
                             - \frac{7540}{27} \zeta_5 \right) n_f^2
           - \left( \frac{3164}{27} - \frac{64}{9} \zeta_3 \right) n_f^3
           + \frac{320}{81} n_f^4 \,,
\end{eqnarray}
while $\beta_0$ and $\beta_1$ are scheme independent in the Landau gauge.

We now turn to the coefficients of the anomalous dimension in the $\MS$
scheme. Our conventions have been given in Sec.~\ref{sec.pert}, 
see in particular Eqs.~(\ref{anodim}) and (\ref{defgamma}). For
notational simplicity the superscript $\MS$ will be omitted. Note that
we assume an anticommuting $\gamma_5$, so the coefficients $\gamma_i$
correspond more precisely to the so-called naive dimensional 
regularization.

While $\cO^V_\mu$ and $\cO^A_\mu$ have of course vanishing anomalous
dimension, we find for $\cO^S$ and $\cO^P$ (see 
Refs.~\cite{vermaseren,chetyrkin})
\begin{eqnarray}
\gamma_0 & = & -8 \,, \\
\gamma_1 & = & - \frac{404}{3} + \frac{40}{9} n_f \,, \\
\gamma_2 & = & - 2498 + \left( \frac{4432}{27} 
           + \frac{320}{3} \zeta_3 \right) n_f + \frac{280}{81} n_f^2 \,, \\
\gamma_3 & = & - \frac{4603055}{81} - \frac{271360}{27}\zeta_3 + 17600 \zeta_5
              + \left( \frac{183446}{27} + \frac{68384}{9} \zeta_3 
                - 1760 \zeta_4 - \frac{36800}{9} \zeta_5 \right) n_f
              \nonumber \\ & & {}
              + \left( -\frac{10484}{243} - \frac{1600}{9} \zeta_3 
                   + \frac{320}{3} \zeta_4 \right) n_f^2
       + \left( \frac{664}{243} - \frac{128}{27} \zeta_3 \right) n_f^3 \,.
\end{eqnarray}
For $\cO^T_{\mu \nu}$ we have~\cite{gracey1}
\begin{eqnarray}
\gamma_0 & = & \frac{8}{3} \,, \\
\gamma_1 & = & \frac{724}{9} - \frac{104}{27} n_f \,, \\
\gamma_2 & = & \frac{105110}{81} - \frac{1856}{27} \zeta_3
             - \left( \frac{10480}{81} 
             + \frac{320}{9} \zeta_3 \right) n_f - \frac{8}{9} n_f^2 \,.
\end{eqnarray}
The operators $\cO_{v_{2,a}}$, $\cO_{v_{2,b}}$, $\cO_{r_{2,a}}$
and $\cO_{r_{2,b}}$ have the same anomalous dimension. From Ref.~\cite{larin}
we get:
\begin{eqnarray}
\gamma_0 & = & \frac{64}{9} \,, \\
\gamma_1 & = & \frac{23488}{243} - \frac{512}{81} n_f \,, \\
\gamma_2 & = & \frac{11028416}{6561} + \frac{2560}{81} \zeta_3
             - \left( \frac{334400}{2187} 
             + \frac{2560}{27} \zeta_3 \right) n_f - \frac{1792}{729} n_f^2 \,.
\end{eqnarray}
For $\cO_{v_3}$, $\cO_{v_{3,a}}$, $\cO_{a_2}$ and $\cO_{r_3}$ we extract
from Ref.~\cite{retey}:
\begin{eqnarray}
\gamma_0 & = & \frac{100}{9} \,, \\
\gamma_1 & = & \frac{34450}{243} - \frac{830}{81} n_f \,, \\
\gamma_2 & = & \frac{64486199}{26244} + \frac{2200}{81} \zeta_3
             - \left( \frac{469910}{2187} 
             + \frac{4000}{27} \zeta_3 \right) n_f - \frac{2569}{729} n_f^2 \,.
\end{eqnarray}
The anomalous dimension of $\cO_{v_4}$ can be found in Ref.~\cite{larin}:
\begin{eqnarray}
\gamma_0 & = & \frac{628}{45} \,, \\
\gamma_1 & = & \frac{5241914}{30375} - \frac{26542}{2025} n_f \,, \\
\gamma_2 & = & \frac{245787905651}{82012500} + \frac{11512}{405} \zeta_3
             - \left( \frac{726591271}{2733750} 
       + \frac{5024}{27} \zeta_3 \right) n_f - \frac{384277}{91125} n_f^2 \,.
\end{eqnarray}
The three-loop anomalous dimension of transversity operators
has been calculated by Gracey.
For $\cO_{h_{1,a}}$ and $\cO_{h_{1,b}}$ we get from Ref.~\cite{gracey2}
\begin{eqnarray}
\gamma_0 & = & 8 \,, \\
\gamma_1 & = & 124 - 8 n_f \,, \\
\gamma_2 & = & \frac{19162}{9} 
             - \left( \frac{5608}{27} 
       + \frac{320}{3} \zeta_3 \right) n_f - \frac{184}{81} n_f^2 \,.
\end{eqnarray}
In Ref.~\cite{gracey3} we find for $\cO_{h_{2,a}}$, $\cO_{h_{2,b}}$, 
$\cO_{h_{2,c}}$ and $\cO_{h_{2,d}}$:
\begin{eqnarray}
\gamma_0 & = &  \frac{104}{9} \,, \\
\gamma_1 & = & \frac{38044}{243} - \frac{904}{81} n_f  \,, \\
\gamma_2 & = & \frac{17770162}{6561} + \frac{1280}{81} \zeta_3
             - \left( \frac{552308}{2187} 
       + \frac{4160}{27} \zeta_3 \right) n_f - \frac{2408}{729} n_f^2 \,.
\end{eqnarray}
Finally, we can take the anomalous dimension of the quark field 
to four loops from Ref.~\cite{cheret1}: 
\begin{eqnarray}
\gamma_0 & = & 0 \,, \\
\gamma_1 & = & \frac{134}{3} - \frac{8}{3} n_f \,, \\
\gamma_2 & = & \frac{20729}{18} - 79 \zeta_3 - \frac{1100}{9} n_f
               + \frac{40}{27} n_f^2 \,, \\
\gamma_3 & = & \frac{2109389}{81} - \frac{565939}{162}\zeta_3 
              + \frac{2607}{2} \zeta_4 - \frac{761525}{648} \zeta_5
              - \left( \frac{324206}{81} + \frac{4582}{27} \zeta_3 
                + 79 \zeta_4 + \frac{320}{3} \zeta_5 \right) n_f
              \nonumber \\ & & {}
           + \left( \frac{7706}{81} + \frac{320}{9} \zeta_3 \right) n_f^2
           + \frac{280}{243} n_f^3 \,.
\end{eqnarray}

Let us now consider the coefficients needed for the conversion from the 
$\RI$ scheme and the MOM scheme to the $\MS$ scheme, as defined in 
Eqs.~(\ref{convri}) and (\ref{convmom}), respectively. Since the 
$\RI$ scheme is in general
not covariant, these coefficients may depend on the direction of the 
momentum $p$. 
In order to keep the paper at a reasonable length we refrain from
giving the bases that are used in the representation of the vertex 
functions and enter the precise definition of the $\MOM$ scheme.
The calculation makes use of the perturbative expressions
for the vertex functions and of the ratio 
$Z_{q,\mathrm {bare}}^\MS / Z_{q,\mathrm {bare}}^\MOM = 
Z_{q,\mathrm {bare}}^\MS / Z_{q,\mathrm {bare}}^\RI$.
This ratio coincides with the quantity $C_2^{\mathrm {RI}^\prime}$ in 
Ref.~\cite{cheret1}, where we can read off the expansion
\begin{equation} \label{zqexp}
\frac{Z_q^\MS}{Z_q^\MOM} = 1
  + b_2 \left(  \frac{g^{\MS} (\mu)^2}{16 \pi^2} \right)^2
  + b_3 \left(  \frac{g^{\MS} (\mu)^2}{16 \pi^2} \right)^3 + \cdots
\end{equation}
with
\begin{eqnarray}
b_2 & = & - \frac{359}{9} + 12 \zeta_3 + \frac{7}{3} n_f \,, \\
b_3 & = & - \frac{439543}{162} + \frac{8009}{6}\zeta_3 
              + \frac{79}{4} \zeta_4 - \frac{1165}{3} \zeta_5
              + \left( \frac{24722}{81} - \frac{440}{9} \zeta_3 \right) n_f
           - \frac{1570}{243} n_f^2 \,.
\end{eqnarray}

From Ref.~\cite{cheret1} we get for $\cO^S$ and $\cO^P$
\begin{eqnarray}
c_1 & = &  \frac{16}{3}\,, \\
c_2 & = &  \frac{4291}{18} - \frac{152}{3}\zeta_3 - \frac{83}{9} n_f \,, \\
c_3 & = &  \frac{3890527}{324} - \frac{224993}{54}\zeta_3 
                                              + \frac{2960}{9} \zeta_5
         - \left( \frac{241294}{243} - \frac{4720}{27} \zeta_3 
                                     + \frac{80}{3} \zeta_4 \right) n_f
              \nonumber \\ & & {}
         + \left( \frac{7514}{729} + \frac{32}{27} \zeta_3 \right) n_f^2 \,.
\end{eqnarray}
With the help of Ref.~\cite{gracey4} we find for $\cO^T_{\mu \nu}$ 
\begin{eqnarray}
c_1 & = &  0 \,, \\
c_2 & = &  -\frac{3847}{54} + \frac{184}{9}\zeta_3 + \frac{313}{81} n_f \,, \\
c_3 & = &   -\frac{9858659}{2916} + \frac{678473}{486} \zeta_3 
                   + \frac{1072}{81} \zeta_4 - \frac{10040}{27} \zeta_5
         + \left( \frac{286262}{729} - \frac{2096}{27} \zeta_3 
                                     + \frac{80}{9} \zeta_4 \right) n_f
              \nonumber \\ & & {}
         - \left( \frac{13754}{2187} + \frac{32}{81} \zeta_3 \right) n_f^2 \,.
\end{eqnarray}
For the vector and axial-vector currents $\cO^V_\mu$ and $\cO^A_\mu$ 
one finds from Ref.~\cite{gracey4}
\begin{eqnarray}
c_1 & = &  0 \,, \\
c_2 & = &  \left( -\frac{134}{3} 
             + \frac{8}{3} n_f \right) R \,, \\
c_3 & = &   \left( -\frac{52321}{18} + 607 \zeta_3 
         + \left( \frac{8944}{27} - 32 \zeta_3 \right) n_f
         - \frac{208}{27} n_f^2 \right) R
\end{eqnarray}
with $R = p_\mu^2/p^2$. If the index $\mu$ is averaged over in the 
renormalization condition (see Eq.~(\ref{calczav})), $R$ takes 
the value $1/4$. 

Why don't $c_2$ and $c_3$ vanish although the vector current is 
conserved in the continuum?
In the continuum the quark propagator $S(p)$ and the vertex function
$\Gamma_\mu (p)$ of the vector current are linked by the Ward identity
\begin{equation}
\mathrm i \Gamma_\mu (p) = \frac{\partial}{\partial p_\mu} S^{-1}(p) \,.
\end{equation}
By Lorentz symmetry the massless inverse propagator must have the form
\begin{equation}
  S^{-1}(p) = \mathrm i A(p^2) \slashed{p} \,,
\end{equation}
where we expect $A$ to depend logarithmically on $p^2/\mu^2$.
Therefore the vertex function has the form
\begin{equation}
\Gamma_\mu(p) = A(p^2) \gamma_\mu + 
     \frac{\mathrm d A}{\mathrm d p^2} 2 p_\mu \slashed{p} \,,
\end{equation}
and the trace with the Born term gives
\begin{equation}
\frac{1}{12} {\rm tr} \left( \gamma_\mu  \Gamma_\mu(p) \right)
= A(p^2) + 2 p^2 \frac{\mathrm d A}{\mathrm d p^2} \frac{p_\mu^2}{p^2}
= A(p^2) +  2 p^2 \frac{\mathrm d A}{\mathrm d p^2} R \,.
\end{equation}
The vector Ward identity therefore requires the existence
of terms proportional to $R$ in the trace, with a coefficient
given by the logarithmic derivative of $A$. The coefficient
$c_1$ vanishes because in the Landau gauge there is no term in the
propagator of the form $g^2 \ln (p^2/\mu^2)$, but in most other
gauges $c_1$ is nonzero and
we already have a term proportional to $R$ at one loop.

In the following cases we express the coefficients $c_1$, $c_2$ and $c_3$
in the form
\begin{eqnarray}
c_1 & = & c^{(1)}_1 + c^{(2)}_1 R \,, \\
c_2 & = & c^{(1)}_2  + b_2 + c^{(2)}_2 R \,, \\
c_3 & = & c^{(1)}_3  + b_2 c^{(1)}_1 + b_3 
          + \left( c^{(2)}_3 + b_2 c^{(2)}_1 \right) R \,,
\end{eqnarray}
where $R$ contains the momentum dependence and is given in 
Table~\ref{tab:Rval}. The expressions $R^{(j)}(p)$ for $j=1,2,3,4$ 
read:
\begin{eqnarray} \label{rp1}
R^{(1)}_{\mu \nu \lambda \rho}(p) & = & 
  \frac{p_\mu^2 (p_\nu^2 - p_\lambda^2)^2}
     {(p_\nu^2 - p_\lambda^2)^2 + 4 p_\mu^2 (p_\nu^2 + p_\lambda^2)}
+ \frac{p_\mu^2 (p_\nu^2 + p_\lambda^2 - 2 p_\rho^2)^2}
     {(p_\nu^2 + p_\lambda^2 - 2 p_\rho^2)^2 
              + 4 p_\mu^2 (p_\nu^2 + p_\lambda^2 + 4 p_\rho^2)} \\
\label{rp2}
R^{(2)}_{\mu \nu \lambda}(p) & = & 
  \frac{p_\mu^2 p_\nu^2 p_\lambda^2}
       {p_\mu^2 p_\nu^2 +p_\mu^2 p_\lambda^2 +p_\nu^2 p_\lambda^2} \\
\label{rp3}
R^{(3)}(p) & = & 
  \frac{(p_1^2 - p_4^2)^2 (p_2^2 - p_3^2)^2}
       {(p_1^2 + p_4^2)(p_2^2 - p_3^2)^2 + (p_2^2 + p_3^2)(p_1^2 - p_4^2)^2} \\
\label{rp4}
R^{(4)}(p) & = & 
  \frac{\left( (p_1^2 + p_4^2)(p_2^2 + p_3^2) - 2 p_1^2 p_4^2 
                                              - 2 p_2^2 p_3^2 \right)^2}
   {p_1^2 (p_2^2 + p_3^2 - 2 p_4^2)^2 + p_2^2 (p_1^2 + p_4^2 - 2 p_3^2)^2 
  + p_3^2 (p_1^2 + p_4^2 - 2 p_2^2)^2 + p_4^2 (p_2^2 + p_3^2 - 2 p_1^2)^2}
\nonumber \\  {} 
\end{eqnarray}

\begin{table}
\caption{Momentum dependent factors $R$. For the definitions of the lengthier
expressions $R^{(j)}(p)$ see Eqs.~(\ref{rp1}) - (\ref{rp4}).}
\label{tab:Rval}
\begin{ruledtabular}
\begin{tabular}{ll}
$\cO_{v_{2,a}}$, $\cO_{r_{2,a}}$ &
   $\displaystyle \frac{2 p_1^2 p_4^2}{p^2 (p_1^2 + p_4^2)}$  \\[0.5cm]
$\cO_{v_{2,b}}$ , $\cO_{r_{2,b}}$ &
   $\displaystyle \frac{\left( p_4^2 - (p_1^2 + p_2^2 + p_3^2)/3 \right)^2}
        {2 p^2 \left( p_4^2 + (p_1^2 + p_2^2 + p_3^2)/9 \right)} $ \\[0.5cm]
$\cO_{v_3}$, $\cO_{r_3}$ & 
   $\displaystyle \frac{-9 p_4^2 \left( p_1^2 - (p_2^2 + p_3^2)/2 \right)^2}
      {p^2 \left( (4 p_1^2 + p_2^2 + p_3^2) p_4^2 + 
        \left( p_1^2 - (p_2^2 + p_3^2)/2 \right)^2 \right)} $ \\[0.5cm]
$\cO_{v_{3,a}}$, $\cO_{a_2}$ &  
   $\displaystyle \frac{-9 p_1^2 p_2^2 p_4^2}
      {p^2 \left( (p_1 p_4)^2 + (p_2 p_4)^2 + (p_1 p_2)^2\right)} $ \\[0.5cm]
$\cO_{v_4}$ & 
   $\displaystyle \frac{64 \left( p_1^2 - p_2^2 \right)^2 
                                  \left( p_4^2 - p_3^2 \right)^2}
      {p^2 \left( \left( p_1^2 + p_2^2 \right) \left( p_4^2 - p_3^2 \right)^2  
 + \left( p_3^2 + p_4^2 \right) \left( p_2^2 - p_1^2 \right)^2 \right)} $ \\[0.5cm]
$\overline{\cO}_{v_{2,a}}$, $\overline{\cO}_{r_{2,a}}$ &
   $\displaystyle \frac{1}{3 p^2} \sum_{\mu < \nu} 
                  \frac{p_\mu^2 p_\nu^2}{p_\mu^2 + p_\nu^2}$ \\[0.5cm]
$\overline{\cO}_{v_{2,b}}$ , $\overline{\cO}_{r_{2,b}}$ &
   $\displaystyle \frac{1}{6 p^2} \left[ 
      \frac{(p_1^2 + p_2^2 - p_3^2 - p_4^2)^2}{p^2} +
      \frac{(p_3^2 - p_4^2)^2}{p_3^2 + p_4^2} +
      \frac{(p_1^2 - p_2^2)^2}{p_1^2 + p_2^2} \right] $ \\[0.5cm]
$\overline{\cO}_{v_3}$, $\overline{\cO}_{r_3}$ &   
   $\displaystyle - \frac{9}{8 p^2} \left( R^{(1)}_{1234}(p) 
+ R^{(1)}_{2134}(p) + R^{(1)}_{3124}(p) + R^{(1)}_{4123}(p) \right) $ \\[0.5cm]
$\overline{\cO}_{v_{3,a}}$, $\overline{\cO}_{a_2}$ &  
   $\displaystyle - \frac{9}{4 p^2} \left( R^{(2)}_{123}(p) + 
   R^{(2)}_{124}(p) +  R^{(2)}_{134}(p) +  R^{(2)}_{234}(p) \right) $ \\[0.5cm]
$\overline{\cO}_{v_4}$ &   
   $\displaystyle  \frac{32}{p^2} \left( R^{(3)}(p) + R^{(4)}(p) \right) $
\end{tabular}
\end{ruledtabular}
\end{table}

For $\cO_{v_{2,a}}$, $\cO_{v_{2,b}}$, $\cO_{r_{2,a}}$ and $\cO_{r_{2,b}}$ 
one extracts from Ref.~\cite{gracey2}
\begin{eqnarray}
c^{(1)}_1 & = &  - \frac{124}{27} \,, \\
c^{(1)}_2 & = &  - \frac{68993}{729} + \frac{160}{9} \zeta_3
                 + \frac{2101}{243} n_f  \,, \\
c^{(1)}_3 & = &   - \frac{451293899}{157464} + \frac{1105768}{2187} \zeta_3
                  - \frac{8959}{324} \zeta_4 - \frac{4955}{81} \zeta_5
         + \left( \frac{8636998}{19683} - \frac{224}{81} \zeta_3 
                                     + \frac{640}{27} \zeta_4 \right) n_f
              \nonumber \\ & & {}
  - \left( \frac{63602}{6561} + \frac{256}{243} \zeta_3 \right) n_f^2 \,, \\
c^{(2)}_1 & = &   - \frac{8}{9} \,, \\
c^{(2)}_2 & = &   - \frac{2224}{27} - \frac{40}{9} \zeta_3
                  + \frac{40}{9} n_f  \,, \\
c^{(2)}_3 & = &  - \frac{136281133}{26244} + \frac{376841}{243} \zeta_3 
                 - \frac{43700}{81} \zeta_5
          + \left( \frac{15184}{27} - \frac{1232}{81} \zeta_3 \right) n_f
          - \frac{9680}{729} n_f^2 \,. 
\end{eqnarray}
In the case of the operators $\cO_{h_{1,a}}$ and $\cO_{h_{1,b}}$ one
obtains from Ref.~\cite{gracey2}
\begin{eqnarray}
c^{(1)}_1 & = &  - \frac{14}{3} \,, \\
c^{(1)}_2 & = &  - \frac{2237}{18} + \frac{62}{3} \zeta_3
                 + \frac{32}{3} n_f  \,, \\
c^{(1)}_3 & = &   - \frac{1852993}{432} + \frac{97391}{108} \zeta_3
                  - \frac{79}{4} \zeta_4 - \frac{7060}{27} \zeta_5
         + \left( \frac{306881}{486} - \frac{122}{9} \zeta_3 
                                     + \frac{80}{3} \zeta_4 \right) n_f
              \nonumber \\ & & {}
  - \left( \frac{1160}{81} + \frac{32}{27} \zeta_3 \right) n_f^2 \,, \\
c^{(2)}_1 & = & c^{(2)}_2 = c^{(2)}_3 = 0 \,. 
\end{eqnarray}
Similarly, Ref.~\cite{gracey3} yields for $\cO_{h_{2,a}}$, 
$\cO_{h_{2,b}}$, $\cO_{h_{2,c}}$ and $\cO_{h_{2,d}}$
\begin{eqnarray}
c^{(1)}_1 & = &  - \frac{218}{27} \,, \\
c^{(1)}_2 & = &  - \frac{669202}{3645} + \frac{452}{15} \zeta_3
                 + \frac{4394}{243} n_f  \,, \\
c^{(1)}_3 & = &   - \frac{1020141085}{157464} + \frac{59050063}{43740} \zeta_3
                  - \frac{7679}{324} \zeta_4 - \frac{12434}{27} \zeta_5
              \nonumber \\ & & {}
         + \left( \frac{98639141}{98415} + \frac{12712}{1215} \zeta_3 
                                     + \frac{1040}{27} \zeta_4 \right) n_f
  - \left( \frac{177970}{6561} + \frac{416}{243} \zeta_3 \right) n_f^2 \,, \\
c^{(2)}_1 & = & c^{(2)}_2 = c^{(2)}_3 = 0 \,.
\end{eqnarray}
The operators $\cO_{v_3}$, $\cO_{v_{3,a}}$, $\cO_{r_3}$ and $\cO_{a_2}$
require the coefficients
\begin{eqnarray}
c^{(1)}_1 & = &  - \frac{214}{27} \,, \\
c^{(1)}_2 & = &  - \frac{4763093}{29160} + \frac{152}{5} \zeta_3
                 + \frac{32363}{1944} n_f  \,, \\
c^{(1)}_3 & = &   - \frac{8619089351}{1574640} + \frac{12125507}{10935} \zeta_3
                  - \frac{8599}{324} \zeta_4 - \frac{2525}{9} \zeta_5
         + \left( \frac{1364405723}{1574640} + \frac{814}{135} \zeta_3 
                                     + \frac{1000}{27} \zeta_4 \right) n_f
              \nonumber \\ & & {}
  - \left( \frac{1227463}{52488} + \frac{400}{243} \zeta_3 \right) n_f^2 \,, \\
c^{(2)}_1 & = &   \frac{4}{9} \,, \\
c^{(2)}_2 & = &   \frac{4432}{135} + \frac{56}{15} \zeta_3
                  - \frac{50}{27} n_f  \,, \\
c^{(2)}_3 & = &  \frac{279011797}{131220} - \frac{1717789}{2430} \zeta_3 
                 + \frac{9370}{27} \zeta_5
          - \left( \frac{1665047}{7290} + \frac{28}{5} \zeta_3 \right) n_f
          + \frac{4210}{729} n_f^2 \,, 
\end{eqnarray}
extracted from Ref.~\cite{gracey3}. For $\cO_{v_4}$ we find
\begin{eqnarray}
c^{(1)}_1 & = &  - \frac{7214}{675} \,, \\
c^{(1)}_2 & = &  - \frac{764724499}{3645000} + \frac{1756}{45} \zeta_3
                 + \frac{5655503}{243000} n_f  \,, \\
c^{(1)}_3 & = &   - \frac{282373048664443}{39366000000} 
                  + \frac{796627067}{546750} \zeta_3
                  - \frac{43507}{1620} \zeta_4 - \frac{38398}{81} \zeta_5
              \nonumber \\ & & {}
         + \left( \frac{1160956742099}{984150000} + \frac{3208}{135} \zeta_3 
                                     + \frac{1256}{27} \zeta_4 \right) n_f
      - \left( \frac{1167227687}{32805000} 
                           + \frac{2512}{1215} \zeta_3 \right) n_f^2 \,, \\
c^{(2)}_1 & = &   - \frac{1}{40} \,, \\
c^{(2)}_2 & = &   - \frac{731129}{432000} - \frac{23}{90} \zeta_3
                  + \frac{119}{1200} n_f  \,, \\
c^{(2)}_3 & = &  - \frac{1047728166241}{9331200000} 
                 + \frac{109467991}{2916000} \zeta_3 
                 - \frac{13111}{648} \zeta_5
  + \left( \frac{232632277}{19440000} + \frac{755}{972} \zeta_3 \right) n_f
              \nonumber \\ & & {}
          - \frac{51959}{162000} n_f^2 \,, 
\end{eqnarray}
from Ref.~\cite{gracey3}. When using Gracey's results given in 
Refs.~\cite{gracey2,gracey3,gracey4} it is important to note that 
Gracey's $\RI$ scheme is not the same as ours. Furthermore, 
Eqs.~(A.1) and (A.9) in Ref.~\cite{gracey3} are not quite correct, 
but we hope that we have worked with properly rectified versions.
Another correction concerns Eq.~(4.4) in Ref.~\cite{gracey2}, where
the coefficient of $T_F N_f$ in the $a^2$ contribution should be
468 and not 486.

The corresponding coefficients $c^\prime_1$, $c^\prime_2$ and $c^\prime_3$
in Eq.~(\ref{convmom}) for the conversion from MOM to $\MS$ are 
obtained by setting $R=0$. So in the cases where the coefficients
$c_i$ are independent of $R$ we have $c^\prime_i = c_i$ and the $\RI$
scheme can be identified with the MOM scheme, at least to the order
considered.

\section{Lattice perturbation theory to two loops}
\label{sec.2loops}

In the two-loop approximation of bare lattice perturbation theory we have
\begin{equation} 
Z_{\mathrm {bare}}^{\MS} (\mu,a)_{\mathrm {pert}}
 = 1 + \frac{g^2}{16 \pi^2} (- \gamma_0 \ln (a \mu) + z_1 ) 
  + \left( \frac{g^2}{16 \pi^2} \right)^2 ( l_1 \ln^2 (a \mu)
       + l_2 \ln (a \mu) + z_2 ) \,.
\end{equation}
For the currents the coefficients $l_1$, $l_2$, $z_1$, $z_2$ 
can be read off from Refs.~\cite{panagopoulos1,panagopoulos2} 
as functions of $\csw$. Note that $z_1 = - C_F \Delta$ in the 
notation of Sec.~\ref{sec.latpert}.

Anticipating that we may want to expand in a coupling $g_{\mathrm {LAT}}$ 
different from the bare lattice coupling $g$, e.g., the boosted coupling
$g_\Box$ (see Eq.~(\ref{boosted})), we express the $\MS$ coupling 
$g_{\MS}$ as a function of $g_{\mathrm {LAT}}$:
\begin{equation} 
\frac{1}{g_{\MS}^2} = \frac{1}{g_{\mathrm {LAT}}^2} 
   + 2 \frac{\beta_0}{16 \pi^2} \ln (a \mu) - t_1^{\mathrm {LAT}} 
   + \left( 2 \frac{\beta_1}{(16 \pi^2)^2} \ln (a \mu) 
                   - t_2^{\mathrm {LAT}} \right) g_{\mathrm {LAT}}^2
   + O(g_{\mathrm {LAT}}^4) \,.
\end{equation}
Here $t_i^{\mathrm {LAT}} = t_i - p_i$ ($i=1,2$), where the constants 
$p_i$ encode the relation between $g$ and $g_{\mathrm {LAT}}$:
\begin{equation} 
\frac{1}{g_{\mathrm {LAT}}^2} = \frac{1}{g^2} - p_1 - p_2 g^2 + O(g^4) \,.
\end{equation}
For $g_{\mathrm {LAT}} = g$ one has $p_i = 0$, hence 
$t_i^{\mathrm {LAT}} = t_i$, and the relation
between $g_{\MS}$ and $g$ takes the usual form~\cite{luwe,bode,christou,bode2}.

For setting up tadpole improvement we need the expansion
\begin{eqnarray} 
u_0 & = & 1 + r_1 \frac{g^2}{16 \pi^2} 
            + r_2 \left( \frac{g^2}{16 \pi^2} \right)^2 + O(g^6) \\
 & = & 1 + r_1 \frac{g_{\mathrm {LAT}}^2}{16 \pi^2} 
    + r_2^{\mathrm{LAT}}
        \left( \frac{g_{\mathrm {LAT}}^2}{16 \pi^2} \right)^2 
                                        + O(g_{\mathrm {LAT}}^6) \,,
\end{eqnarray} 
where $r_2^{\mathrm{LAT}} = r_2 - 16 \pi^2 r_1 p_1$. 
The coefficients $r_1$ and $r_2$ can be found from Ref.~\cite{athen}.
For an operator with $n_D$ covariant derivatives the
tadpole-improved two-loop expression for the renormalization
factor then takes the form
\begin{equation} \begin{array}{l} \displaystyle
Z_{\mathrm {bare}}^{\MS} (\mu,a)_{\mathrm {ti}}
= u_0^{1-n_D} \bigg[1 + \frac{g_{\mathrm {LAT}}^2}{16 \pi^2}  
     \left( - \gamma_0 \ln (a \mu) + z_1 + (n_D-1) r_1 \right)
\\ \displaystyle {}
  + \left( \frac{g_{\mathrm {LAT}}^2}{16 \pi^2} \right)^2
    \left( l_1 \ln^2 (a \mu) + (l_2 + 16 \pi^2 p_1 \gamma_0 
              - (n_D-1) r_1 \gamma_0 ) \ln (a \mu) 
\right. \\ \displaystyle \left. {}
    + z_2^{\mathrm {LAT}} + (n_D-1) r_2^{\mathrm {LAT}}
    + \half (n_D-1)(n_D-2) r_1^2 + (n_D-1) r_1 z_1 \right) 
                                 + O(g_{\mathrm {LAT}}^6) \bigg] \,,
\end{array}
\end{equation}
where $z_2^{\mathrm {LAT}} = z_2 - 16 \pi^2 p_1 z_1$. Of course, for
the currents we have $n_D = 0$. As in Sec.~\ref{sec.latpert}, 
the corresponding estimates for $Z^{\mathrm {RGI}}$ are finally given by 
$\Delta Z ^{\MS} (\mu_0) Z_{\mathrm {bare}}^{\MS} (\mu_0,a)_{\mathrm {pert}}$
and 
$\Delta Z ^{\MS} (\mu_0) Z_{\mathrm {bare}}^{\MS} (\mu_0,a)_{\mathrm {ti}}$
with $\mu_0 = 1/a$. For the expansion parameter we take
$g_{\mathrm {LAT}} = g_\Box = g/u_0^2$. Therefore we have
\begin{eqnarray} 
p_1 & = & - \frac{r_1}{4 \pi^2} = \frac{1}{4} C_F = \frac{1}{3} \,, \\
p_2 & = & - \frac{2 r_2 + 3 r_1^2}{128 \pi^4} \,.
\end{eqnarray} 

In order to implement TRB perturbation theory at the two-loop level we
start from Eq.~(\ref{zrgilat}). Using the three-loop expressions for 
$\gamma^{\mathrm {LAT}}$ and $\beta^{\mathrm {LAT}}$ we obtain
\begin{equation} 
Z^{\mathrm {RGI}} 
  = \left( 2 \beta_0 \frac {g_{\mathrm {LAT}}^2}{16 \pi^2} \right)
              ^{-\frac{\gamma_0}{2 \beta_0}} 
    \exp \{ F(g_{\mathrm {LAT}}^2/(16 \pi^2)) \}
\end{equation}
with
\begin{equation} 
F(x) = 
 \frac{\beta_2^{\mathrm {LAT}} \gamma_0 - \beta_0 \gamma_2^{\mathrm {LAT}}}
      {4 \beta_0 \beta_2^{\mathrm {LAT}}} f_1(x) + 
 \frac{\beta_1 \beta_2^{\mathrm {LAT}} \gamma_0 
       + \beta_0 \beta_1 \gamma_2^{\mathrm {LAT}} 
       - 2 \beta_0 \beta_2^{\mathrm {LAT}} \gamma_1^{\mathrm {LAT}}}
      {2 \beta_0^2 \beta_2^{\mathrm {LAT}}} f_2(x) \,,
\end{equation}
where
\begin{equation} 
f_1(x) = \ln \left( 1 + \frac{\beta_1}{\beta_0} x 
                      + \frac{\beta_2^{\mathrm {LAT}}}{\beta_0} x^2 \right)
\end{equation}
and
\begin{equation} 
f_2(x) = \frac{1}
    {\sqrt{4 \beta_2^{\mathrm {LAT}}/\beta_0 - (\beta_1/\beta_0)^2}}
 \arctan \left( 
   \frac{\sqrt{4 \beta_2^{\mathrm {LAT}}/\beta_0 - (\beta_1/\beta_0)^2}x}
        {2 + (\beta_1/\beta_0)x} \right) \,.
\end{equation}
The explicit expression for $\gamma_1^{\mathrm {LAT}}$ has been given in
Eq.~(\ref{g1lat}). For the tadpole-improvement factor we make the ansatz 
\begin{equation}
 u_0^{1-n_D} \exp \left( c_1 f_1(g_{\mathrm{LAT}}^2/(16 \pi^2)) 
                       + c_2 f_2(g_{\mathrm{LAT}}^2/(16 \pi^2)) \right)
\end{equation}
determining the coefficients $c_1$ and $c_2$ such that
\begin{equation}
\exp \left( c_1 f_1(g_{\mathrm{LAT}}^2/(16 \pi^2)) 
             + c_2 f_2(g_{\mathrm{LAT}}^2/(16 \pi^2)) \right) =
u_0^{n_D-1} + O(g_{\mathrm{LAT}}^6) \,.
\end{equation}
Then the final result in TRB perturbation theory reads
\begin{equation} 
Z^{\mathrm {RGI}}_{\mathrm {TRB}} =
u_0^{1-n_D}   \left( 2 \beta_0 \frac {g_{\mathrm{LAT}}^2}{16 \pi^2}\right)
                             ^{-\frac{\gamma_0}{2 \beta_0}}
  \exp \left\{ \tilde{c}_1 f_1(g_{\mathrm{LAT}}^2/(16 \pi^2)) 
            + \tilde{c}_2 f_2(g_{\mathrm{LAT}}^2/(16 \pi^2)) \right\}
\end{equation}
with
\begin{eqnarray} 
\tilde{c}_1 & = & 
 \frac{\beta_2^{\mathrm {LAT}} \gamma_0 - \beta_0 \gamma_2^{\mathrm {LAT}}}
      {4 \beta_0 \beta_2^{\mathrm {LAT}}} 
  + (n_D-1) \frac{\beta_0}{\beta_2^{\mathrm{LAT}}} 
   \left( r_2^{\mathrm{LAT}} + \frac{r_1}{2} \frac{\beta_1}{\beta_0} 
              - \frac{r_1^2}{2} \right) \,, \\[0.5cm]
\tilde{c}_2 & = & 
 \frac{\beta_1 \beta_2^{\mathrm {LAT}} \gamma_0 
       + \beta_0 \beta_1 \gamma_2^{\mathrm {LAT}} 
       - 2 \beta_0 \beta_2^{\mathrm {LAT}} \gamma_1^{\mathrm {LAT}}}
      {2 \beta_0^2 \beta_2^{\mathrm {LAT}}} \nonumber
\\ & & \hspace*{3.0cm} {} 
  + 2 (n_D-1) \left( r_1 - \frac{\beta_1}{\beta_2^{\mathrm{LAT}}} 
    \left( r_2^{\mathrm{LAT}} + \frac{r_1}{2} \frac{\beta_1}{\beta_0}
              - \frac{r_1^2}{2} \right) \right) \,.
\end{eqnarray} 

Finally we have to decide how to deal with $\csw$. We insert 
the one-loop expression~\cite{naik,impropert,cswpert}
\begin{equation} 
\csw = 1 + 0.268588 g^2 + O(g^4) 
\end{equation}
in the above expansions and re-expand the result in the coupling constant.
This means that we set $\csw = 1$ in the one-loop coefficients and
the two-loop coefficients get additional contributions proportional to
the one-loop coefficient in the expansion of $\csw$.

\section{Fit details}
\label{sec.fits}
 
In this Appendix we give the details of the fits applied in the
cases where perturbatively subtracted data are available.

Working with the $\ell$-loop approximation of the conversion factor
$Z_{\RI}^{\mathcal S} (M)$ we write generically
\begin{equation} 
Z_{\RI}^{\mathcal S} (M) = 1 
 + \sum_{i=1}^{\ell} c_i^{\mathcal S \mathcal S^\prime} 
  \left( \frac{g^{\mathcal S^\prime} (M)^2}{16 \pi^2} \right)^i
 + R^c_\ell (M)^{\mathcal S \mathcal S^\prime} 
= Z^c_\ell (M)^{\mathcal S \mathcal S^\prime} 
  + R^c_\ell (M)^{\mathcal S \mathcal S^\prime} 
\end{equation}
with the remainder $R^c_\ell (M)^{\mathcal S \mathcal S^\prime} 
       = O ( g^{\mathcal S^\prime}(M)^{2\ell + 2})$.
The scheme $\mathcal S^\prime$ chosen for the coupling in which
$Z_{\RI}^{\mathcal S}$ is expanded could be the $\MS$ scheme
as in Eqs.~(\ref{convri}), (\ref{convmom}), (\ref{convmom2}), but
another option would be the $\momt$ scheme. 

Unfortunately, in many cases the perturbative expansion of 
$Z_{\RI}^{\mathcal S}$ is not very well convergent, in particular
for $\mathcal S = \mathcal S^\prime = \MS$. However, if one chooses 
$\mathcal S = \MOM$, $Z^c_3 $ turns out to be equal to 1 for 
$\cO^S$, $\cO^P$, $\cO^T_{\mu \nu}$, $\cO_{h_{1,a}}$, $\cO_{h_{1,b}}$, 
$\cO_{h_{2,a}}$, $\cO_{h_{2,b}}$, $\cO_{h_{2,c}}$, $\cO_{h_{2,d}}$.
Generally, the use of the $\momt$ coupling seems to improve the
convergence for $\mathcal S = \MS$ as well as for $\mathcal S = \MOM$.
Working with $\mathcal S = \MOM$ instead of $\mathcal S = \MS$
seems to have the additional advantage that at least some of the 
effects of $Z_{\RI}^{\mathcal S}$ are shifted to the factor 
$\Delta Z ^{\mathcal S}$ via the anomalous dimension.
In $\Delta Z ^{\mathcal S}$ we can then exploit renormalization 
group improvement.

Similarly, using the $n$-loop approximation of the $\beta$ function
and the anomalous dimension we express $\Delta Z ^{\mathcal S} (M)$ 
as
\begin{eqnarray} 
\Delta Z ^{\mathcal S} (M) & = & 
  \left( 2 \beta_0 \frac{g^{\mathcal S} (M)^2}{16 \pi^2}
                          \right) ^{-\frac{\gamma_0}{2 \beta_0}}
  \exp \left \{ - \int_0^{g^{\mathcal S} (M)^2 / 16 \pi^2} \! \mathrm d x
   \; \frac{ \sum_{i=0}^{n-2} \left( \beta_0 \gamma^{\mathcal S}_{i+1}  
                     - \gamma_0 \beta^{\mathcal S}_{i+1} \right) x^i}
       {2 \beta_0 \sum_{i=0}^{n-1} \beta^{\mathcal S}_i x^i} 
                                        + R_n^{\mathcal S} (M) \right \} 
\nonumber \\
  & = & \Delta_n^{\mathcal S} (M) \, \mathrm e ^{R_n^{\mathcal S} (M)} 
\end{eqnarray}
with $R_n^{\mathcal S} (M) = O (g^{\mathcal S}(M)^{2n})$. 
From the $\MS$ anomalous dimension one can compute the anomalous
dimension in the scheme $\mathcal S$ to $n$ loops, provided
the conversion factor $Z_{\mathcal S}^{\MS}$ (and the $\beta$ 
function) is known to $n-1$ loops, see Eqs.~(\ref{conv1}) -- (\ref{conv3})
for $\mathcal S = \MOM$. 

With the help of the above representations of $Z_{\RI}^{\mathcal S}$
and $\Delta Z ^{\mathcal S}$ we get from Eq.~(\ref{factor})
\begin{equation} 
\left( Z^c_\ell (\mu_p)^{\mathcal S \mathcal S^\prime} 
       + R^c_\ell (\mu_p)^{\mathcal S \mathcal S^\prime} \right) 
Z_{\mathrm {bare}}^{\RI} (\mu_p,a) = 
Z^{\mathrm {RGI}}(a) \Delta_n^{\mathcal S} (\mu_p)^{-1} \, 
                                   \mathrm e ^{- R_n^{\mathcal S} (\mu_p)} \,.
\end{equation}
In this relation as well as in Eq.~(\ref{factor}) lattice artefacts
vanishing like a power of $a$ have been neglected. For larger values
of $\mu_p$ this is not justified any more, even after perturbative subtraction
of lattice artefacts. Therefore we write
\begin{equation} 
Z_{\mathrm {bare}}^{\RI} (\mu_p,a) = 
Z_{\mathrm {bare}}^{\RI} (\mu_p,a)_{\mathrm {MC}} - A(a^2 \mu_p^2)
\end{equation}
subtracting the (remaining) lattice artefacts $A$ from the Monte Carlo data
$Z_{\mathrm {bare}}^{\RI} (\mu_p,a)_{\mathrm {MC}}$. Of course, $A$
could be much more complicated than a simple function of $a^2 \mu_p^2$,
but in the end we have to restrict ourselves to a polynomial in
$a^2 \mu_p^2$ anyway. So we use this simplified expression already 
here for notational convenience and obtain
\begin{equation} 
Z^c_\ell (\mu_p)^{\mathcal S \mathcal S^\prime} 
        Z_{\mathrm {bare}}^{\RI} (\mu_p,a)_{\mathrm {MC}}
= \frac{Z^{\mathrm {RGI}}(a) \Delta_n^{\mathcal S} (\mu_p)^{-1} \, 
        \mathrm e ^{- R_n^{\mathcal S} (\mu_p)}}
{1 + R^c_\ell (\mu_p)^{\mathcal S \mathcal S^\prime} 
           /Z^c_\ell (\mu_p)^{\mathcal S \mathcal S^\prime}}
+ Z^c_\ell (\mu_p)^{\mathcal S \mathcal S^\prime} A(a^2 \mu_p^2) \,.
\end{equation}
Given that 
$Z^c_\ell (\mu_p)^{\mathcal S \mathcal S^\prime} = 
                                 1 + O (g^{\mathcal S}(\mu_p)^2)$
we approximate this relation by 
\begin{equation} \label{fitfunction}
Z^c_\ell (\mu_p)^{\mathcal S \mathcal S^\prime} 
         Z_{\mathrm {bare}}^{\RI} (\mu_p,a)_{\mathrm {MC}}
= \frac{Z^{\mathrm {RGI}}(a) \Delta_n^{\mathcal S} (\mu_p)^{-1} \, 
        \mathrm e ^{- R_n^{\mathcal S} (\mu_p)}}
{1 + R^c_\ell (\mu_p)^{\mathcal S \mathcal S^\prime}}  + A(a^2 \mu_p^2) \,.
\end{equation}
As we shall parameterize 
$\left( R^c_\ell \right)^{\mathcal S \mathcal S^\prime}$, 
$R_n^{\mathcal S}$ and $A$ in the following
and fit the corresponding parameters, the above approximation 
should not be problematic. On the left-hand side we have our (possibly 
subtracted) Monte Carlo results for the renormalization factors, 
extrapolated to the chiral limit and converted to the intermediate 
scheme $\mathcal S$ using the $\ell$-loop approximation.
These numbers are fitted with the expression on the 
right-hand side, where the values of $Z^{\mathrm {RGI}}(a)$ 
at our four values of $a$ are the desired numbers. 

More precisely, we set 
\begin{eqnarray}
\label{corrn}
R_n^{\mathcal S} (\mu_p) & = & f_1 \, g^{\mathcal S}(\mu_p)^{2n} 
                     + f_2 \, g^{\mathcal S}(\mu_p)^{2n + 2} + \cdots \,,   \\
\label{corrd}
R^c_\ell (\mu_p)^{\mathcal S \mathcal S^\prime} 
       & = & b_1 \, g^{\mathcal S^\prime}(\mu_p)^{2\ell + 2}
                          + b_2 \, g^{\mathcal S^\prime}(\mu_p)^{2\ell + 4}
                          + \cdots \,,   \\
\label{corrarte}
A(a^2 \mu_p^2) & = & g_1 a^2 \mu_p^2 + g_2 (a^2 \mu_p^2)^2 + \cdots \,,
\end{eqnarray}
where $b_1$, $f_1$, $\ldots$ are the (fit) parameters. Remember that in 
Sec.~\ref{sec.method} we have argued that $O(a)$ lattice artefacts
are absent. 
In principle, the coefficients $g_1$, $g_2$, {\ldots} could depend on
the coupling, i.e., on $\beta$. However, in the range of couplings
we have at our disposal the variation of powers of $a$ is much larger
than the possible variation (logarithmic in $a$) of the coefficients.
Therefore it seems justified to neglect this dependence and to treat
$g_1$, $g_2$, {\ldots} as constants. This might also help in 
disentangling lattice artefacts from truncation errors.

For the operators considered here, the $\MS$ anomalous dimension is
known to three loops, in some cases even to four loops, see 
Appendix~\ref{sec.coeff}. Upon combination with the four-loop $\beta$
function one can thus reach at least $n=3$. The conversion factors
$Z_{\RI}^{\MS}$ and $Z_{\MOM}^{\MS}$, on the other hand, are known 
to three loops in all cases.

There are quite a few parameters that can be varied in the analysis: 
\begin{enumerate}
\item
the intermediate scheme $\mathcal S$,
\item
the scheme $\mathcal S^\prime$ chosen for the coupling in which 
$Z_{\RI}^{\mathcal S}$ is expanded,
\item
the orders of the perturbative expansions used in (\ref{fitfunction}), 
i.e., the numbers $n$ and $\ell$,
\item
the order $O_\beta$ of the perturbative expansion of the $\beta$ 
function inserted in (\ref{alphacalc}) when computing the running coupling 
$g^{\mathcal S} (M)$,
\item
the number of terms $N_1$, $N_2$, $N_a$ taken into account in the 
correction terms (\ref{corrn}), (\ref{corrd}), (\ref{corrarte}), 
respectively, 
\item
the fit interval.
\end{enumerate}

Ideally, the results should be independent of all these choices.
Moreover, one would expect that a significant deviation from the 
``continuum limit'' curve (obtained by setting $A=0$) appears 
only for $\mu_p$ values where the data show lattice artefacts,
e.g., in the form of a violation of the scaling property (\ref{scal}).
Unfortunately, the fits of unsubtracted data do not follow this 
expectation, and this is the main reason why we consider the 
corresponding results as unreliable and do not apply
our fit procedure to these data.

Our final choices are motivated by the following observations.
The plateaus in $Z^{\mathrm {RGI}}$ look better for the
choice $\mathcal S = \MOM$ than for $\mathcal S = \MS$. This may
be due to the above mentioned fact that the perturbative expansion
of $Z_{\RI}^{\mathcal S}$ seems to be better behaved for $\mathcal S = \MOM$. 
For the scheme $\mathcal S^\prime$ the choice
$\mathcal S^\prime = \momt$ seems to be favorable. For the operator
$\overline{\cO}_{h_{1,a}}$ a comparison between 
$\mathcal S = \mathcal S' = \MS$ and 
$\mathcal S = \MOM$, $\mathcal S' = \momt$ is shown in 
Fig.~\ref{fig.2scheme}.

Not surprisingly, the maximal values for $n$ and $\ell$ lead to
the best plateaus. For the number of terms taken into account
in (\ref{corrn}), (\ref{corrd}), and (\ref{corrarte}) only 0 and
1 are reasonable choices. 

As already mentioned, the perturbative behavior, i.e., the 
plateau starts only at rather large values of the scale, typically 
around $\mu_p^2 \approx 5 \, \mbox{GeV}^2$.
So the lower limit $\mu^2_{\mathrm {min}}$ of the fit interval
should be at least $5 \, \mbox{GeV}^2$. However, the precise
value does not seem to be too crucial, both 
$\mu^2_{\mathrm {min}} = 5 \, \mbox{GeV}^2$ and
$\mu^2_{\mathrm {min}} = 10 \, \mbox{GeV}^2$ look reasonable.

So we arrive at the following choices. In the expansions 
originating from continuum perturbation theory which enter the fit 
formulae we use as many terms as are
available, i.e., we take for $n$ and $\ell$ the largest values possible.
The same applies to $O_\beta$, so we set $O_\beta = 4$.
We choose $\mathcal S = \MOM$ and 
$\mathcal S^\prime = \momt$. All data for 
$\mu_p^2 \geq 10 \, \mbox{GeV}^2$ are included in the fit. 
It is clear that the perturbative corrections (\ref{corrn}) 
and (\ref{corrd}) are hard to distinguish when inserted in 
(\ref{fitfunction}) and it does not make much sense to include both of them.
So we set $N_1=0$ and $N_2=1$. Furthermore we choose $N_a=1$.
Hence we end up with six fit parameters: the four values 
$Z^{\mathrm {RGI}}(a)$ along with the coefficients $b_1$ and $g_1$.
The data points are weighted by their statistical errors
although the deviations from the fit curves are mostly of systematic 
origin. 

\section{Going from RGI results to values in the $\MS$ scheme}
\label{sec.conv}

In this Appendix we collect the factors by which one has to multiply
$Z^{\mathrm {RGI}}$ in order to obtain the corresponding number
in the $\MS$ scheme. They are given in Tables~\ref{tab.conv1}, 
\ref{tab.conv2} and \ref{tab.conv3} for various values of $r_0$ 
and $r_0 \Lambda_\MS $. 

\begin{table}[h]
\caption{Factors for converting $Z^{\mathrm {RGI}}$ to the $\MS$ scheme
obtained with $r_0 = 0.467 \, \mbox{fm}$ and $r_0 \Lambda_\MS = 0.617$.}
\label{tab.conv1}
\begin{ruledtabular}
\begin{tabular}{lcc}
Op. & $\mu^2 = 4 \, \mbox{GeV}^2$ & $\mu^2 = 5 \, \mbox{GeV}^2$ \\
\hline
$\cO^S$, $\cO^P$        
                                      & 1.40701 & 1.44044 \\
$\cO^T_{\mu \nu}$    
                                      & 0.91926 & 0.91089 \\
$Z_q$   
                                      & 1.04534 & 1.04291 \\
$\cO_{v_{2,a}}$, $\cO_{v_{2,b}}$, $\cO_{r_{2,a}}$, $\cO_{r_{2,b}}$
                                      & 0.71544 & 0.70183 \\
$\cO_{h_{1,a}}$, $\cO_{h_{1,b}}$ 
                                      & 0.69538 & 0.68009 \\
$\cO_{v_3}$, $\cO_{v_{3,a}}$, $\cO_{a_2}$, $\cO_{r_3}$      
                                      & 0.58648 & 0.56943 \\
$\cO_{h_{2,a}}$, $\cO_{h_{2,b}}$, $\cO_{h_{2,c}}$, $\cO_{h_{2,d}}$ 
                                      & 0.57878 & 0.56107 \\
$\cO_{v_4}$ 
                                      & 0.50844 & 0.49008 \\
\end{tabular}
\end{ruledtabular}
\end{table}

\begin{table}[ht]
\caption{Factors for converting $Z^{\mathrm {RGI}}$ to the $\MS$ scheme
obtained with $r_0 = 0.467 \, \mbox{fm}$ and $r_0 \Lambda_\MS = 0.662$.}
\label{tab.conv2}
\begin{ruledtabular}
\begin{tabular}{lcc}
Op. & $\mu^2 = 4 \, \mbox{GeV}^2$ & $\mu^2 = 5 \, \mbox{GeV}^2$ \\
\hline
$\cO^S$, $\cO^P$        
                                      & 1.38514 & 1.41952  \\
$\cO^T_{\mu \nu}$    
                                      & 0.92491 & 0.91609  \\
$Z_q$   
                                      & 1.04704 & 1.04441  \\
$\cO_{v_{2,a}}$, $\cO_{v_{2,b}}$, $\cO_{r_{2,a}}$, $\cO_{r_{2,b}}$
                                      & 0.72464 & 0.71029  \\
$\cO_{h_{1,a}}$, $\cO_{h_{1,b}}$ 
                                      & 0.70575 & 0.68959  \\
$\cO_{v_3}$, $\cO_{v_{3,a}}$, $\cO_{a_2}$, $\cO_{r_3}$      
                                      & 0.59809 & 0.58001  \\
$\cO_{h_{2,a}}$, $\cO_{h_{2,b}}$, $\cO_{h_{2,c}}$, $\cO_{h_{2,d}}$ 
                                      & 0.59085 & 0.57205  \\
$\cO_{v_4}$ 
                                      & 0.52100 & 0.50145  \\
\end{tabular}
\end{ruledtabular}
\end{table}

\begin{table}[ht]
\caption{Factors for converting $Z^{\mathrm {RGI}}$ to the $\MS$ scheme
obtained with $r_0 = 0.5 \, \mbox{fm}$ and $r_0 \Lambda_\MS = 0.617$.}
\label{tab.conv3}
\begin{ruledtabular}
\begin{tabular}{lcc}
Op. & $\mu^2 = 4 \, \mbox{GeV}^2$ & $\mu^2 = 5 \, \mbox{GeV}^2$ \\
\hline
$\cO^S$, $\cO^P$                                                               
                                      & 1.42764 & 1.46022 \\
$\cO^T_{\mu \nu}$                                                              
                                      & 0.91406 & 0.90608 \\
$Z_q$                                                                          
                                      & 1.04382 & 1.04155 \\
$\cO_{v_{2,a}}$, $\cO_{v_{2,b}}$, $\cO_{r_{2,a}}$, $\cO_{r_{2,b}}$
                                      & 0.70698 & 0.69402 \\
$\cO_{h_{1,a}}$, $\cO_{h_{1,b}}$                                               
                                      & 0.68587 & 0.67134 \\
$\cO_{v_3}$, $\cO_{v_{3,a}}$, $\cO_{a_2}$, $\cO_{r_3}$                         
                                      & 0.57587 & 0.55972 \\
$\cO_{h_{2,a}}$, $\cO_{h_{2,b}}$, $\cO_{h_{2,c}}$, $\cO_{h_{2,d}}$             
                                      & 0.56775 & 0.55100 \\
$\cO_{v_4}$                                                                    
                                      & 0.49699 & 0.47969 \\
\end{tabular}
\end{ruledtabular}
\end{table}


\end{document}